\documentclass[11pt]{article}
\usepackage{graphicx}
\usepackage{apacite}
\usepackage{natbib}
\setcitestyle{authoryear,round}
\usepackage{color}
\usepackage[dvipsnames]{xcolor}
\usepackage{epsfig,lscape}
\usepackage{amssymb,amsfonts,amsmath,amsthm}
\usepackage{rotating}
\usepackage{bm}
\usepackage{bbm}
\usepackage{hyperref}
\usepackage{mathtools}
\usepackage{physics}
\usepackage{float}

\usepackage{threeparttable}
\usepackage{booktabs}
\usepackage{subcaption}
\usepackage{tikz}
\usepackage{amssymb}
\usepackage{longtable}

\newcommand{\indep}{\perp \!\!\! \perp}

\usepackage[left=1in,right=1in,top=.9in,bottom=1.1in]{geometry}

\def\su{\mathcal U}
\def\sl{\mathcal L}
\def\st{\mathcal T}

\def\sr{\mathbb{R}}

\def\E{\mathbb{E}}
\def\P{\mathbb{P}}

\def\te{\Tilde{\mathbb{E}}}

\def\me{\mathrm e}

\def\dif{\mathrm d}

\def\N{\mathrm{N}}

\def\T{ {\mathrm{\scriptscriptstyle T}} }
\def\psit{\psi(\bar{\alpha}^{\T}\bm{G})}
\def\psie{\psi(\hat{\alpha}^{\T}\bm{G})}
\def\psin{\psi(\alpha^{\T}\bm{G})}

\def\mt{\psi(\beta^{*\T}\bm{Z})}
\def\me{\psi(\hat{\beta}^{\T}\bm{Z})}

\def\m{\psi({\beta}^{\T}\bm{Z})}

\def\pt{\pi(\bm{X};\bar{\gamma})}
\def\pe{\pi(\bm{X};\hat{\gamma})}
\def\p{\pi(\bm{X};\gamma)}

\def\wt{w(\bm{X};\bar{\gamma})}

\def\we{w(\bm{X};\hat{\gamma})}

\def\ta{\bar{\alpha}}
\def\tb{\beta^{*}}
\def\tg{\bar{\gamma}}

\def\esa{\hat{\alpha}}
\def\esb{\hat{\beta}}

\def\esg{\hat{\gamma}}

\def\argmin{\mathrm{argmin}}

\newenvironment{prf}
{\noindent \textbf{Proof.}}{\hfill $\Box$ \vspace{.1in}}

\newtheorem{thm}{Theorem}
\newtheorem{lem}{Lemma}
\newtheorem{pro}{Proposition}

\newtheorem{ass}{Assumption}

\theoremstyle{definition}

\theoremstyle{definition}
\newtheorem{rem}{Remark}

\allowdisplaybreaks
\begin{document}

\begin{titlepage}
    \begin{center}
        {\Large Semi-supervised Regression Analysis with \\
Model Misspecification and High-dimensional Data}

        \vspace{.15in} Ye Tian,\footnotemark[1] Peng Wu,\footnotemark[2] and Zhiqiang Tan\footnotemark[3]
        \footnotetext[1]{Department of Statistics, Rutgers University,
    Piscataway, NJ 08854, USA (E-mail yt334@stat.rutgers.edu).}

    \footnotetext[2]{Department of Applied Statistics, Beijing Technology and Business University, Beijing, 102488, China\\ \indent \indent(E-mail: pengwu@btbu.edu.cn).}

    \footnotetext[3]{Department of Statistics, Rutgers University,
    Piscataway, NJ 08854, USA (E-mail: ztan@stat.rutgers.edu).}
        \vspace{.1in}
        \today
    \end{center}

    \paragraph{Abstract.}
    The accessibility of vast volumes of unlabeled data has sparked growing interest in semi-supervised learning (SSL) and covariate shift transfer learning (CSTL). In this paper, we present an inference framework for estimating regression coefficients in conditional mean models within both SSL and CSTL settings, while allowing for the misspecification of conditional mean models. We develop an augmented inverse probability weighted (AIPW) method, employing regularized calibrated estimators for both propensity score (PS) and outcome regression (OR) nuisance models, with PS and OR models being sequentially dependent.
We show that when the PS model is correctly specified, the proposed estimator achieves consistency, asymptotic normality, and valid confidence intervals, even with possible OR model misspecification and high-dimensional data. Moreover, by suppressing detailed technical choices, we demonstrate that previous methods can be unified within our AIPW framework. Our theoretical findings are verified through extensive simulation studies and a real-world data application.

\end{titlepage}

\section{Introduction}

In recent years, vast volumes of unlabeled data have become increasingly accessible, sparking growing interest in how to leverage these data in both academic research and industry applications. One of the active areas of research is semi-supervised learning (SSL).
In addition, covariate shift transfer learning (CSTL) also exploits information from unlabeled target data. Both have various application scenarios like computer vision~\citep{sohn2020fixmatch,NEURIPS2021_07ac7cd1,zheng2022simmatch}, natural language process~\citep{chen-huang-2016-semi,ruder-etal-2019-transfer, zhao-etal-2022-semi}, causal inference~\citep{alvari2019less, pmlr-v216-aloui23a,zhang2023semi}, healthcare data analysis~\citep{castro2020causality,liu2023,TANG2024110020}, etc.

In both the SSL and CSTL settings,  we have access to a labeled dataset $\sl$ and an unlabeled dataset $\su$, where the labeled dataset $\sl$ contains observations with both the covariates $\bm{X}$ and the outcome $Y$, while the unlabeled dataset $\su$ consists solely of observations with the covariates $\bm{X}$. The training set $\st$ is the union of $\sl$ and $\su$.
Nevertheless, there is a key distinction between the classical SSL and CSTL setups~\citep{semimit, liu2023}.
In CSTL, the conditional distributions of $Y$ given $\bm{X}$ in the labeled and unlabeled datasets 
 are assumed to be the same, whereas the marginal distributions of $\bm{X}$ are different (hence the term covariate shift), and the estimator is ultimately evaluated on unlabeled data.  However, under the classical SSL setup, it is assumed that the distributions of labeled data, unlabeled data, and population are the same, making no difference in evaluating the estimator on which distribution.
To accommodate both SSL and CSTL, we consider a more general setting. We only assume the conditional distribution of $Y$ given $\bm{X}$ is the same in labeled and unlabeled data, while marginal distributions of $\bm{X}$ are permitted to be different. Estimators evaluated on the population and unlabeled data are both considered.

While there is a long history of SSL~\citep{semimit, xiaojin2008semi} and CSTL~(\citeauthor{cstl},\citeyear{cstl}), a growing literature has considered inference procedures only recently. Notable advancements have been made in estimating the population mean $\E(Y)$~\citep{ANRU2019, zhang2021} and regression coefficients in (generalized) linear models~\citep{cha2016,cha2018,liu2023}. Inferences of quantile regression~\citep{cha2022semi}, explained variance~\citep{Tony2020}, and model performance metrics such as true and false positive rates~\citep{Gron2017} have also attracted interest.

In this article, we focus on the inference of regression coefficients in (conditional) mean models in SSL settings, hence called
semi-supervised regression analysis.
We demonstrate a unified framework for estimating and inferring these coefficients, particularly in cases where the (conditional) mean model and outcome regression (OR) model $\E[Y|\bm{X}]$ may be misspecified. Previous SSL and CSTL methods that  considered the same goal,  such as \citet{cha2016}, \citet{cha2018},
\citet{ANRU2019},
\citet{zhang2021}
and \citet{liu2023}, can largely be accommodated in the augmented inverse probability weighting (AIPW) framework~\citep{rubin1994, tan2020a, Wu2021}.
See Section \ref{sec:comp} and Section \ref{sec:ext} for further details.

Despite significant advancements made, there remain limitations in the previous AIPW methods.
Methods developed in SSL settings usually treat the problem as the one where data are missing completely at random (MCAR)~\citep{cha2016, cha2018, ANRU2019, zhang2021}.  
This restricts their application scenarios and overlooks the significance of constructing the propensity score (PS) model. In contrast, our setting is in general
a missing-at-random (MAR) problem~\citep{little2019statistical}, and the estimation of the PS model is no longer negligible.
In the setting of MCAR, the PS remains a constant, whereas in the setting of MAR, the PS varies with the covariates $\bm{X}$.
In MAR problems, similarly as in \citet{tan2020a}, the estimation of PS and OR models needs to be carefully handled in a way different from regularized least squares or maximum likelihood as in previous papers,
so that $\sqrt{N}$-consistent estimation can be achieved with possible misspecification of the OR model, where $N$ is the sample size of $\st$.

In summary, we mainly make the following two contributions.
First, we present an inference framework that consolidates several previous settings, including the estimation of population mean and regression coefficients in conditional mean models of $Y$ given any sub-vector of $\bm{X}$ under both SSL and CSTL setups.
Second, we propose a novel AIPW method that enables
$\sqrt{N}$-consistent and asymptotically normal estimation and achieves valid confidence intervals (CIs),
when the PS model is correctly specified but the OR model may be misspecified. This robustness to model misspecification is achieved by carefully exploiting the connection between PS and OR models and designing estimating equations for nuisance parameters, differently from regularized least-squares or maximum-likelihood estimation.
The desirable properties of our method, $\sqrt{N}$-consistency, asymptotical normality, and valid confidence intervals, are formally established in the setting of sparse high-dimensional PS and OR models, while allowing that the estimator of PS model has convergence rates slower than $1/\sqrt{N}$ and the OR model may be misspecified.

This work is also related to the causal inference problem under the strong ignorability assumption~\citep{Rosenbaum-Rubin-1983}.
Specifically, the SSL problem is similar to estimating the average treatment effect (ATE) and the conditional ATE (CATE)~\citep{wager2018estimation, Athey-Tibsirani-Wager-2019, zimmert2019, fan2022, Wu2021}. The CSTL problem can be viewed as an analog to the estimation of the average treatment effect on the treated (ATT).

The rest of this paper is organized as follows. In section \ref{sec:setup}, we present our setup and define the target parameters of interest. In Section \ref{sec:method}, we construct a novel AIPW estimator for the target parameter under the SSL setting.
We show theoretical properties of the proposed estimator in Section \ref{sec:analysis}, and compare them with the previous literature in Section \ref{sec:comp}.
Simulation results are shown in Section \ref{sec:simulation}, and an application to crime study is presented in Section \ref{sec-application}.
Extension of proposed methods to the CSTL setting is given in Section \ref{sec:ext} followed by concluding discussions in Section \ref{sec:summary}. Proofs of all theoretical results and associated
technical materials are presented in the Supplementary Material.

\section{Setup and preliminaries}\label{sec:setup}

\subsection{Data and target parameters}\label{sec:dtp}
Let $Y \in \mathbb{R}$ be a response variable and $\bm{X} =(1, X_1, \ldots, X_d)^{\mathrm{\scriptscriptstyle T}} \in \mathbb{R}^{d + 1}$
be a covariate vector with the first element being the constant 1.
In addition, let $R \in \{0, 1\}$ be the indicator of whether $Y$ is observed:
$R=1$ if observed and $R = 0$ if missing.
Assume that $\{(\bm{X}_{i},  Y_{i}, R_{i}): i = 1, \ldots, N\}$ is an independent and identically distributed (i.i.d.) sample from
a joint distribution of $(\bm{X}, Y, R)$, denoted as $\mathbb{P}$. The observed dataset,
$\{(\bm{X}_{i},  R_i Y_{i}, R_{i}): i = 1, \ldots, N\}$, can be split into a labeled dataset and an unlabeled dataset as follows:
\begin{align*}
    \sl =  \{ (\bm{X}_i,  Y_i, R_i = 1), ~ i = 1, \ldots, n\},  \quad
    \su = \{ (\bm{X}_i, R_i = 0), i = n+1, \ldots, N\}.
\end{align*}

For $\bm{Z}$ a sub-vector of $\bm{X}$, it is of interest to fit a regression model for the conditional mean $\E(Y|\bm{Z})$:
\begin{equation}\label{eq:def-m}
 \E(Y| \bm{Z})  = \psi(\beta^{*\mathrm{\scriptscriptstyle T}} \bm{Z}),
\end{equation}
where $\E(\cdot)$ denotes the expectation under $\P$,
$\beta^*$ is a parameter vector, and  $\psi(\cdot)$ is an (increasing) inverse link function, such as the identity function $\psi(u) = u$ and logit function $\psi(u) = 1/\{1+\exp(-u)\}$.
The model \eqref{eq:def-m} is allowed to be misspecified, that is, $\E(Y| \bm{Z})$ may not be put in the form $ \psi(\beta^{\mathrm{\scriptscriptstyle T}} \bm{Z})$.
With possible model misspecification, $\beta^*$ is defined as the solution to the estimating equations:
\begin{equation} \label{eq:def-tb}
  \E \left [ \left \{Y - \psi(\beta^{\mathrm{\scriptscriptstyle T}}\bm{Z}) \right \} \bm{Z} \right ]=0.
\end{equation}
For a generalized linear model with $\psi(\cdot)$ as the canonical inverse link, the estimating equation \eqref{eq:def-tb}
leads to maximum likelihood estimation, so that
$\psi(\beta^{*\mathrm{\scriptscriptstyle T}} \bm{Z})$ can be interpreted as the best likelihood-based approximation to $\E(Y| \bm{Z})$ using model \eqref{eq:def-m}.  

The regression model \eqref{eq:def-m} is flexible.
The target parameter $\beta^*$ accommodates a variety of estimands in the previous literature.
\begin{itemize}
    \item[(a)] If we set $\bm{Z} = 1$ and $\psi(u) = u$, then $\beta^*= \E(Y)$.
     The problem corresponds to the semi-supervised estimation of the population mean~\citep{ANRU2019, zhang2021}.

    \item[(b)]  If we set $\bm{Z}$ to be a univariate covariate, for example, $\bm{Z}=X_1$, then $\beta^*$ corresponds to the regression coefficient in the
    regression model of $Y$ given the particular covariate $X_1$. This problem was studied by
    \citeauthor{liu2023}~(\citeyear{liu2023}) and \citeauthor{Wu2021}~(\citeyear{Wu2021}).

    \item[(c)] If we set $\bm{Z} = \bm{X}$, then $\beta^*$ corresponds to the coefficient vector in the regression model
    of $Y$ given the full covariate vector $\bm{X}$~\citep{cha2016, cha2018}.
\end{itemize}

In addition to the parameter vector $\beta^*$, it may also be of interest to consider target parameters defined within unlabeled data
corresponding to the CSTL setting, as studied in several recent papers~\citep{liu2023, he2024}.
To incorporate this setting, we consider a regression model for the conditional mean in the unlabeled data:
\begin{equation} \label{eq:def-m-0}
 \E(Y| R=0, \bm{Z})  = \psi(\beta^{0*\mathrm{\scriptscriptstyle T}} \bm{Z}).
\end{equation}
With possible misspecification of  model \eqref{eq:def-m-0}, the parameter $\beta^{0*}$ is defined as the solution to
\begin{equation}\label{eq:def:b0}
  \quad \E \left [ (1-R) \left \{Y - \psi(\beta^{\mathrm{\scriptscriptstyle T}}\bm{Z}) \right \} \bm{Z} \right ]=0.
\end{equation}
To illustrate the main ideas, we focus on the estimation of $\beta^*$, and defer the associated results for the estimation of $\beta^{0*}$ to Section \ref{sec:ext}.

\subsection{General assumptions}
Without imposing any assumption, we cannot obtain a consistent estimator of $\beta^*$ due to the missingness of $Y$ in the unlabeled data.
Below, we introduce the identifiability assumption.

\begin{ass} \label{ass:model}
$Y \indep R \mid \bm{X}$, i.e., $Y$ and $R$ are conditionally independent given $\bm{X}$.
\end{ass}

Assumption \ref{ass:model} is crucial for the identifiability of $\beta^*$ and $\beta^{0*}$.  It ensures $\E(Y|\bm{X}, R=1) = \E(Y|\bm{X}, R=0)$, indicating that the conditional mean of $Y$ is the same for both unlabeled and labeled data after accounting for the full covariates $\bm{X}$. This establishes the connection between the labeled and unlabeled data.
Moreover, Assumption \ref{ass:model} implies that $\P(R=1|\bm{X}, Y) = \P(R=1|\bm{X})$, meaning that the label indicator $R$ depends solely on the covariates $\bm{X}$, i.e., $R$ is missing at random~\citep{Molenberghs-etal2015, Imbens-Rubin2015}.
It should be noted that Assumption \ref{ass:model} does not imply $Y \indep R \mid \bm{Z}$ when $\bm{Z} \neq \bm{X}$. If $Y \indep R \mid \bm{Z}$ does hold and models (\ref{eq:def-m}) and (\ref{eq:def-m-0}) are correctly specified, then $\beta^* = \beta^{0*}$.
Otherwise, $\beta^*$ and $\beta^{0*}$ may differ from each other. 


Despite bearing many similarities, Assumption \ref{ass:model} differs from the classic SSL setup in the missing mechanism
as represented by the probabilistic behavior of $R$. SSL assumes the missing-completely-at-random mechanism (MCAR)~\citep{cha2016,cha2018,ANRU2019, zhang2021},
that is, $R \indep (\bm{X}, Y)$ and thus $\P(R=1|\bm{X})$ is a constant, independent of $\bm{X}$.
In contrast, we allow $R$ to probabilistically depend on the covariates $\bm{X}$. In addition, we make the following technical assumption.

\begin{ass} \label{assump2}
    $\pi^*(\bm{X})= \P(R=1 | \bm{X}) > 0$.  
\end{ass}

This condition is introduced to ensure that each unit has a positive probability of belonging to $\mathcal{L}$.
Then the labeled dataset $\mathcal{L}$ is of a non-negligible size compared with $N$.

By Assumption~\ref{assump2}, as $N \to \infty$, the ratio $n/N$ may randomly fluctuate
but converges to a constant within the interval $(0,1)$, equal to $\E(R)=\pi^*(\bm{X})$.
This distinguishes our sampling process from the stratified sampling process widely used in the previous literature~\citep{cha2016, cha2018,ANRU2019,zhang2021}, where the sizes of labeled and unlabeled datasets,
$n$ and $N-n$, are deterministic.
For the asymptotic analysis, they assume that both $n$ and $N$ tend to infinity
such that $n/N$ converges to a constant in $[0,1)$, including zero. For details, see the discussion in Section \ref{sec:comp}.

\subsection{AIPW estimating equations} \label{sec:aipw-construction}

For estimating $\beta^*$, we introduce the augmented inverse probability weighting (AIPW) estimating equations.
Essentially the same AIPW estimating equation has been used in the previous literature, albeit under somewhat different settings than ours. See Section \ref{sec:comp} for a connection and comparison of our method with the previous methods.


With the true PS $\pi^{*}(\bm{X}) = \P(R=1|\bm{X})$, then under Assumption \ref{ass:model}, we have
\[   \E \left [ \frac{R}{\pi^*(\bm{X})} \left \{Y - \psi(\beta^{\mathrm{\scriptscriptstyle T}}\bm{Z}) \right \} \bm{Z} \right ] = \E \left [ \left \{Y - \psi(\beta^{\mathrm{\scriptscriptstyle T}}\bm{Z}) \right \} \bm{Z} \right ].   \]
Then a sample estimating equation for $\beta^*$ is
\begin{align}\label{eq:def-wesb}
    \tilde \E \left [\frac{R}{\hat{\pi}(\bm{X})} \{ Y - \m \} \bm{Z} \right ] = 0,
\end{align}
where $\hat{\pi}(\bm{X})$ is an estimator of $\pi^*(\bm{X})$ and $\tilde \E$ denotes the sample mean, defined as $\tilde \E(U) = N^{-1}\sum_{i=1}^N $
 
\noindent $U_i$ for a variable $U$.

Let $\hat{\beta}_{\mathrm{IPW}}$ be a solution to equations \eqref{eq:def-wesb}.
If the PS model is correctly specified, then under certain regularity conditions, $\hat{\pi}(\bm{X}) \xrightarrow{\P}\pi^{*}(\bm{X})$ and $\hat{\beta}_{\mathrm{IPW}} \xrightarrow{\P}\beta^{*}$. However, if the PS model is misspecified, then $\hat{\pi}(\bm{X}) \not\xrightarrow{\P}\pi^{*}(\bm{X})$ and $\hat{\beta}_{\mathrm{IPW}} \not \xrightarrow{\P}\beta^{*}$.
To mitigate the possible inconsistency of $\hat{\beta}_{\mathrm{IPW}}$, the AIPW method introduces an augmented term. Specifically, let
$\phi^{*}(X) = \E[Y|\bm{X}]$ be the true
OR function and $\hat{\phi}(\bm{X})$ be a corresponding estimator. Then the AIPW estimating equation is
\begin{align}\label{eq:def-emp-AIPW-p}
\tilde \E \left [\frac{R}{\hat{\pi}(\bm{X})} \left \{Y - \m\right \} \bm{Z}  + \left \{1 - \frac{R}{\hat{\pi}(\bm{X})} \right \}\left \{\hat{\phi}(\bm{X}) - \m \right\} \bm{Z} \right ]= 0.
\end{align}
Let $\hat{\beta}_{\mathrm{AIPW}}$ be the solution to equation \eqref{eq:def-emp-AIPW-p}.
If the PS model is misspecified, the augmented term
 \begin{align*}
  \tilde \E \left [\left \{1 - \frac{R}{\hat{\pi}(\bm{X})}\right \}\left \{\hat{\phi}(\bm{X}) - \m \right\} \bm{Z} \right ]
 \end{align*}
corrects the bias of $\tilde \E \left [\frac{R}{\hat{\pi}(\bm{X})} \{ Y - \m \} \bm{Z} \right ]$ by introducing the estimator $\hat \phi(\bm{X})$. In addition, if the PS model is correctly specified, the augmented term improves the estimation efficiency of $\beta^*$ by leveraging the association between $\bm{X}$ and $Y$.
It can be shown that the left side of equation (\ref{eq:def-emp-AIPW-p}) converges  in probability to that of equation (\ref{eq:def-tb}), if either
$\hat{\pi}(\bm{X}) \xrightarrow{\P} \pi^{*}(\bm{X})$ or $\hat{\phi}(\bm{X}) \xrightarrow{\P} \phi^{*}(\bm{X)}$, which is the property of double robustness.

In the classic SSL setup, estimating $\beta^{*}$ is considered to be an MCAR problem, where $\hat{\pi}(\bm{X})$ is a constant, independent of $\bm{X}$,
and the estimator $\hat\phi(\bm{X})$ is usually defined using an OR model by
(unweighted) least squares, maximum likelihood, or variations~\citep{cha2016, cha2018, ANRU2019}.
However, our semi-supervised regression is formulated as a MAR problem, where $\hat{\pi}(\bm{X})$ depends on $\bm{X}$.
In such a scenario, as shown in the next section, the estimators $\hat{\pi}(\bm{X})$ and $\hat{\phi}(\bm{X})$
for the PS and OR functions can be defined in a sequential manner, different from least squares or maximum likelihood,
in order to obtain desirable properties with possible model misspecification.

\section{Method}\label{sec:method}

We develop a novel AIPW method that achieves $\sqrt{N}$-consistency in the setting of sparse high-dimensional PS and OR models, even if the estimation of the PS model exhibits convergence rates slower than $1/\sqrt{N}$ and the OR model is misspecified.

\subsection{Model specification for nuisance parameters}\label{sec:cho-nui}

AIPW estimation based on the estimating equation \eqref{eq:def-emp-AIPW-p} requires
constructing the estimators $\hat{\pi}(\bm{X})$ and $\hat{\phi}(\bm{X})$ for $\pi^*(\bm{X})$ and $\phi^*(\bm{X})$, using some PS and OR models.
In contrast with the previous literature, we introduce a dependency between $\hat{\pi}(\bm{X})$ and $\hat{\phi}(\bm{X})$
by carefully specifying basis functions and incorporating weighted estimation.

Specifically, let $\bm{F(X)} = \{1, f_{1}(\bm{X}), \ldots, f_{p}(\bm{X}) \}^{\mathrm{\scriptscriptstyle T}}$ be a vector of known functions of $\bm{X}$. We allow $p$ to be high-dimensional, tending to infinity as $N$ increases.
As in~\citet{tan2020a},
we propose using logistic regression as a working model for the PS function $\pi^*(\bm{X})$,
\begin{equation}\label{eq:def-ps}
    \P(R=1|\bm{X}) = \pi(\bm{X};\gamma) = [1 + \exp\{-\gamma^{\mathrm{\scriptscriptstyle T}} \bm{F(X)}\}]^{-1},
\end{equation}
{where $\gamma$ is an unknown coefficient parameter.

\begin{rem} In several related works on classic SSL and stratified sampling setups, making efforts to estimate the PS may not be necessary.  Firstly, in the classic SSL setup, the true PS is a constant, leading to a constant PS model.
Secondly, in the stratified sampling setup, the proportion $n/N$ is fixed and known, which corresponds to a known PS function.
Thus, researchers concentrate on specifying OR models.
However, for our setup, both the PS model and the OR model must be carefully formulated to achieve desirable properties.
\end{rem}

Next, we turn to modeling the OR function $\phi^*(\bm{X})$. The working model for $\phi^*(\bm{X})$ is specified as
\begin{equation} \label{eq:def-or}
    \E[Y|\bm{X}]  =  \phi(\bm{X}; \alpha) = \psi\{\alpha^{\mathrm{\scriptscriptstyle T}} \bm{G(X)}\},
\end{equation}
where $\bm{G(X)} = \{1, g_{1}(\bm{X}), \ldots, g_{q}(\bm{X}) \}^{\mathrm{\scriptscriptstyle T}}$ be a vector of known functions of $\bm{X}$ and $q$ can be high-dimensional.
In contrast to the previous literature, to ensure valid inference even when the OR model is misspecified, we carefully specify a choice of $\bm{G(X)}$ as follows:
   \begin{equation} \label{eq-gx}  \bm{G(X)} = [ \bm{F(X)}^\mathrm{\scriptscriptstyle T}, \{\bm{Z} \otimes \bm{F(X)}\}^{\mathrm{\scriptscriptstyle T}} ]^\mathrm{\scriptscriptstyle T},  \end{equation}
where $\bm{Z} \otimes \bm{F(X)}$ consists of all interactions between $\bm{Z}$ and $\bm{F(X)}$
(i.e., all products of individual components from $\bm{Z}$ and $\bm{F(X)}$).
Equation \eqref{eq-gx} represents the minimal choice for $\bm{G(X)}$, and additional covariates can also be incorporated,
such as nonlinear terms of $\bm{Z}$ and $\bm{F(X)}$. Under sparsity conditions, these additional terms can be readily accommodated.

%


\begin{rem}\label{rem:tec-dif}
Under the classic SSL setups, various OR working models are used for estimating $\phi^*(\bm{X})$. For example, \citet{cha2016}  recommended using a partial linear OR model.
In addition,
\citet{zhang2021} proposed using flexible OR working models such as Lasso, elastic net, etc., provided that they meet certain rates of convergence.
 \end{rem}

\subsection{Estimation procedures}  \label{sec3-2}

The proposed method consists of the following three steps: (a) estimating the parameter $\gamma$ in the PS model~\eqref{eq:def-ps}; (b) estimating the parameter $\alpha$ in the OR model~\eqref{eq:def-or}; (c) estimating the target parameter $\beta$. Below, we give details for the three steps.

For estimating $\gamma$, rather than employing a regularized maximum likelihood estimator, we utilize a regularized calibrated estimator~\citep{tan2020b}, defined as
\begin{equation}\label{eq:def-reg-gamma-h}
\esg = \mathop{\argmin}_{\gamma \in \mathbb{R}^{p+1}} L_{\mathrm{RCAL}}(\gamma) = \mathop{\argmin}_{\gamma \in \mathbb{R}^{p+1}} \{\ell_{\mathrm{CAL}}(\gamma) + \lambda_{\gamma} \| \gamma_{1:p}\|_{1}\},
\end{equation}
where
  \[  \ell_{\mathrm{CAL}}(\gamma) =\tilde \E[ R \exp{-\gamma^\mathrm{\scriptscriptstyle T} \bm{F(X)}} + (1- R)\gamma^\mathrm{\scriptscriptstyle T} \bm{F(X)} ],  \]
 $\lambda_{\gamma}$ is a pre-specified tuning parameter, $||\cdot||_1$ denotes the $L_1$-norm, and for any vector $\nu$, $\nu_{i:j}$ is the sub-vector of $\nu$ consisting of its $i$--th to $j$--th elements (both ends included).

 For a possibly misspecified model $\p$, under suitable regularity conditions, $\hat \gamma$ converges in probability to its target value $\tg$ defined by
\begin{equation}
\tg = \mathop{\argmin}_{\gamma \in \mathbb{R}^{p+1}}  \E [ R \exp{-\gamma^\mathrm{\scriptscriptstyle T} \bm{F(X)}} + (1- R)\gamma^\mathrm{\scriptscriptstyle T} \bm{F(X)}  ].
\end{equation}

For estimating $\alpha$, we adopt a regularized weighted maximum likelihood estimator~\citep{tan2020a}, which is defined as
\begin{equation}\label{eq:def-reg-alp-est}
\esa = \mathop{\argmin}_{\alpha \in \mathbb{R}^{q+1}} L_{\mathrm{RWL}}(\alpha; \esg) =  \mathop{\argmin}_{\alpha \in \mathbb{R}^{q+1}} \{\ell_{\mathrm{WL}}(\alpha; \esg) + \lambda_{\alpha} \| \alpha_{1:q} \|_{1}\},
\end{equation}
where
$$\ell_{\mathrm{WL}}(\alpha; \esg) = \te ( R \we [ -Y \alpha^{\mathrm{\scriptscriptstyle T}}\bm{G(X)} +  \Psi\{\alpha^{\mathrm{\scriptscriptstyle T}} \bm{G(X)}\} ]),$$
  $w(X, \hat \gamma) = \{1 - \pi(\bm{X}, \hat \gamma)\}/\pi(\bm{X}, \hat \gamma)$, $\Psi(u) = \int_{0}^u \psi(t)dt$ is the antiderivative of $\psi$, $\lambda_{\alpha}$ is a pre-specified tuning parameter. Similarly to the target value $\bar \gamma$, we define the target value of $\alpha$ as follows 
\begin{equation}\label{eq:def-reg-alp-tar}
\ta = \mathop{\argmin}_{\alpha \in \mathbb{R}^{q+1}} \E \left( R \wt \left [-Y \alpha^{\mathrm{\scriptscriptstyle T}}\bm{G(X)} + \Psi\{\alpha^{\mathrm{\scriptscriptstyle T}}\bm{G(X)}\} \right ] \right ).
\end{equation}

Similarly as in \cite{tan2020a}, the
construction of the loss function $L_{\mathrm{RWL}}(\alpha; \esg)$ differs from the regularized maximum likelihood estimator in two aspects: (a) it utilizes a weight function $w(X, \hat \gamma)$ for each labeled observation; (b) the weight function depends on the fitted PS  $\pi(\bm{X}, \hat \gamma)$.
This design is crucial for the proposed method to achieve desirable properties with possible misspecification of the OR model.
A subtle difference from \cite{tan2020a} is that the OR basis functions $\bm{G}$ are here explicitly allowed to differ from the PS basis functions $\bm{F}$.}


After obtaining the estimators of $\gamma$ and $\alpha$, the proposed calibrated AIPW estimator of $\beta$, denoted as $\hat \beta$, is the solution to the following estimating equation
\begin{equation}\label{eq:rand}
\tilde \E \{\tau(\bm{O}, \esa, \beta, \esg)\} = 0,
\end{equation}
where $\bm{O} = (\bm{X}, \bm{Z}, Y,R)$ and
\begin{align}
\tau(\bm{O}, \alpha, \beta, \gamma) = & \frac{R}{\p} \{ Y - \m  \} \bm{Z} + \left \{1 - \frac{R}{\p}\right \} \left \{  \psin - \m \right \} \bm{Z} \nonumber\\
= &  \frac{R}{\p} \{ Y - \psin  \} \bm{Z} + \left \{  \psin - \m \right \} \bm{Z}
\end{align}
Our estimating equations \eqref{eq:rand} share the same form as the AIPW estimating equations \eqref{eq:def-emp-AIPW-p}.
However, there exists a crucial distinction between our method and previous related methods based on \eqref{eq:def-emp-AIPW-p}.
In previous methods, the PS and OR models are specified and fitted independently of each other, typically both by least squares, maximum likelihood, or variations.
In our method, the PS and OR models are specified and fitted in a sequentially dependent manner.
This design allows our estimator $\hat\beta$ to achieve $\sqrt{N}$-consistency in the presence of misspecified OR models.

\section{Theoretical analysis}\label{sec:analysis}

In this section, we present the theoretical analysis of the proposed estimator $\hat \beta$. In Section \ref{sec:pro_nui_par}, we 
examine the theoretical properties of estimators $\esg$ and $\esa$ in the PS and OR models.
Then, we study the asymptotic properties of the proposed estimator $\esb$ in Section \ref{sec:pro_aipw}.
Finally, in Section \ref{sec4-3}, we extend our analysis to the classical SSL setting (stratified sampling with constant PS).


\subsection{Properties of the estimators for nuisance parameters} \label{sec:pro_nui_par}

We analyze the theoretical properties of $\esg$ and $\esa$, which provide the basis for investigating properties of $\hat \beta$. For ease of presentation, we denote $\bm{F(X)}$ and $\bm{G(X)}$ as $\bm{F}$ and $\bm{G}$, respectively.

We first present the properties of $\esg$ based on \citet{tan2020a} [Theorem 1 and Theorem 3].
The following assumptions are taken from \citet{tan2020a}.

\begin{ass}[Regularity conditions for $\hat \gamma$] \label{ass:p}Suppose that the following conditions are satisfied:
\begin{itemize}
\item[$(\mathrm{i})$] $\max_{j = 0, \ldots, p} |f_j(\bm{X})| \leq C_{0} $ a.s. for a constant $C_{0} \geq 1$;

\item[$(\mathrm{ii})$]  $\tg^{\mathrm{\scriptscriptstyle T}} \bm{F} \geq B_{0}$ a.s. for a constant $B_{0} \in \mathbb{R}$, that is, $\pi(\bm{X}; \bar \gamma)$ is bounded from below by $\{1 + \exp(-B_{0})\}^{-1}$;

\item[$(\mathrm{iii})$] the compatibility condition holds for $\bm{\bm{\Sigma}}_{\tg}$ with the subset $S_{\tg} = \{0\} \cup \{j : \tg_{j} \neq 0, j =
1, \ldots , p\}$ and some constants $\nu_{0} > 0$ and $\xi_{0} > 1$, where $\bm{\bm{\Sigma}}_{\tg} = \E \{R \wt \bm{F} \bm{F}^{\mathrm{\scriptscriptstyle T}}\}$ is the Hessian of $\E \{ \ell_{\mathrm{CAL}}(\gamma)\}$ at $\gamma = \tg$;

\item[$(\mathrm{iv})$]
$|S_{\tg}|\lambda_{0} \leq \zeta_{0}$ for a sufficiently small constant $\zeta_{0} > 0$, depending only on $(A_{0}, C_{0},$  $\xi_{0}, \nu_{0})$,
where $|\cdot|$ denotes the cardinality of a set, $\lambda_{0} =  c_{\gamma} \sqrt{\ln\{(1 + p) / \epsilon\} / N}$, $c_{\gamma}$ is a constant only depending on $(B_{0}, C_{0})$ and $A_{0} > (\xi_{0} + 1)/(\xi_{0} - 1)$ is a constant.\footnote{$c_{\gamma}$ is defined in Section \ref{sup:pp1} of Supplement.}
\end{itemize}
\end{ass}

The conditions in Assumption \ref{ass:p} are plausible as discussed in \citet{tan2020a}. Based on Assumption \ref{ass:p}, we have the following Proposition \ref{prop:tan2017}.

\begin{pro} \label{prop:tan2017}
Suppose that Assumption \ref{ass:p} is satisfied, and $\lambda_{\gamma}$ in \eqref{eq:def-reg-gamma-h} is specified by $\lambda_{\gamma} = A_{0} \lambda_{0}$. Then, with probability at least $1 - 8\epsilon$,
\begin{equation}\label{eq:gamma-rate}
    D_{\mathrm{CAL}}^{\dagger}(\esg^{\T}\bm{F}, \tg^{\T}\bm{F}) + (A_{0} - 1)\lambda_{0} \|\esg - \tg  \|_{1} \leq M_{0}|S_{\tg}|\lambda^{2}_{0},
\end{equation}
where $M_{0} > 0$ is a constant depending only on $(A_{0}, C_{0}, B_{0}, \xi_{0}, \nu_{0}, \zeta_{0})$, and $D_{\mathrm{CAL}}^{\dagger}(\esg^{\T}F, \tg^{\T}F)$ is the symmetrized Bregman Divergence w.r.t $\ell_{\mathrm{CAL}}(\gamma)$, i.e.
\begin{equation*}
    D^{\dag}_{\mathrm{CAL}}(\esg^{\T}\bm{F}, \tg^{\T}\bm{F}) = -\tilde \E [R \{\exp( -\esg^{\T}\bm{F} ) - \exp( -\tg^{\T}\bm{F} )\} ( \esg^{\T}\bm{F} - \tg^{\T}\bm{F})].
\end{equation*}
\end{pro}

Note that $D^{\dag}_{\mathrm{CAL}}(\esg^{\T}\bm{F}, \tg^{\T}\bm{F}) \geq 0$, then equation \eqref{eq:gamma-rate} implies that
  \[  
 \|\esg - \tg  \|_{1} \leq \frac{M_{0}}{A_0 - 1} |S_{\tg}|\lambda_0,     \]
 which indicates that the $L_{1}$-convergence rate of the proposed regularized calibrated estimator $\esg$ is $|S_{\tg}|\lambda_0$,
where $|S_{\tg}|$ is the nonzero size of $\tg$ and  $\lambda_{0}=c_{\gamma} \sqrt{\ln\{(1 + p) / \epsilon\} / N}$.
For example, taking $\epsilon = 1/(1+p)$ gives $\lambda_{0} = c_{\gamma} \sqrt{2\ln(1 + p) / N}$, which leads to $\|\esg - \tg  \|_{1} = O(|S_{\tg}| \sqrt{\ln(1 + p)/ N})$.

For studying the properties of $\esa$, we make the following Assumption \ref{ass:o}.

\begin{ass}[Regularity conditions for $\hat \alpha$] \label{ass:o}

Let $\psi_{1}(u)$ denote the derivative of $\psi(u)$. Suppose that the following conditions are satisfied:

\begin{itemize}

\item[$(\mathrm{i})$]  $C_{1} \leq \psi_{1}(\ta^{\T}\bm{G}) \leq C_{2}$ a.s. for positive constants $(C_{1},
 C_{2})$;

\item[$(\mathrm{ii})$] $\psi_{1}(u) \leq \psi_{1}(u')\exp(C_{3}|u - u'|)$ for any $(u, u')$ and certain constant $C_{3} \geq 0$;

\item[$(\mathrm{iii})$] $\max_{j = 0, \ldots, q} |g_{j}(\bm{X})| \leq C_{4} $ a.s. for a constant $C_{4} \geq 1$;

\item[$(\mathrm{iv})$] $Y - \psit$ is uniformly sub-Gaussian given $\bm{X}$: $D^{2}_{0} \E (\exp[\{Y - \psit\}^{2}/D^{2}_{0}]-1|\bm{X}) \leq D^{2}_{1}$ for some positive constants $(D_{0}, D_{1})$;

\item[$(\mathrm{v})$] the compatibility condition holds for $\bm{\Sigma}_{\ta}$ with the subset $S_{\ta} = \{ 0 \} \cup \{j:\ta_{j} \neq 0, j = 0, \ldots, q
 \}$ and some constant $\nu_{1} > 0$ and $\xi_{1} > 1$, where $\bm{\Sigma}_{\ta} = \E \left[R \wt \psi_{1}(\ta^{\T}\bm{G}) \bm{G}\bm{G}^{\T}\right]$;

\item[$(\mathrm{vi})$] $(1 + \xi_{1})^{2} \nu_{1}^{-2}|S_{\ta}| \lambda_{1} \leq \zeta_{1}$ for a sufficiently small constant $\zeta_{1} > 0$, where
$\lambda_{1} = \max[\lambda_{0},  c_{\alpha}$

\noindent $\sqrt{\ln\{(1 + q) / \epsilon\} / N}]$, $c_{\alpha}$ is a constant  depending on $(B_{0}, C_{2}, C_{4}, D_{0}, D_{1})$;\footnote{$c_{\alpha}$ is defined in Section \ref{sup:pp2} of Supplement.} 

\item[$(\mathrm{vii})$] let $A_{1} > (\xi_{1} + 1) / (\xi_{1} - 1)$ be a constant. There exist $0 \leq \eta_{2}, \eta_{3} < 1$, such that $\Tilde{c}_{\alpha}|S_{\ta}| \lambda_{1} \leq \eta_{2}$  and  $\Tilde{c}_{\gamma}|S_{\tg}| \lambda_{0} \leq \eta_{3}$, where $\Tilde{c}_{\alpha}$ and $\Tilde{c}_{\gamma}$ are both constants.\footnote{ $\Tilde{c}_{\alpha}$ depends on $(A_{0}, A_{1}, C_{0}, C_{3}, C_{4}, M_{0}, \nu_{1}, \xi_{1}, \zeta_{0}, \zeta_{1})$ and $\Tilde{c}_{\gamma}$ depends on $(A_{0}, A_{1}, C_{0}, C_{1}, C_{3}, C_{4}, D_{0}, D_{1}, M_{0}, \xi_{1}, \zeta_{0})$. They are defined in  Section \ref{sup:pp2} of Supplement.}
\end{itemize}
\end{ass}

Assumptions \ref{ass:o}$(\mathrm{i})$--$(\mathrm{ii})$ are mild conditions on the smoothness of the inverse link function $\psi$. Commonly used functions like the identity and logit functions satisfy these requirements.
Assumptions \ref{ass:o}$(\mathrm{iii})$--$(\mathrm{vii})$ are similar to those used in related analysis by~\citet{tan2020a}.
A subtle difference is that
the compatibility condition in~\citet{tan2020a} [Assumption 2 $(\mathrm{ii})$] is assumed for $\bm{\Sigma}_{\tg}$ with $\bm{F} = \bm{G}$, whereas our compatibility condition is assumed for $\bm{\Sigma}_{\ta}$. In our setting, $\bm{G}$ has a higher dimension than $\bm{F}$ (except in the case where $\bm{Z} = 1$).

\begin{pro}\label{prop:main-mix-p}
Suppose Assumptions  \ref{ass:p} and \ref{ass:o}  are satisfied. If $\ln\{(1 + p)/\epsilon\}/N < 1$ and $\lambda_{\alpha}$ in \eqref{eq:def-reg-alp-est} is specified as $A_{1}\lambda_{1}$, then with probability at least $1 - 10\epsilon$,
\begin{equation}\label{eq:a-ub}
\begin{split}
     & D^{\dagger}_{\mathrm{WL}}(\esa^{\T}\bm{G},  \ta^{\T}\bm{G}, \tg) + \exp(\eta_{01})(A_{1} - 1)\lambda_{1}\| \esa - \ta \|_{1}
    \leq M_{11} |S_{\tg}|\lambda^{2}_{0} + M_{12 }|S_{\ta}|\lambda_{1}^{2},
\end{split}
\end{equation}
where $\eta_{01} = (A_{0}-1)^{-1}M_{0}C_{0}\zeta_{0}$, $M_{11}$ and $M_{12}$ are also constants;\footnote{Constant $M_{11}$ depends on $(A_{0}, C_{0}, C_{1}, D_{0}, D_{1}, M_{0}, \eta_{3}, \zeta_{0})$ and $M_{12}$ depends on $(A_{0}, A_{1}, C_{0}, M_{0}, \nu_{1}, \zeta_{0}, \zeta_{1}, \xi_{1}, \eta_{2})$. They are defined in Section \ref{sup:pp2} of Supplement.} $D^{\dagger}_{\mathrm{WL}}(\esa^{\T}\bm{G} ,\ta^{\T}\bm{G}, \tg)$ is the symmetrized Bregman divergence with respect to $\ell_{\mathrm{WL}}(\alpha; \tg)$, which is given by
\begin{equation}\label{eq:bd_or}
D^{\dagger}_{\mathrm{WL}}(\esa^{\T}\bm{G}, \ta^{\T}\bm{G}, \tg) = \tilde \E \left [ R \wt \{\psi(\esa^{\T}\bm{G}) - \psi(\ta^{\T}\bm{G})\} ( \esa^{\T}\bm{G} - \ta^{\T}\bm{G} ) \right].
\end{equation}
\end{pro}

Proposition \ref{prop:main-mix-p} gives the convergence rate of $\esa$. Since $D^{\dagger}_{\mathrm{WL}}(\esa^{\T}\bm{G},  \ta^{\T}\bm{G}, \tg) \geq 0$ and  $\lambda_{1} \geq \lambda_{0}$ (Assumption \ref{ass:o}(vi)),
$$\| \esa - \ta \|_{1} \leq c_{m} (|S_{\tg}| + |S_{\ta}|)\lambda_{1},$$
for some constant $c_{m}$. Compared with an unweighted lasso estimator, the convergence rate of our estimator $\esa$ not only depends on the non-sparsity size of $\phi(\bm{X}; \bar \alpha)$ itself but also on that of $\pi(\bm{X}; \bar \gamma)$.


\subsection{Large sample properties of the proposed estimator} \label{sec:pro_aipw}

In this subsection, we present our main result: the properties of the proposed estimator $\esb$. We show that the proposed estimator is consistent if either the PS model or the OR model is correctly specified. In addition, the proposed confidence intervals are valid when the PS model is correctly specified, irrespective of whether the OR model is correctly specified or not.

\begin{ass} \label{ass:x}
Suppose that the following conditions are satisfied:
\begin{itemize}

   \item [$(\mathrm{i})$] assume $\bm{Z} = \{ Z_{0,}, \ldots, Z_{m-1} \}  \in \sr^{m}$, and $\max_{j = 0, \ldots, m-1} |Z_{j}| \leq C_{5} $ almost surely for a constant $C_{5} \geq 1$, where $m$ is fixed as $N$ increases;

   \item [$(\mathrm{ii})$] assume that $\beta \in \Theta_\beta \subset \sr^{m}$, and for all  $\epsilon > 0$, $\inf_{\beta \in \Theta_{\beta}:\| \beta - \tb \|_{1} \geq \epsilon} \E \{\| \tau(\ta, \beta, \tg) \|_{1}\} > 0$;

   \item [$(\mathrm{iii})$] $\E \{ \sup_{\beta \in \Theta_{\beta}} \|\tau(\ta, \beta, \tg)\|_{1} \} < \infty$;

   \item [$(\mathrm{iv})$] $\psi_{1}(\beta^{*\T}\bm{Z}) \leq C_{6}$ a.e., for some constant $C_{6} >0$;

   \item [$(\mathrm{v})$] $|S_{\tg}| \sqrt{\ln(1+p) \ln(1 + q)} = o(N^{1/2})$ and $|S_{\ta}| \ln(1 + q) = o(N^{1/2})$.

\end{itemize}
\end{ass}

Assumptions \ref{ass:x}$(\mathrm{ii})$--$(\mathrm{iii})$ are standard conditions in the asymptotic theory of estimating equations. Assumption \ref{ass:x}$(\mathrm{iv})$ is a mild condition on the smoothness of the inverse link function at the value of $\beta^{*\T}\bm{Z}$. Assumption \ref{ass:x}$(\mathrm{v})$ is comparable to the sparsity requirement in \cite{tan2020a}.

\begin{thm} \label{thm:psc-h-p}
Suppose Assumptions \ref{ass:model}--\ref{ass:x} are satisfied, and the PS model \eqref{eq:def-ps} is correctly specified with $\pi(\cdot;\bar\gamma) = \pi^*(\cdot)$.
If $\ln((1 + p)/\epsilon)/N < 1$, then the following results hold.

(i) The estimator $\hat \beta$ is consistent and asymptotically normal, and
\begin{equation*}
    \sqrt{N}(\esb - \tb) \xrightarrow{d} \N(0, \bm{\Sigma}),
\end{equation*}
where $\xrightarrow{d}$ denotes convergence in distribution,
$\bm{\Sigma} = \bm{\bm{\Gamma}}^{-1}\bm{\bm{\Lambda}}\bm{\bm{\Gamma}}^{-1}$ with $\bm{\bm{\Gamma}} = \E \{ \psi_{1}(\beta^{*\T}\bm{Z}) \bm{Z} \bm{Z}^{\T}\}$, and
\begin{align}
\bm{\bm{\Lambda}} = & \E\left \{ \tau(\bm{O}, \bar \alpha, \beta^*, \bar \gamma) \tau(\bm{O}, \bar \alpha, \beta^*, \bar \gamma)^\mathrm{\scriptscriptstyle T} \right \} \nonumber \\
= & \E \left ( \left [ \frac{1}{\pt} \{Y - \psit\}^{2} + \{\psit - \mt  \}^{2} \right ] \bm{Z} \bm{Z}^{\T} \right )
\nonumber \\
& +2 \E [ \{Y - \psit\} \{\psit - \mt\} \bm{Z} \bm{Z}^{\T} ].
\end{align}

(ii) A consistent estimator of $\bm{\Sigma}$ is $\hat{\bm{\Sigma}}= \hat{\bm{\bm{\Gamma}}}^{-1}\hat{\bm{\bm{\Lambda}}}\hat{\bm{\bm{\Gamma}}}^{-1}$, where
$ \hat{\bm{\bm{\Gamma}}} = \tilde \E \{ \psi_{1}(\esb^{\T}\bm{Z})\bm{Z} \bm{Z}^{\T}\}$ and
 \begin{align*}
 \hat{\bm{\bm{\Lambda}}} = & \tilde \E \left \{ \tau(\bm{O}, \hat \alpha, \hat\beta, \hat \gamma) \tau(\bm{O}, \hat \alpha, \hat\beta, \hat  \gamma)^\mathrm{\scriptscriptstyle T}\right \}.
 \end{align*}
Thus, for a constant vector $\bm{c}$ with the same dimension of $\beta$, an asymptotic $(1-\eta)$ confidence interval for $\bm{c}^{\T}\beta^{*}$ is $\bm{c}^{\T} \hat \beta \pm z_{\eta/2}\sqrt{\bm{c}^\mathrm{\scriptscriptstyle T} \hat{\bm{\Sigma}} \bm{c}/N}$, where $z_{\eta/2}$ is the $(1-\eta/2)$ quantile of the standard normal distribution.
\end{thm}

Theorem \ref{thm:psc-h-p} shows that if the PS model is correct, regardless of whether the OR working model is correct or not, the proposed estimator $\hat{\beta}$ is consistent and  asymptotically normal, and the proposed confidence intervals based on $\hat{\bm{\Sigma}}$ are valid.
 In contrast, the estimator $\hat{\beta}_{\mathrm{IPW}}$ defined by equations \eqref{eq:def-wesb} is not $\sqrt{N}$-consistent in general, even with a correctly specified PS model. This is because in high-dimensional settings, the convergence rate of $\hat{\pi}(\bm{X})$ is typically slower than $1/\sqrt{N}$, leading to a slower convergence rate of $\hat{\beta}_{\mathrm{IPW}}$.
Similarly, when the OR working model is misspecified and the estimator $\hat \phi(\bm{X}) $ is inconsistent, the convergence rate of the AIPW estimator with nuisance  parameters being estimated using conventional regularized maximum likelihood may still be slower than $1/\sqrt{N}$.

In Theorem \ref{thm:psc-h-p}, we present our results in high-dimensional settings. The conclusions in Theorem~\ref{thm:psc-h-p} also hold in low-dimensional settings with a reduced form of Assumptions \ref{ass:p}, \ref{ass:o} and \ref{ass:x}.

\subsection{Extension to stratified sampling with constant PS} \label{sec4-3}
To facilitate comparison with existing methods described in Section \ref{sec:comp}, we extend the theoretical analysis of the proposed estimator to the classical SSL setting
(stratified sampling with constant PS), where the sizes of labeled and unlabeled datasets, $n$ and $N - n$, are deterministic.
For fixed $n$ and $N$, the observed data are generated as follows:
    \begin{itemize} \addtolength{\itemsep}{-.1in}
        \item The labeled dataset $(\bm{X}_{1}, Y_{1}), \ldots, (\bm{X}_{n}, Y_{n}) \overset{\text{i.i.d.}}{\sim}\P(\bm{X}, Y | R=1)$. 
        \item The unlabeled dataset
         $(\bm{X}_{n+1}, \ldots, \bm{X}_{N})  \overset{\text{i.i.d.}}{\sim}\P(\bm{X}, Y |R=0)$.
\end{itemize}
Moreover, by letting $\bm{F}=1$, $\hat \pi (\bm{X}) = \pi(\bm{X};\hat\gamma)= n/N$ (constant PS) and allowing a general choice of $\bm{G}$ instead of \eqref{eq-gx},
our AIPW estimator, denoted as $\hat{\beta}^{s}$, can be rewritten as
a solution to the following estimating equation:
\begin{equation}  \label{eq-21}
    \frac{1}{n}\sum^{n}_{i=1} \left \{ Y_{i} - \psi(\esa^{\T}\bm{G}_{i}) \right \}\bm{Z}_{i} + \frac{1}{N}\sum_{i=1}^{N} \{\psi(\esa^{\T}\bm{G}_{i}) - \psi(\beta^{\T}\bm{Z}_{i})\} \bm{Z}_{i} = 0,
\end{equation}
where $\bm{Z}_{i}$ denotes the $j$-th observation of $\bm{Z}$ and $\bm{G}_{i}$ is the abbreviation of $\bm{G}(\bm{X}_{i})$.
Due to the constant $\hat\pi(\bm{X})$, our estimator $\hat\alpha$ reduces to the regularized unweighted maximum likelihood or least squares estimator.
The following proposition for $\hat{\beta}^{s}$ can be readily derived.

\begin{pro}\label{prop:ss}
Suppose that the conditions of Theorem \ref{thm:psc-h-p} are satisfied with $\bm{F}=1$, $\pi^*(\bm{X}) \equiv n/N$, and a general choice of $\bm{G}$,
where Assumption \ref{ass:x}(v) reduces to $|S_{\ta}|\ln(q+1)  = o_{p}(\sqrt{n})$. Then we have
\begin{equation*}
    \sqrt{n}(\hat{\beta}^{s} - \tb) \xrightarrow{d} \N(0, \bm{\Sigma^{s}}),
\end{equation*}
where $\bm{\Sigma^{s}} = \bm{\Gamma}^{-1}\bm{\Lambda}^{s}\bm{\Gamma}^{-1}$, with $\bm{\Gamma}$ being defined as in Theorem \ref{thm:psc-h-p} and
\begin{align*}
\bm{\Lambda}^{s} = & \E \left [ \left \{Y - \frac{N-n}{N}\psi(\ta^{\T}\bm{G}) - \frac{n}{N} \psi(\beta^{*\T}\bm{Z})\right \}^{2} \bm{Z} \bm{Z}^{\T} \right ] \\
& + \frac{(N - n)n}{N^{2}} \E \left [ \{ \psi(\ta^{\T}\bm{G}) - \psi(\beta^{*\T}\bm{Z}) \}^{2} \bm{Z} \bm{Z}^{\T} \right ].
\end{align*}
\end{pro}
For comparison, by letting $\bm{F}=1$ and $\pi^* (\bm{X}) \equiv n/N$ in Theorem \ref{thm:psc-h-p}, the variance matrix $\bm{\Sigma}$ for $\esb$ such that $ \sqrt{N}(\esb - \tb) \xrightarrow{d} \N(0, \bm{\Sigma})$
is $\bm{\Sigma} = \bm{\Gamma}^{-1}\bm{\Lambda}  \bm{\Gamma}^{-1}$, where $\bm{\Lambda}$ is simplified as
\begin{align*}
&\bm{\Lambda} =  \E\left ( \left [ \frac{N}{n} \{ Y - \psit \}^{2} + \{ \psit - \mt  \}^{2}\right ] \bm{Z} \bm{Z}^{\T} \right )\\
& +2 \left ( \left [ \{ Y - \psit \}  \{ \psit - \mt \}  \right ] \bm{Z} \bm{Z}^{\T} \right )
\end{align*}
By direct calculation (see Section \ref{sup:ess} of Supplement),
it is shown that $\bm{\Sigma}/N = \bm{\Sigma}^{s}/n$, which means the unscaled asymptotic variances of $\hat{\beta}^{s}$ and $\esb$ are the same.
Hence, in the classical SSL with constant PS, the asymptotic variances of our estimators, $\esb$ under random sampling or $\hat{\beta}^{s}$ under stratified sampling,
are equivalent to each other.

\section{Comparison with previous methods} \label{sec:comp}
In this subsection, we first briefly summarize characteristics of previous related methods.
All of them  can be integrated into the AIPW estimation framework. We also compare asymptotic variances of our methods with previous methods.

\subsection{Unified framework}

Various methods have been proposed in the classical SSL setting, i.e., stratified sampling with constant PS ($\bm{F}=1$) as in Section \ref{sec4-3}~\citep{cha2016, cha2018, ANRU2019, zhang2021}.
From an AIPW point of view, the major difference among previous methods lies in the choices of OR working models.
For example,
\citeauthor{ANRU2019}~(\citeyear{ANRU2019}) and  \citeauthor{zhang2021}~(\citeyear{zhang2021}) proposed linear OR working models for the estimation of $\E(Y)$,
i.e., with $\bm{Z}=1$.
\citeauthor{cha2016}~(\citeyear{cha2016}) and \citeauthor{cha2018}~(\citeyear{cha2018}) proposed using non-parametric or semi-parametric OR working models,
such as kernel smoothing or partially linear model, for regression analysis with $\bm{Z}$ a sub-vector of $\bm{X}$.


If we disregard the specific choice of the OR working model, the previous methods can be incorporated into the AIPW estimating framework.  In our notation, the estimators of \citeauthor{cha2016}~(\citeyear{cha2016}), \citeauthor{ANRU2019}~(\citeyear{ANRU2019}) and \citeauthor{zhang2021}~(\citeyear{zhang2021}) can be reformulated as solutions of the following equations:
\begin{align}\label{eq:eme-oth-f}
\frac{1}{n} \sum_{i=1}^{n}\{Y_{i} - \psi(\esa^{\T}\bm{G}_{i})\} \bm{Z}_{i} + \frac{1}{N}\sum_{i=1}^{N}\{\psi(\esa^{\T}\bm{G}_{i}) - \psi(\beta^{\T}\bm{Z}_{i})\}\bm{Z}_{i}=0,
\end{align}
for different choices of $\bm{Z}_i$ and $\psi(\cdot)$. Specifically,~\citeauthor{ANRU2019}~(\citeyear{ANRU2019}) and \citeauthor{zhang2021}~(\citeyear{zhang2021}) correspond to the case of $\bm{Z}=1$ and $\psi(\cdot)$ is the identity function in equation \eqref{eq:eme-oth-f}, while \citeauthor{cha2016}
(\citeyear{cha2016}) corresponds to the case where $\bm{Z}$ is any sub-vector of $\bm{X}$ and $\psi(\cdot)$ is an arbitrary inverse link function.
Suppose a constant PS model \eqref{eq:def-ps} is used with $ \pi^{*} (\bm{X}) \equiv n/N$ and a general choice of $\bm{G}$ is used in the OR model \eqref{eq:def-or}.
Then the AIPW estimating equation \eqref{eq:rand} or, in the simplified form,  \eqref{eq-21} in Section \ref{sec4-3}, coincides with \eqref{eq:eme-oth-f}.


In addition,
\citeauthor{cha2018}~(\citeyear{cha2018}) adopted a variation of AIPW estimating equations. By the assumption $\lim_{n,N\rightarrow \infty} n/N \rightarrow 0$ and
controlling kernel smoothing in fitting OR working models, they made it possible to drop the labeled part and only retain the augmented term of unlabeled data in \eqref{eq:eme-oth-f}. Their estimating equations can be reformulated in our notation as
    \begin{align*}
    \frac{1}{N-n}\sum_{i=n+1}^{N}\{\psi(\esa^{\T}\bm{G}_{i}) - \psi(\beta^{\T}\bm{Z}_{i})\}\bm{Z}_{i}=0,
    \end{align*}
with $\psi(\cdot)$ to be identity function and $\bm{Z} =\bm{X}$, corresponding to full linear regression.

\subsection{Variance comparison}

Under stratified sampling with constant PS,
both estimators of \citeauthor{ANRU2019}~(\citeyear{ANRU2019}) and \citeauthor{zhang2021}~(\citeyear{zhang2021}) of $\E(Y)$ achieve asymptotic normality and their asymptotic variance is Var($Y - \ta^{\T}\bm{G}$) + $(n/N)$Var($\ta^{\T}\bm{G}$).
Under this setting, by Proposition \ref{prop:ss} with $\bm{Z}=1$,
our AIPW estimator has the asymptotic variance
\begin{align}\label{eq:savm}
 \E \left \{(Y -\ta^{\T}\bm{G})^{2} + \frac{n}{N} (\ta^{\T}\bm{G} - \beta^{*})^2 + 2\frac{n}{N}(Y -\ta^{\T}\bm{G})(\ta^{\T}\bm{G} - \beta^{*})\right\},
\end{align}
where $\beta^{*} =\E(Y) = \E(\ta^{\T}\bm{G})$ and $\E\{(Y -\ta^{\T}\bm{G})\bm{G}\}=0$ by definition of $\ta$ and the fact that $\bm{G}$ includes 1. Then
\begin{align*}
 \E \left \{(Y -\ta^{\T}\bm{G})(\ta^{\T}\bm{G} - \beta^{*})\right\} =   \E\{ (Y -\ta^{\T}\bm{G})(\ta^{\T}\bm{G} ) \}=0,
\end{align*}
and \eqref{eq:savm} reduces to Var($Y - \ta^{\T}\bm{G}$) + $(n/N)$Var($\ta^{\T}\bm{G}$) matching results in \citeauthor{ANRU2019}~(\citeyear{ANRU2019}) and \citeauthor{zhang2021}~(\citeyear{zhang2021}).

Under stratified sampling with constant PS, the estimators of regression coefficients in conditional mean models proposed by \cite{cha2016} and \cite{cha2018}  achieve asymptotic normality under the assumption that
$\lim_{n,N\rightarrow \infty} n/N \rightarrow 0$.  Their asymptotic variances satisfy the following formula:
\begin{align}\label{eq:vc1}
    \bm{\Gamma}^{-1}\text{Var}[\{ Y - \psit \}\bm{Z}] \bm{\Gamma}^{-1}.
\end{align}
Under this setting, by Proposition \ref{prop:ss}, our AIPW estimator has the asymptotic variance
\begin{align*}
&  \bm{\Gamma}^{-1}\text{Var}[\{ Y - \psit \}\bm{Z}] \bm{\Gamma}^{-1} + \frac{n}{N}\bm{\Gamma}^{-1}\E [ \{\psit - \mt\}^{2} \bm{Z}\bm{Z}^{\T}]\bm{\Gamma}^{-1}\\
& + 2\frac{n}{N}\bm{\Gamma}^{-1}\E [\{Y - \psit  \}\{ \psit - \mt \}\bm{Z}\bm{Z}^{\T}] \bm{\Gamma}^{-1},
\end{align*}
which, compared with \eqref{eq:vc1}, in general has additional term
\begin{align*}
\frac{n}{N}  \bm{\Gamma}^{-1}\Bigl \{\E ( [ \{\psit - \mt\}^{2} + 2\{Y - \psit  \}\{ \psit - \mt \}] \bm{Z}\bm{Z}^{\T} )\Bigr\} \bm{\Gamma}^{-1}.
\end{align*}
The additional term reduces to $0$ under the condition $\lim_{n,N\rightarrow \infty} n/N \rightarrow 0$, implying that our result aligns with those of~\citet{cha2016}and \citet{cha2018} with the same condition.

\section{Simulation study}\label{sec:simulation}
In this section, we design experiments to evaluate the finite-sample performance of the proposed method and compare it with the competing $\mathrm{IPW}$ and $\mathrm{AIPW}$ methods. We consider the estimators of population mean for $\bm{Z}=1$, regression coefficients in the mean model for $\bm{Z}=X_1$ and $\bm{Z} =\bm{X}$, respectively, and $\psi(\cdot)$ is assumed to be the identity function.

\subsection{Data generating process} \label{sec6-1}

Throughout the simulation, we generate the covariates $\bm X $ as follows.
We first generate a random vector from $\N(0, \bm{\Sigma})$, where the variance matrix $\bm{\Sigma} \in \mathbb{R}^{3 \times 3}$ has elements $\bm{\Sigma}_{i,j}$ defined as $2^{-|i-j|}$ for $i, j =1, 2, 3$.
Then we clamp each of its coordinates within $[-3, 3]$ to obtain $(X_{1}, X_{2}, X_{3})$ and $\bm{X} = (1, X_{1}, X_{2}, X_{3})$.
In addition, the data source indicator $R$ follows a Bernoulli distribution with success probability $\pi(\bm X)$, where $\pi(\bm X ) = \{1 + \exp(-\gamma^{\mathrm{\scriptscriptstyle T}}\bm F)\}^{-1}$, the parameter $\gamma = (-1.5, -0.8, -0.2, 0.3,0,\ldots,0)^{\mathrm{\scriptscriptstyle T}}$ and the basis functions $\bm F$ are described in Section \ref{sec:choice-bf}.

\textbf{Study I.} We first focus on the estimation of the population mean and consider two data-generating mechanisms:
\begin{itemize}
      \item \textbf{Case 1.} The outcome $Y =  -0.2 + 0.1\Tilde{X}_{1} + 0.4\Tilde{X}_{2}+ 0.7\Tilde{X}_{3} + \epsilon$, where $\epsilon \sim \mathcal{N}(0, 0.1)$ and $\Tilde{X}_{j} = X_{j}\cdot |X|^{0.1}_{j} + X_{j}\cdot |X|^{0.3}_{j} + X_{j}\cdot |X|^{0.5}_{j}$, for $j = 1, 2, 3$. We set $\bm Z =  1$.

 \item \textbf{Case 2.}
    The outcome $Y =  -0.2 + 0.1\Tilde{X}_{1} + 0.4\Tilde{X}_{2}+ 0.7\Tilde{X}_{3} + \epsilon$, where $\epsilon \sim \mathcal{N}(0, 0.1)$ and $\Tilde{X}_{j} = |X_{j}| \exp(|X_{j}|^{0.1} + |X_{j}|^{0.3})$ for $j = 1, 2, 3$. We set $\bm Z =  1$.
\end{itemize}

In many scenarios, estimating the conditional mean given a subset of variables in $\bm{X}$ garners statistical interest. Accordingly, we design experiments in Study II to evaluate the performance of the proposed estimator in such setups.

  \textbf{Study II.} We further consider three additional cases for estimating regression coefficients in the conditional mean outcome model.
\begin{itemize}

    \item \textbf{Case 3.} The outcome $Y = 0.4\Tilde{X}_{1} + 0.2\Tilde{X}_{2} + \epsilon$, where $\epsilon \sim \mathcal{N}(0, 0.1)$,  $\Tilde{X}_{1} = X_{1} + X^{2}_{1}$ and $\Tilde{X}_{2} = \cos(\pi/9 \cdot X_{1} \cdot X_{3})$. We set $\bm Z =  X_{1}$.

    \item  \textbf{Case 4.} The outcome $Y = 0.4\Tilde{X}_{1} + 0.2\Tilde{X}_{2} + \epsilon$, where $\epsilon \sim \mathcal{N}(0, 0.1)$,  $\Tilde{X}_{1} = X_{1}\cdot \mathbb{I}\{ X_{1} > 0\}\sqrt{|X_{1}|}$ and $\Tilde{X}_{2} = X_{1} \cdot X_{2}$. We set $\bm Z =  X_{1}$.

   \item  \textbf{Case 5.} The outcome $Y = -0.2 + 0.1\Tilde{X}_{1} + 0.4\Tilde{X}_{2} + 0.7\Tilde{X}_{3} +  \epsilon$, where $\epsilon \sim \mathcal{N}(0, 0.1)$, $\Tilde{X}_{1} = X_{1}\cdot X_{2}$, $\Tilde{X}_{2} = X_{2} \cdot X_{3}$ and $\Tilde{X}_{3} = X_{1} \cdot X_{3}$. We set $\bm Z = \bm X$.
\end{itemize}
Cases 3 and 4 involve $\bm{Z}$ as a specific covariate $X_1$, while Case 5 involves $\bm{Z}$ as the full set of covariates $\bm{X}$. In addition, for all Cases 1--5, OR models are misspecified.
\subsection{Implementation details and competing methods}\label{sec:choice-bf}

We compare the proposed method ($\mathrm{AIPW_{\mathrm{RCAL}}}$) with  several alternative methods:  the $\mathrm{IPW}$ method
with Lasso-regularized maximum likelihood estimation for the PS model,
$\mathrm{AIPW}$ methods with Lasso-regularized maximum likelihood estimation for both the PS and OR models without cross-fitting as in \cite{tan2020a},\footnote{
For Lasso-regularized maximum likelihood estimation, the loss functions in fitting PS models and OR models are $-\te [Y \gamma^{\T}\bm{F} - \ln\{ 1 + \exp(\gamma^{\T}\bm{F}) \}] + \lambda_{\gamma}\| \gamma_{1:p} \|_{1}$ and $\te [(Y - \alpha^{\T}\bm{G})^{2}] + \lambda_{\alpha}\| \alpha_{1:q} \|_{1}$, respectively.}
and AIPW methods with cross-fitting and Lasso-regularized maximum likelihood estimation for both the PS and OR models~\citep{chernozhukov2018, zhang2021}.
 These competing estimators are denoted as $\mathrm{IPW}$,  $\mathrm{AIPW_{RML}}$, and $\mathrm{AIPW_{CF}}$, respectively.
For the PS and OR models,  the basis functions $\bm{F}$ and $\bm{G}$ are specified as follows.
\begin{itemize}
    \item $\mathrm{AIPW_{\mathrm{RCAL}}}$: Given $\bm{X}=(1, X_1,\ldots,X_d)^{\mathrm{\scriptscriptstyle T}}$ in Section \ref{sec6-1}, let $\{\xi_{i}\}_{i=1}^{k}$ be the $k$ points equally spaced within $(-a, a)$, where $d=3$, $k=49$, and $a=3$.
    Let $f_{ij}(\bm{X})= (X_{i}- \xi_{j})_{+}$, $i = 1, \ldots, d, j = 1, \ldots, k$.
    Let $\bm{F} =  \{1, f_{11}(\bm{X}), \ldots, f_{1n_{k}}(\bm{X}), \ldots,  f_{d1}(\bm{X}), \ldots, f_{dn_{k}}(\bm{X})\}^{\mathrm{\scriptscriptstyle T}}$ be the basis functions in the PS model, and $\bm{G} = \{\bm{F}^{\mathrm{\scriptscriptstyle T}}, (\bm Z \bigotimes \bm{F})^{\mathrm{\scriptscriptstyle T}}\}^{\mathrm{\scriptscriptstyle T}}$ be the basis functions in the OR model.
    Then the dimension of $\bm{F}$ is 148. For $\bm{Z}=1,\; X_{1} \;, \bm{X}$, the dimensions of $\bm{G}$ are 148, 285 and 589, respectively.
    \item $\mathrm{IPW}$:  Let $\bm{F}$ be the basis functions for the PS model.
    \item $\mathrm{AIPW_{RML}}$: Let $\bm{F}$ and $\bm{G} = \bm{F}$ be the basis functions for both PS and OR models, respectively.
    \item $\mathrm{AIPW_{CF}}$: Let $\bm{F}$ and $\bm{G} = \bm{F}$  be the basis functions for both PS and OR models, respectively.
\end{itemize}

Both the Lasso-regularized calibrated and maximum likelihood estimators for the PS and OR models can be implemented using the R package \texttt{RCAL}~\citep{RCAL}. We employ 5-fold cross-fitting to select the optimal tuning parameters.
In addition, by equation \eqref{eq:def-tb}, $\beta^{*} = \E(\bm{Z}\bm{Z}^{\T})^{-1}\E(Y\bm{Z})$.
Thus the true value of $\beta^*$ is calculated as
$\te(\bm{Z}\bm{Z}^{\T})^{-1}\te(Y\bm{Z})$
through a simulation with a sample size of 100,000.
For $\bm{Z} = X_{1}$ and $\bm{Z} = \bm{X}$, we denote $\beta^* = \beta_1$ and $\beta^* = (\beta_0, \beta_1, \beta_2, \beta_3)^{\T}$, respectively.

\subsection{Summary of results}

We present and analyze the simulation results for Study I and Study II. In the following tables, we compare various methods in terms of five metrics: Bias (Monte Carlo bias), $\sqrt{\mathrm{Var}}$ (Monte Carlo standard deviation), $\sqrt{\mathrm{EVar}}$ (square root of the  mean of variance estimates), CP90 (coverage proportions of the 90\%  CIs), and CP95 (coverage proportions of the 95\% CIs). As discussed in the paragraph below Theorem  \ref{thm:psc-h-p}, under high-dimensional settings, the $\mathrm{IPW}$ estimator is not $\sqrt{N}$-consistent, and its asymptotic normality is not well established. Therefore, we do not report its numerical results for $\sqrt{\mathrm{EVar}}$, CP90, and CP95.

{\bf Results for Study I.} Table \ref{tab:mean}  shows the numerical results for the estimation of population mean $\E(Y)$. 
From the table, the proposed method $\mathrm{AIPW}_{\mathrm{RCAL}}$ has the smallest $\sqrt{\text{Var}}$ and $\sqrt{\text{EVar}}$, 
and  Bias. 
Moreover, CP90 and CP95 of the proposed method are more aligned with their nominal values of 0.90 and 0.95, respectively. This indicates the effectiveness of the proposed method in terms of estimating the population mean.

\begin{table}[ht!]
\centering
\hspace{-.1in}
\caption{Summary of estimates of population mean for Study I.} \label{tab:mean} \vspace{-.1in}
\footnotesize
\begin{center}
\begin{tabular*}{\textwidth}{@{\extracolsep\fill} l@{\hskip -0.05in} c l@{\hskip 0.05in} ccc @{\hskip 0.05in} c l@{\hskip 0.05in} ccc} \hline
      &
      \multicolumn{3}{c}{\textbf{Case 1}} &&
      \multicolumn{3}{c}{\textbf{Case 2}}&\\
      \cline{2-4}\cline{6-9}
      & $\mathrm{AIPW_{\mathrm{RCAL}}}$ & $\mathrm{AIPW_{RML}}$ & $\mathrm{AIPW_{CF}}$ && $\mathrm{AIPW_{\mathrm{RCAL}}}$ & $\mathrm{AIPW_{RML}}$ & $\mathrm{AIPW_{CF}}$ & \\
     \cline{2-4}\cline{6-8}
    Bias & 0.004 & 0.004 & 0.006  && 0.002 & -0.006  & 0.031\\
    $\sqrt{\text{Var}}$ & 0.078  & 0.078  & 0.079   &   &  0.139  & 0.143 &   0.166 \\
    $\sqrt{\text{EVar}}$ & 0.079 & 0.079  & 0.082 & & 0.140  &  0.143  &  0.161\\
    $\text{CP90}$ & 0.904 & 0.908  & 0.918  &  &0.898&0.886 & 0.896 \\
    $\text{CP95}$ & 0.948  & 0.954 & 0.956 && 0.948& 0.946 & 0.956\\\hline
\end{tabular*} \\[.01in]
\end{center}  \vspace{-.1in}
\end{table}

\begin{figure}[H]
\begin{subfigure}{0.49\textwidth}
\centering    \includegraphics[width=.9\textwidth]{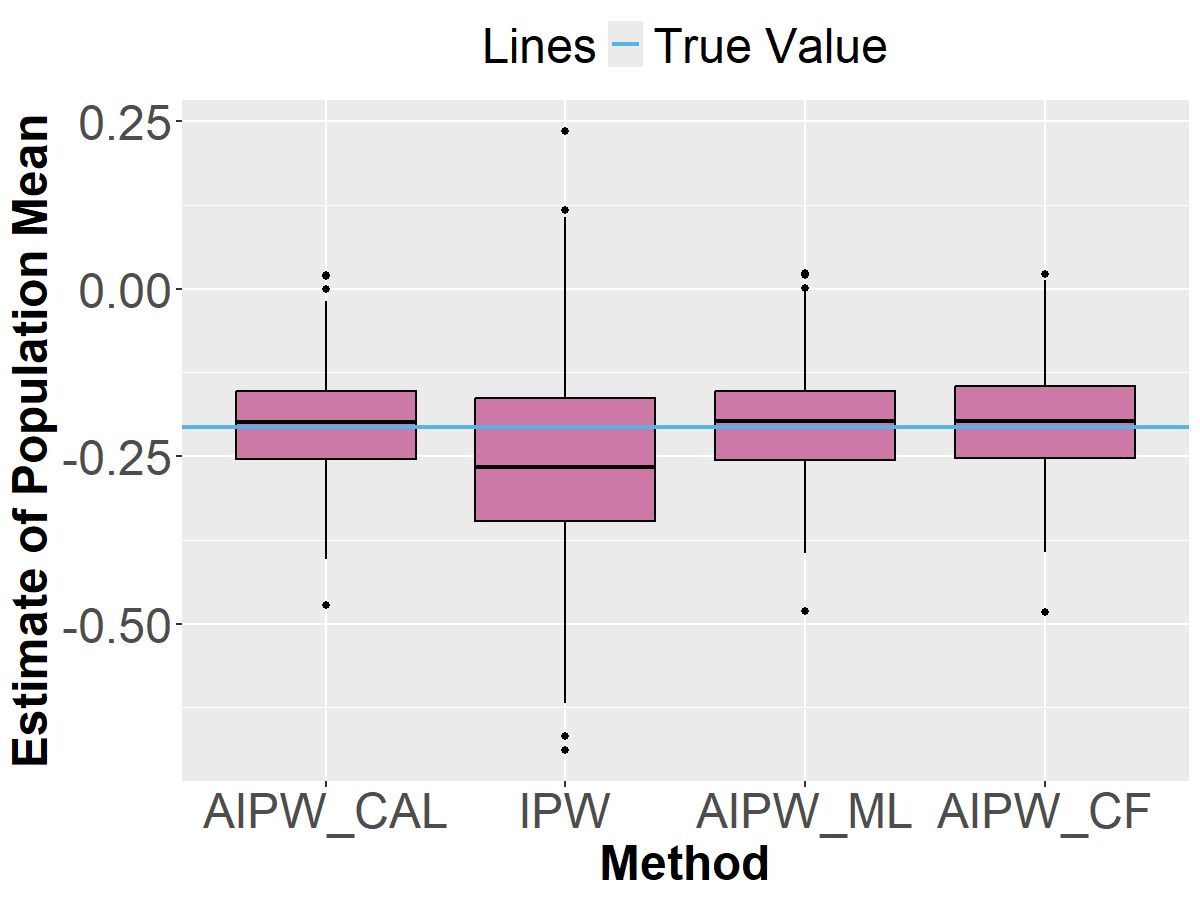}
\caption{Case 1}
\end{subfigure}
\begin{subfigure}{0.49\textwidth}
\centering    \includegraphics[width=.9\textwidth]{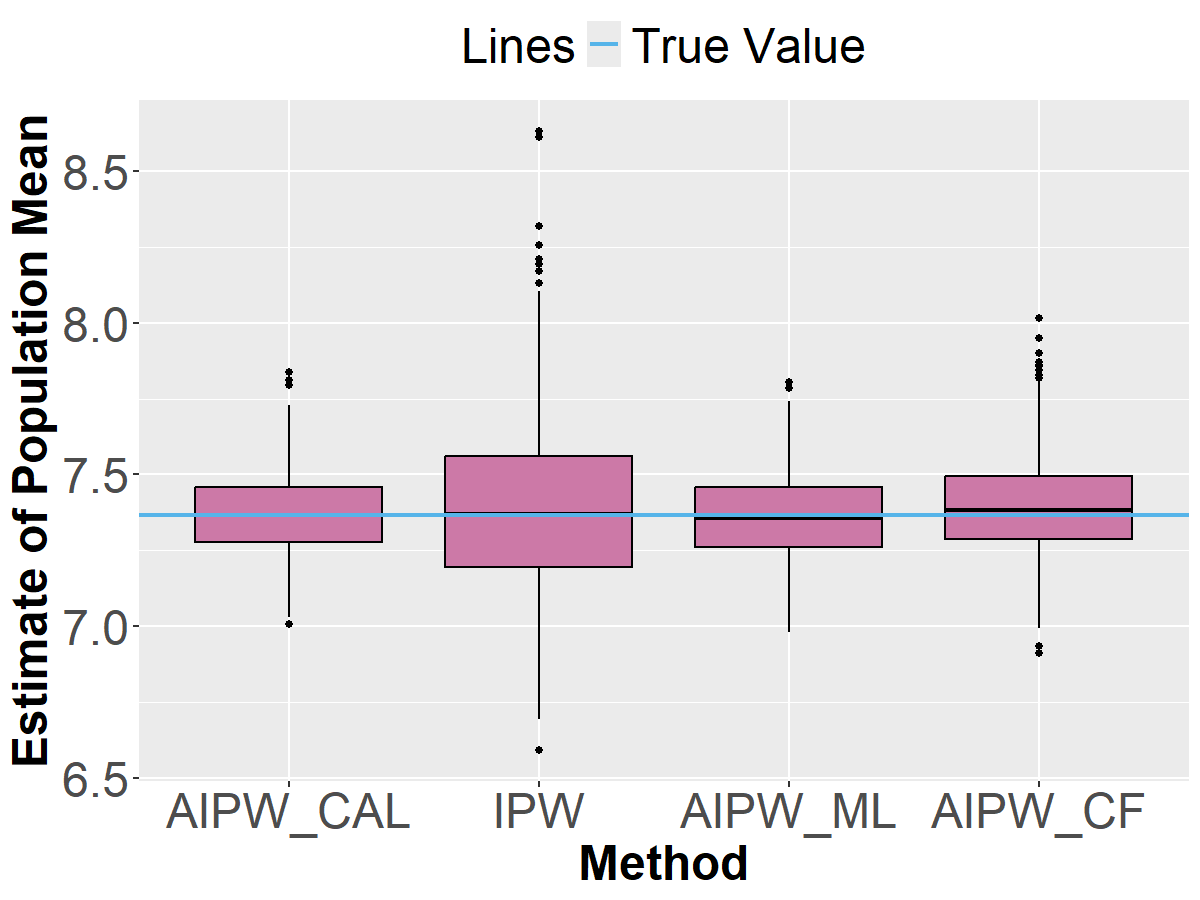}
\caption{Case 2}
\end{subfigure}
\caption{Box plots for estimates of population mean.}\label{fig:bp-1}
\end{figure}

Figure \ref{fig:bp-1} depicts the box plots of the estimates for Case 1 and Case 2, where the blue horizontal line indicates the true value. In both cases,  our method $\mathrm{AIPW}_{\mathrm{RCAL}}$ exhibits the smallest biases, interquartile ranges, and whiskers, indicating the smallest variances compared to the other methods.
In addition, $\mathrm{AIPW_{CF}}$ shows more outliers than the other two methods, which is more apparent in the results of Study $\mathrm{II}$, suggesting that cross-fitting may cause instability for the estimates.

\begin{figure}[H]
\begin{subfigure}{0.49\textwidth}
\centering    \includegraphics[width=.9\textwidth]{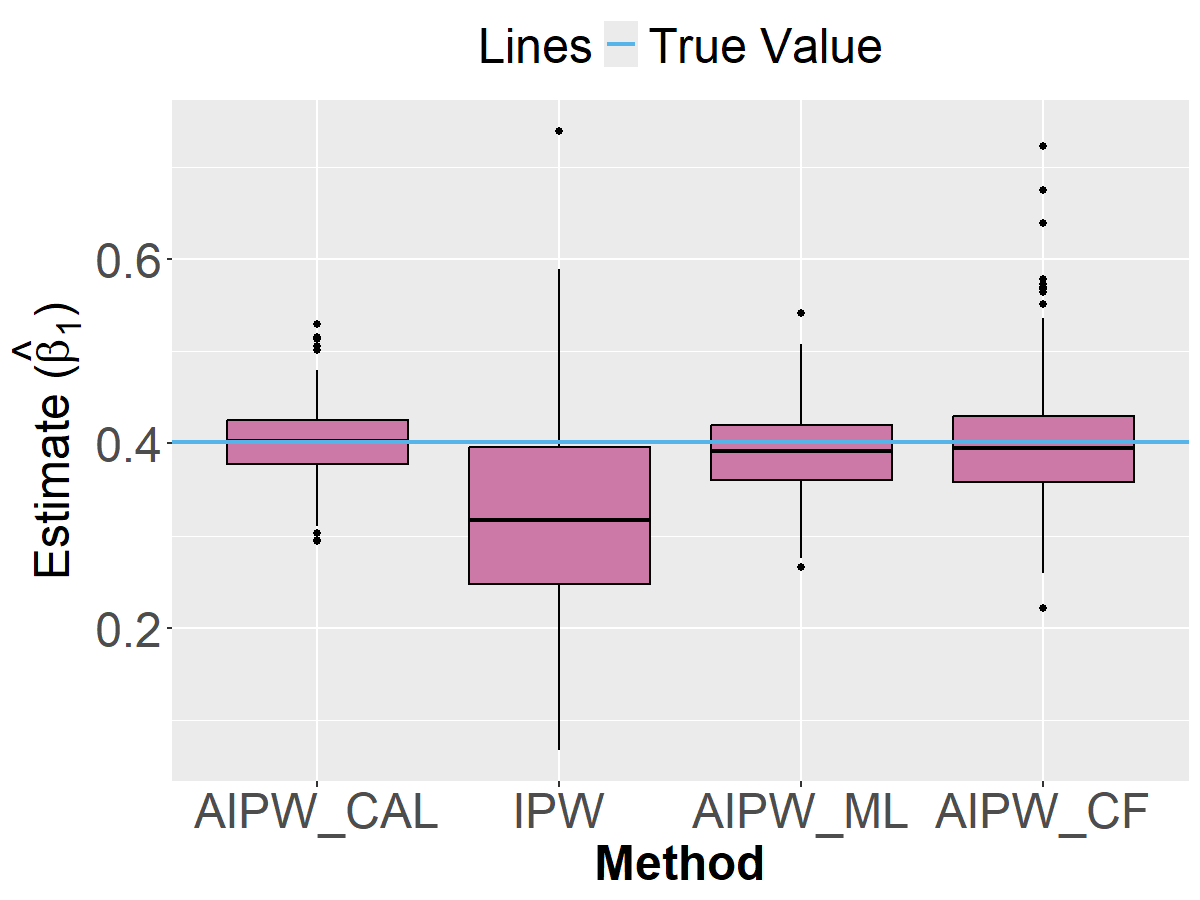}
    \caption{Case 3}
\end{subfigure}
\begin{subfigure}{0.49\textwidth}
\centering    \includegraphics[width=.9\textwidth]{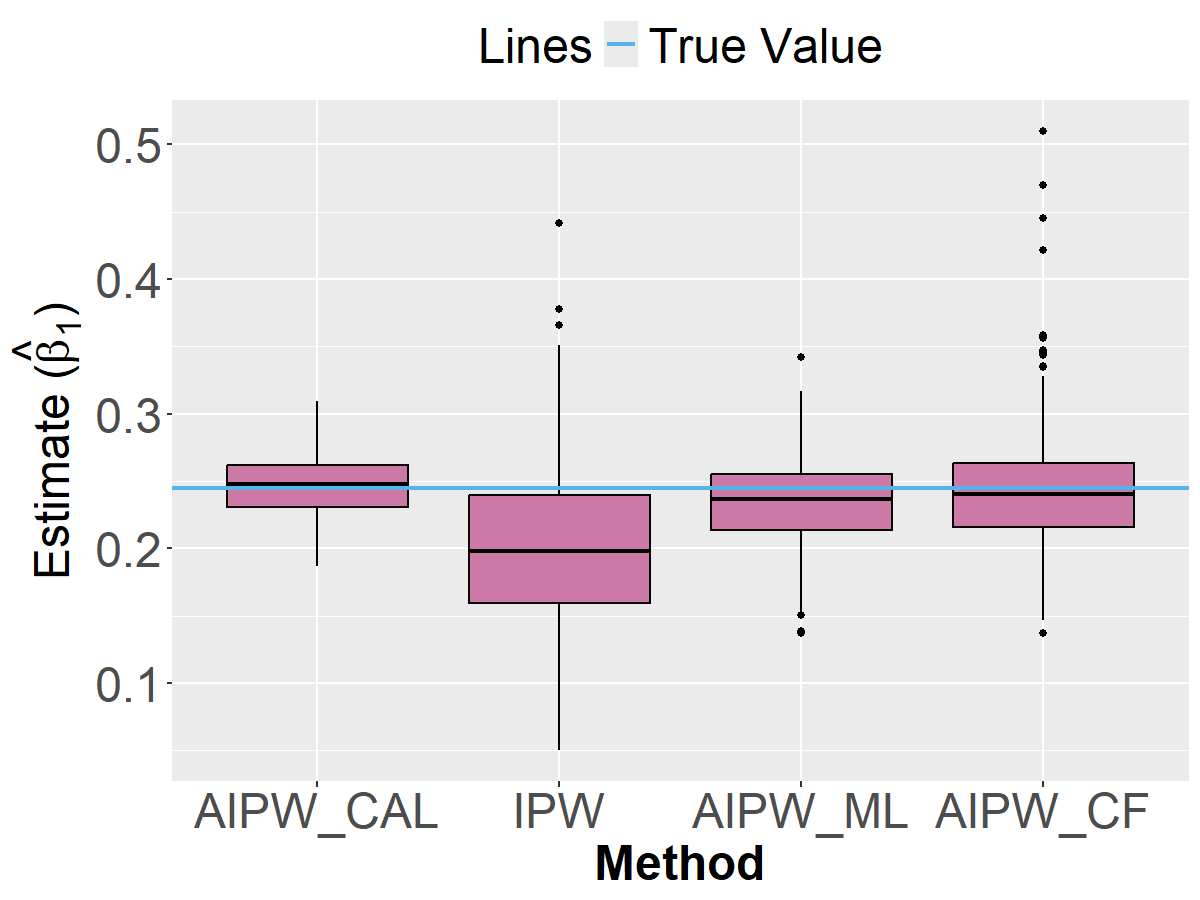}
    \caption{Case4}
\end{subfigure}
\caption{Box plots of estimates of $\beta_1$ in Cases 3 and 4.}\label{fig:bp-2}
\end{figure}

\begin{figure}[H]
\centering
\begin{subfigure}{0.49\textwidth}
\centering
    \includegraphics[width=.9\textwidth]{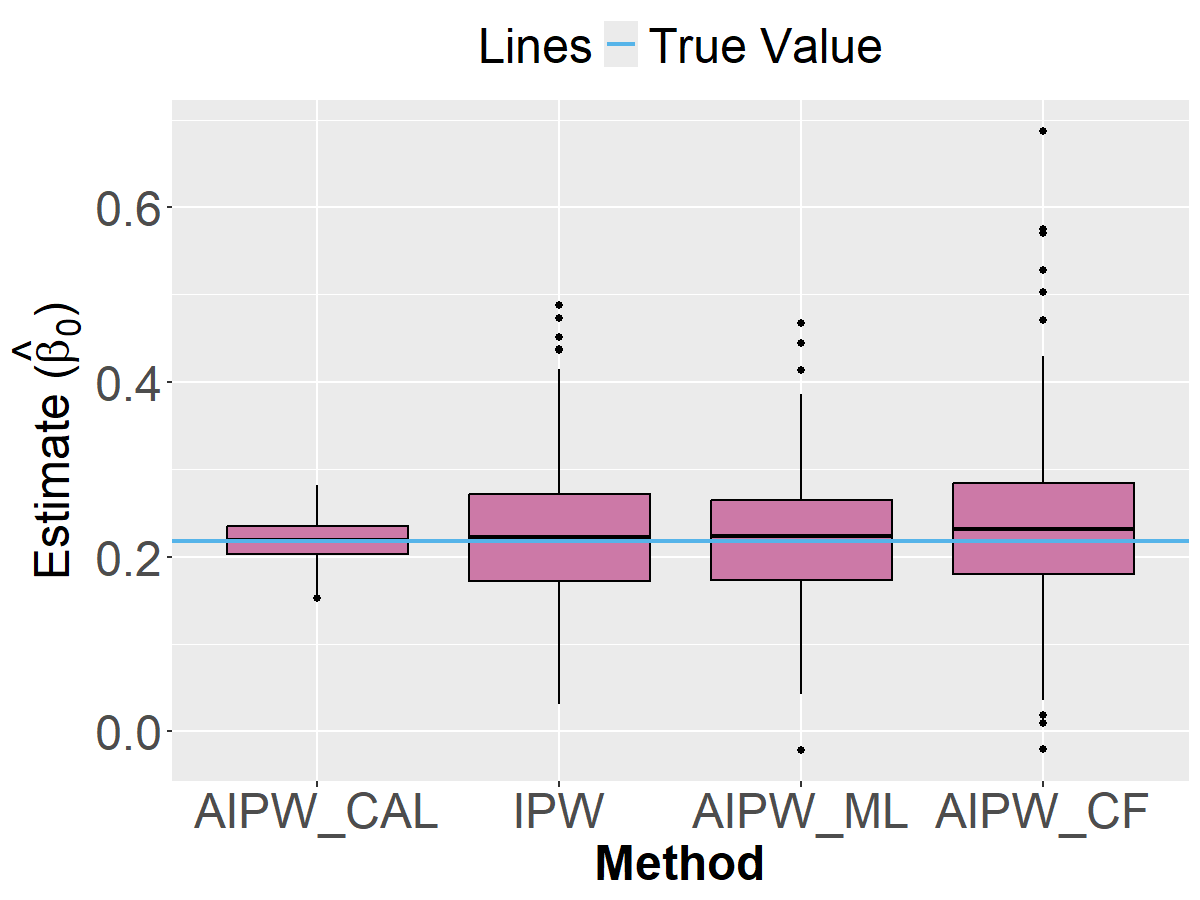}
    \caption{$\hat{\beta}_{0}$}
\end{subfigure}
\begin{subfigure}{0.49\textwidth}
\centering
    \includegraphics[width=.9\textwidth]{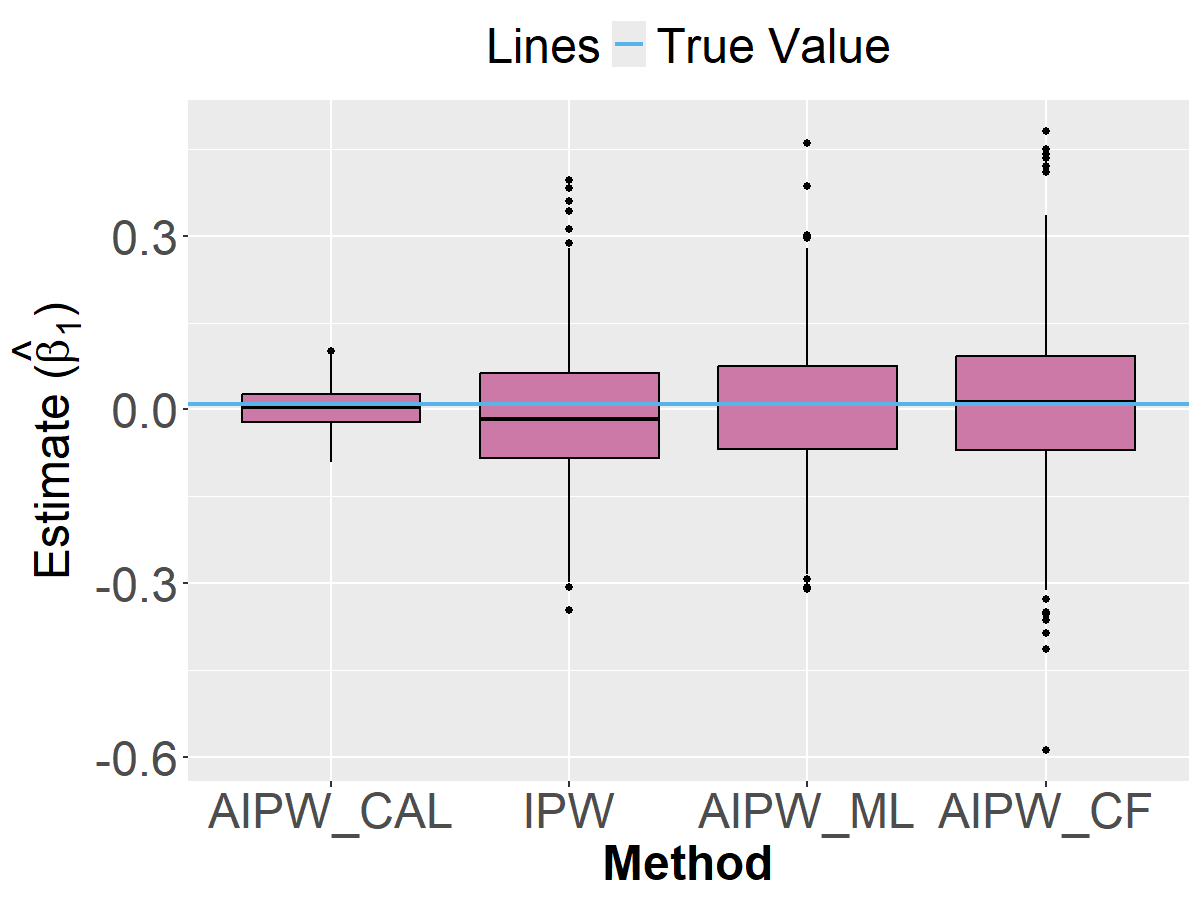}
    \caption{$\hat{\beta}_{1}$}
\end{subfigure}
\centering
\begin{subfigure}{0.49\textwidth}
\centering
    \includegraphics[width=.9\textwidth]{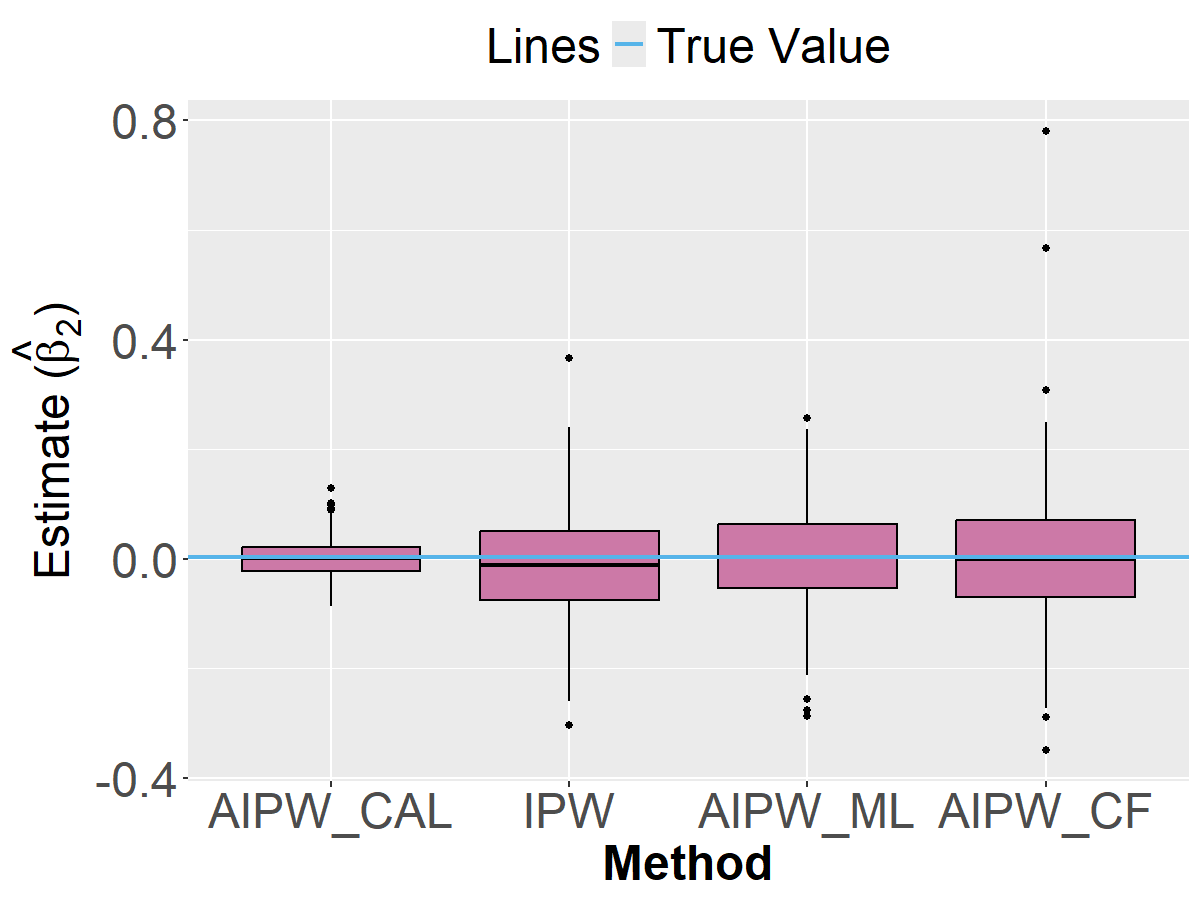}
    \caption{ $\hat{\beta}_{2}$}
\end{subfigure}
\begin{subfigure}{0.49\textwidth}
\centering   \includegraphics[width=.9\textwidth]{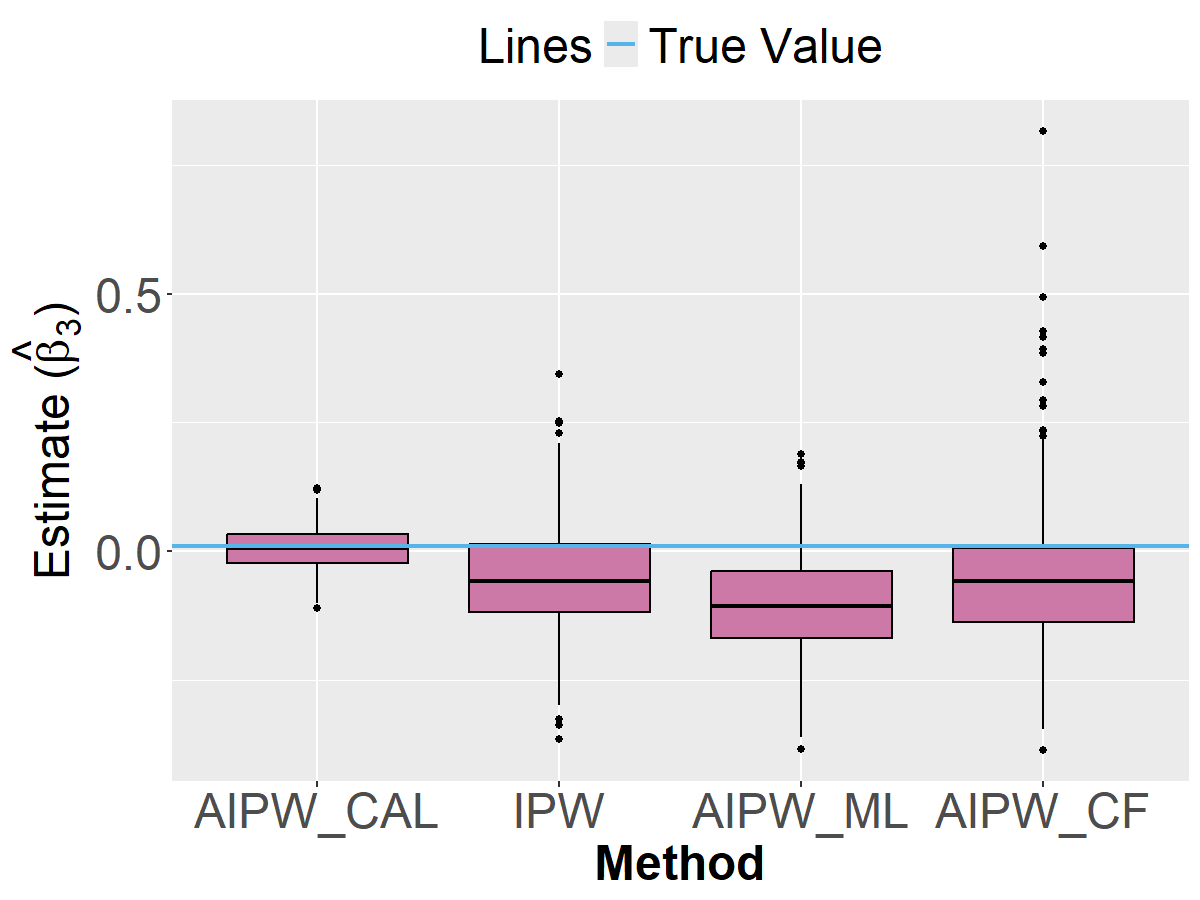}
    \caption{$\hat{\beta}_{3}$}
\end{subfigure}
\caption{Box plots of estimates of $\beta_0,\beta_1,\beta_2,\beta_3$ in Case 5}\label{fig:bp-3}
\end{figure}

{\bf Results for Study II.}  The simulation results for Cases 3--5 are presented in Tables \ref{tab:x1}--\ref{tab:X}, and the corresponding box plots are displayed in Figures \ref{fig:bp-2}--\ref{fig:bp-3}. We observe similar patterns as those in Cases 1 and 2: the proposed method performs well in terms of all metrics.

\begin{table}[ht!]
\centering
\caption{Summary of estimates of $\beta_1$ in Cases 3 and 4.} \label{tab:x1}  \vspace{-.1in}
\footnotesize
\begin{center}
\begin{tabular*}{\textwidth}{@{\extracolsep\fill} l@{\hskip 0.05in} c l@{\hskip 0.05in} ccc @{\hskip 0.05in} c l@{\hskip 0.05in} ccc} \hline
      &
      \multicolumn{3}{c}{\textbf{Case 3}} &&
      \multicolumn{3}{c}{\textbf{Case 4}}&\\
      \cline{2-4}\cline{6-8}
      & $\mathrm{AIPW_{\mathrm{RCAL}}}$ & $\mathrm{AIPW_{RML}}$ & $\mathrm{AIPW_{CF}}$ && $\mathrm{AIPW_{\mathrm{RCAL}}}$ & $\mathrm{AIPW_{RML}}$ & $\mathrm{AIPW_{CF}}$ & \\
     \cline{2-4}\cline{6-8}
    Bias & 0.000  & -0.011 & -0.004 && 0.002 & -0.010 & -0.002  \\
    $\sqrt{\text{Var}}$ & 0.036 & 0.043 & 0.058 &   & 0.023 & 0.031 &  0.042 \\
    $\sqrt{\text{EVar}}$ & 0.036 & 0.037 & 0.056 & & 0.021 &
 0.025 &  0.042 \\
    $\text{CP90}$ & 0.884&0.824 &0.840  & & 0.866 &0.774& 0.838\\
    $\text{CP95}$ &  0.946  & 0.886 & 0.900 &&0.930 &0.834 & 0.906\\\hline
\end{tabular*} \\[.01in]
\end{center}  \vspace{-.1in}
\end{table}

\begin{table}[ht!]
\centering
\caption{Summary of estimates of $\beta_0,\beta_1,\beta_2,\beta_3$ in Case 5} \label{tab:X}  \vspace{-.1in}
\footnotesize
\begin{center}
\begin{tabular*}{\textwidth}{@{\extracolsep\fill} l@{\hskip 0.05in} c l@{\hskip 0.05in} ccc @{\hskip 0.05in} c l@{\hskip 0.05in} ccc} \hline
      &
      \multicolumn{3}{c}{$\beta_{0}$} &&
      \multicolumn{3}{c}{$\beta_{1}$}&\\
      \cline{2-4}\cline{6-8}
      & $\mathrm{AIPW_{\mathrm{RCAL}}}$ & $\mathrm{AIPW_{RML}}$ & $\mathrm{AIPW_{CF}}$ && $\mathrm{AIPW_{\mathrm{RCAL}}}$ & $\mathrm{AIPW_{RML}}$ & $\mathrm{AIPW_{CF}}$ & \\
     \cline{2-4}\cline{6-8}
    Bias &  0.001 & 0.001 &  0.013&& -0.007  & -0.004 & 0.000 \\
    $\sqrt{\text{Var}}$ & 0.024 & 0.068  & 0.084 && 0.036 & 0.114  & 0.139  \\
    $\sqrt{\text{EVar}}$ & 0.025 & 0.055  & 0.077&& 0.036  & 0.089  & 0.131 \\
    $\text{CP90}$ & 0.910 & 0.800 & 0.846  && 0.886 & 0.774  & 0.830\\
    $\text{CP95}$ &  0.958 &0.874 & 0.914 &&0.942 & 0.842 & 0.900 \\\hline
      &
      \multicolumn{3}{c}{$\beta_{2}$} &&
      \multicolumn{3}{c}{$\beta_{3}$}&\\
      \cline{2-4}\cline{6-8}
      & $\mathrm{AIPW_{\mathrm{RCAL}}}$ & $\mathrm{AIPW_{RML}}$ & $\mathrm{AIPW_{CF}}$ && $\mathrm{AIPW_{\mathrm{RCAL}}}$ & $\mathrm{AIPW_{RML}}$ & $\mathrm{AIPW_{CF}}$ & \\
     \cline{2-4}\cline{6-8}
    Bias & -0.001  & 0.002  & 0.001  && -0.006 & 0.114 & -0.060  \\
    $\sqrt{\text{Var}}$ & 0.033  & 0.086 & 0.111  && 0.040  & 0.095 & 0.132 \\
    $\sqrt{\text{EVar}}$ &  0.034  & 0.075  & 0.108  && 0.041  & 0.095  &0.133 \\
    $\text{CP90}$ & 0.920   & 0.818  & 0.866 && 0.898  & 0.592&0.774 \\
    $\text{CP95}$ & 0.958 & 0.902 & 0.938 && 0.940 &0.682  & 0.842\\\hline
\end{tabular*} \\[.01in]
\end{center}  \vspace{-.1in}
\end{table}

\section{Application}\label{sec-application}
\subsection{Data description}

The Communities and Crime dataset comprises 1994 records of crime-related information from communities in the United States, which combine socio-economic data from the 1990 US Census, law enforcement data from the 1990 US LEMAS survey, and crime data from the 1995 FBI UCR.
Each record includes a response variable \texttt{ViolentCrimesPerPop}\footnote{total number of violent crimes per 100,000 population} and 127 covariates, encompassing both location information (such as \texttt{state} and \texttt{county}) and socio-economic factors (such as \texttt{PctTeen2Par}\footnote{percent of kids age 12-17 in two parent households}, \texttt{HousVacant}\footnote{number of vacant households}, etc.). In this study, we are interested in examining the influence of univariate covariates on the response. We consider the case where $\bm{Z} =(1, X_{i})^{\T}$ for a particular univariate covariate $X_{i}$ as discussed in Section~\ref{sec:dtp}, and we denote $\beta^{*} = (\beta_{0}, \beta_{1})$.

Due to the presence of numerous missing values in high-dimensional covariates, we eliminate covariates with high missing ratios. See details of the pre-rocessing procedure in Section \ref{sec:pre-p} of Supplement.
After pre-processing, the analytical dataset consists of 1993 observations and 26 covariates (i.e., $d=26$). The shift in covariates $\bm{X}$ is naturally introduced by the different states where the communities are located.\footnote{Notice that no covariate related to locations of communities is included in $\bm{X}$.}
We set label indicators $R$ for communities in New Jersey (Code 34) to be 1 and those for communities in other states to be 0 and remove the associated response data if $R=0$, resulting in 211 labeled observations and 1782 unlabeled observations. The covariate shift of the joint distribution of $\bm{X}$ was confirmed to exist using a Gaussian kernel two-sample test with maximum mean discrepancy~\citep{You2020ToolsFM}. Additionally, we assess the shift of each individual covariate by a bootstrap version of the Kolmogorov--Smirnov test \citep{JSSv042i07}. For results of those tests, please see Section \ref{sec:test}  of Supplement.

We randomly take 90\% of labeled data and 90\% of unlabeled data to form the training set with the remaining data used for the testing set.
From the remaining 26 covariates, we select four representative ones: \texttt{PctTeen2Par}, \texttt{HousVacant},  \texttt{PctHousNoPhone}\footnote{percent of occupied housing units without phone} and \texttt{PopDens} \footnote{population density in persons per square mile}, which illustrate different aspects of the socio-economic characteristics of communities. Notice that the covariate shift exists in all four covariates.

We compare the proposed method with $\mathrm{IPW}$, $\mathrm{AIPW_{RML}}$ and $\mathrm{AIPW_{CF}}$ methods
with piecewise linear basis functions introduced in Section~\ref{sec:choice-bf}. The PS and OR working models are estimated the same way as described there. For details of the procedures for designing basis functions, please see  Section \ref{sec:dbf} of Supplement.

\subsection{Results}
Table~\ref{tab:a1} presents the estimates of the regression coefficients $\hat{\beta}_{1}$ along with the prediction mean squared error (MSE),
which are calculated using the test data. It reveals that the point estimates of the regression coefficient are similar across the different methods. Notably, our estimators achieve the lowest prediction MSE except \texttt{PctTeen2Par}, highlighting the superior performance of our methods in minimizing predictive errors.
\begin{table}[ht!]
\centering
\caption{Summary of $\hat{\beta}_{1}$ and prediction MSE} \label{tab:a1}  \vspace{-.1in}
\footnotesize
\begin{center}
\begin{tabular*}{\textwidth}{@{\extracolsep\fill} l@{\hskip 0.05in} c l@{\hskip 0.05in} cccc @{\hskip 0.05in} c l@{\hskip 0.05in} cccc} \hline
    &
    \multicolumn{4}{c}{$\hat{\beta}_{1}$} &&
    \multicolumn{4}{c}{prediction MSE}&\\
    \cline{2-5}\cline{7-10}
    & $\mathrm{AIPW_{\mathrm{RCAL}}}$ &$\mathrm{IPW}$ &$\mathrm{AIPW_{RML}}$ & $\mathrm{AIPW_{CF}}$ && $\mathrm{AIPW_{\mathrm{RCAL}}}$ &$\mathrm{IPW}$ & $\mathrm{AIPW_{RML}}$ & $\mathrm{AIPW_{CF}}$ & \\
     \cline{2-5}\cline{7-10}
      PctTeen2Par & -0.137  & -0.172 & -0.147 &-0.256 &&0.034& 0.032 &0.034 &0.044 \\
   HousVacant &0.107 & 0.261 & 0.073 &0.133 && 0.046 & 0.085  & 0.046& 0.047 \\
   PctHousNoPhone   & 0.123 & 0.245 &0.103 &0.050 && 0.036 &0.058  &0.039 & 0.047 \\
    PopDens& 0.045 &0.069  &0.048 & 0.039 &&0.050 & 0.052 &0.052 & 0.055\\
    \hline
\end{tabular*} \\[.01in]
\end{center}  \vspace{-.1in}
\end{table}

Moreover, signs of estimates of coefficients are the same among different methods for each covariate $Z$ of interest.
and they coincide with common sense and previous studies.  For example, the coefficients of \texttt{PctTeen2Par} are negative, since it is believed to have protective effects in assaults~\citep{LUO201783}; the coefficients of HousVacan is positive, and criminological theories predict a positive association between vacancy and crime since
empty structures of houses could provide locations for some crimes (e.g., prostitution, drug dealing), and the absence of residents may prevent social organization and reduce
guardianship~\citep{roth2019empty}. Moreover, $\mathrm{AIPW_{RML}}$ and ours are close in all cases, while $\mathrm{IPW}$ estimators and $\mathrm{AIPW_{CF}}$ estimators are far from others in some cases.

In Figure~\ref{fig:A}, we compare the 95\%  CIs of $\mathrm{AIPW_{\mathrm{RCAL}}}$, $\mathrm{AIPW_{RML}}$ and $\mathrm{AIPW_{CF}}$. From the CIs, we see that for $\mathrm{AIPW_{RML}}$ and our estimators all four single effects are significant. CIs of our estimators and of $\mathrm{AIPW_{RML}}$'s have similar lengths and are overlapped, except HousVacant. The reason of the small difference is that the estimates of $\beta_{0}$ are a bit different. CIs of $\mathrm{AIPW_{CF}}$ are much longer; for PctHousNoPhone and PopDens, the estimates are not significant.  Both phenomenons show that $\mathrm{AIPW_{CF}}$ is not as efficient as other two methods.

\begin{figure}[H]
\centering
\begin{subfigure}{0.49\textwidth}
\centering
    \includegraphics[width=.9\textwidth]{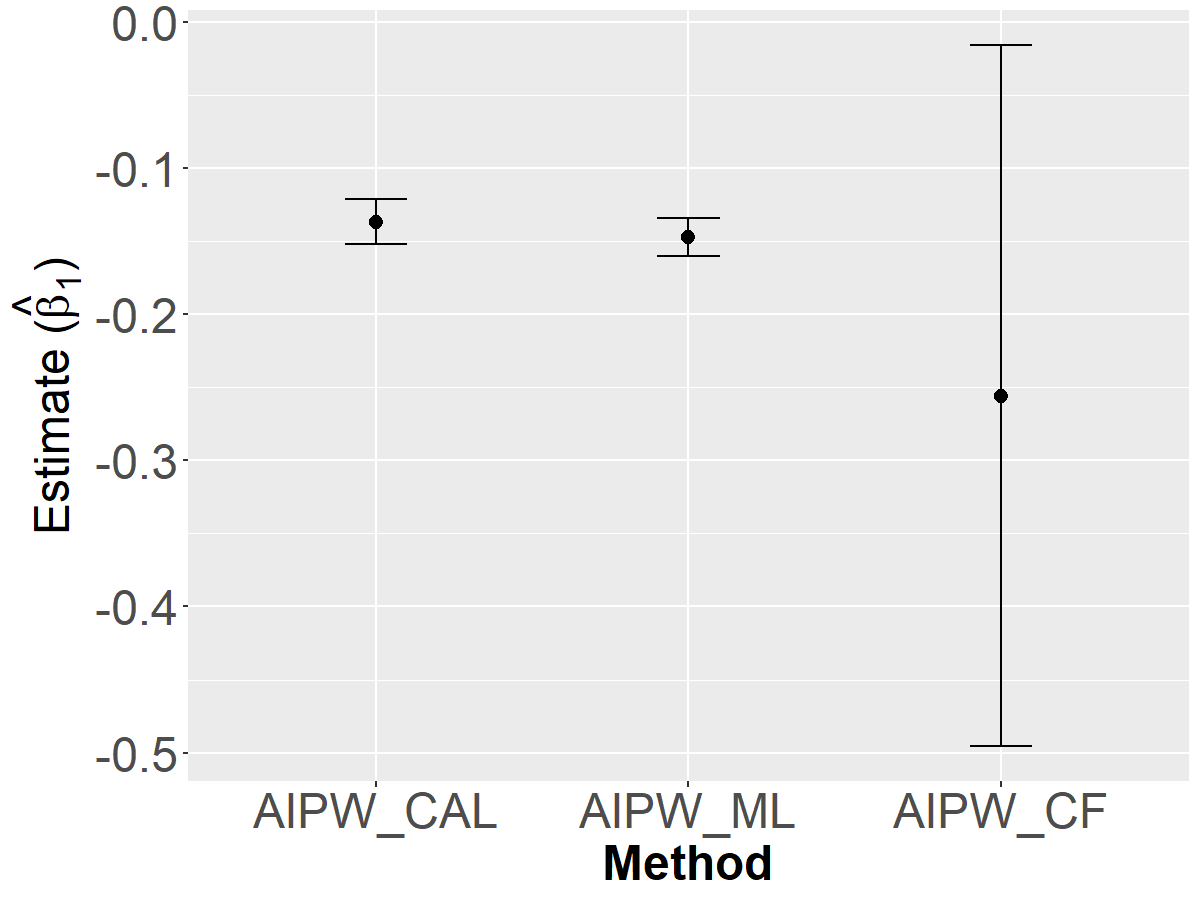}
    \caption{PctTeen2Par}
\end{subfigure}
\begin{subfigure}{0.49\textwidth}
\centering
    \includegraphics[width=.9\textwidth]{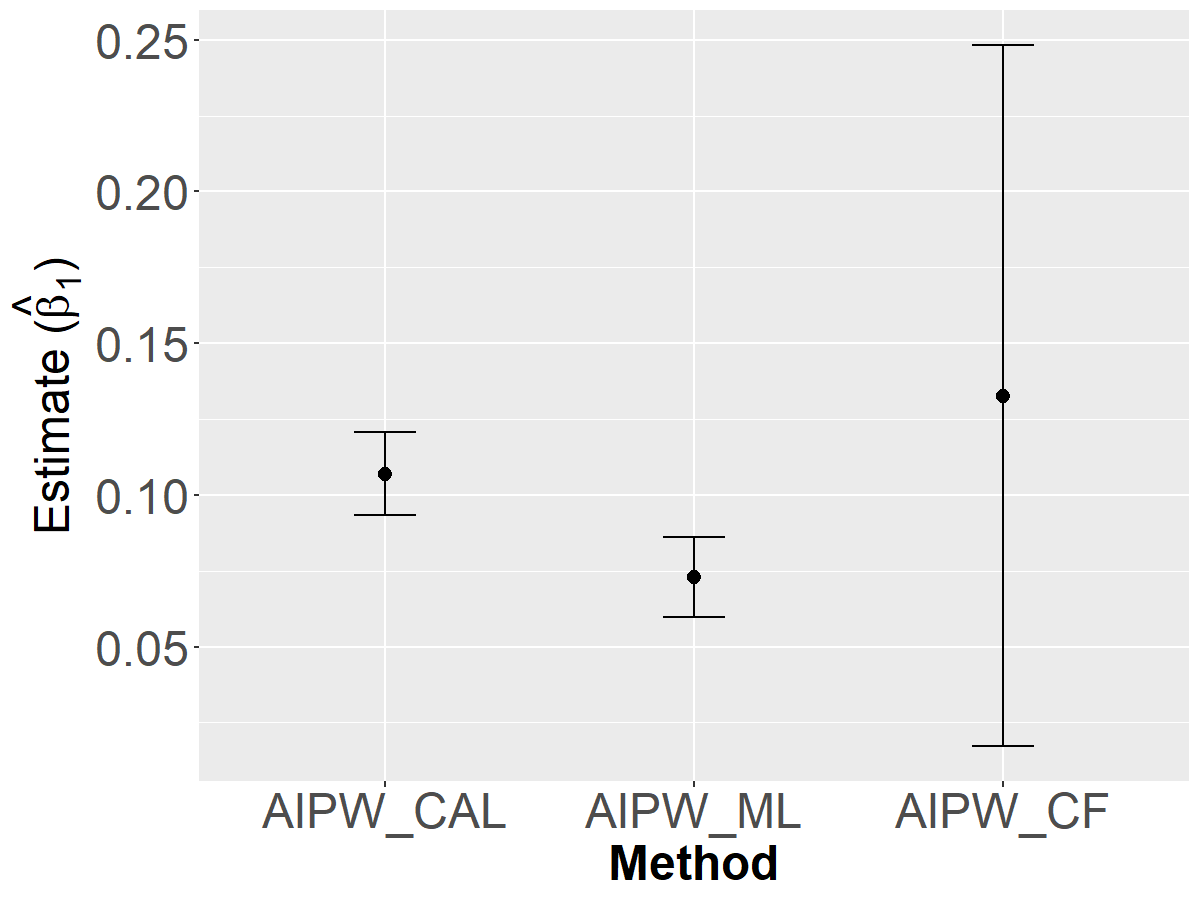}
    \caption{HousVacant}
\end{subfigure}
\end{figure}
\begin{figure}[H]
\ContinuedFloat
\centering
\begin{subfigure}{0.49\textwidth}
\centering
    \includegraphics[width=.9\textwidth]{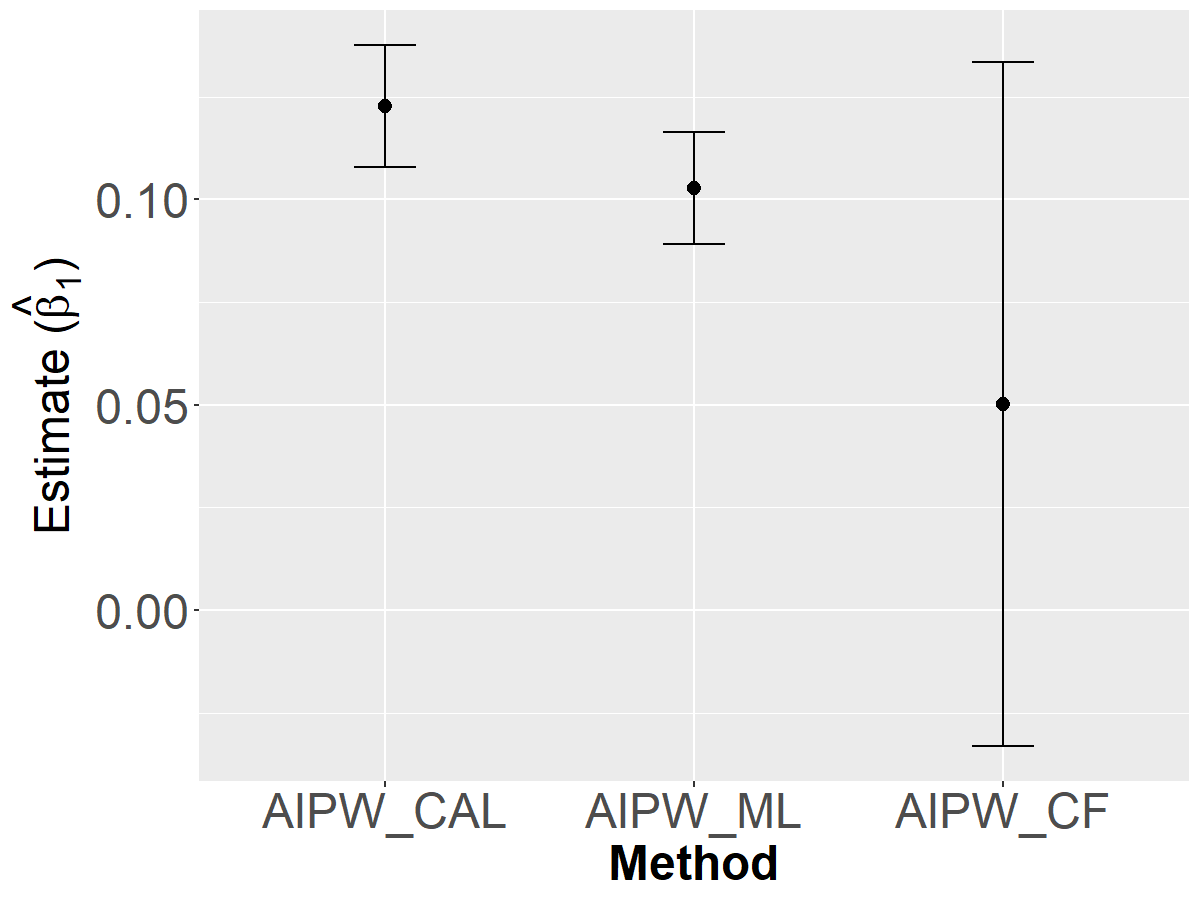}
    \caption{PctHousNoPhone}
\end{subfigure}
\begin{subfigure}{0.49\textwidth}
\centering   \includegraphics[width=.9\textwidth]{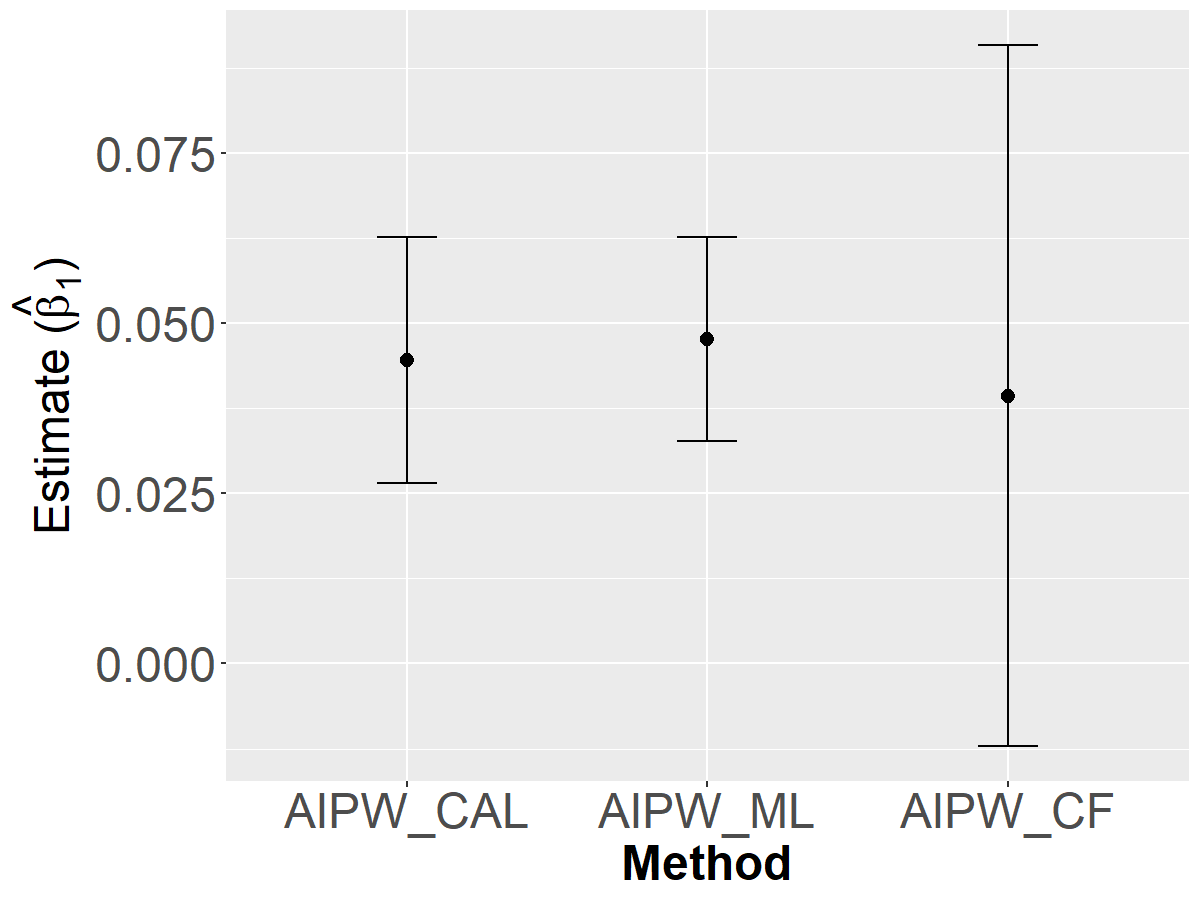}
    \caption{PopDens}
\end{subfigure}
\caption{Comparison of 95\% CIs of $\mathrm{AIPW_{\mathrm{RCAL}}}$, $\mathrm{AIPW_{RML}}$ and $\mathrm{AIPW_{CF}}$}\label{fig:A}
\end{figure}

\section{Extension to estimation of $\beta^{0*}$}  \label{sec:ext}

Consider the estimation of $\beta^{0*}$, defined as a solution to estimating equation \eqref{eq:def:b0}.
Under Assumption \ref{ass:model},
 $  \E \left [ R\{1-\pi^{*}(\bm{X})\}/\pi^*(\bm{X}) \left \{Y - \psi(\beta^{\mathrm{\scriptscriptstyle T}}\bm{Z}) \right \} \bm{Z} \right ] = \E \left [ \{1-R\}\left \{Y - \psi(\beta^{\mathrm{\scriptscriptstyle T}}\bm{Z}) \right \} \bm{Z} \right ]. $
Then a natural sample estimating equation for $\beta^{0*}$ is $
    \tilde \E \left [R\{1-\hat{\pi}(\bm{X})\}/\hat{\pi}(\bm{X}) \{ Y - \m \} \bm{Z} \right ] = 0$. We augment the estimating equations similarly as described in Section \ref{sec:aipw-construction} and obtain the sample AIPW estimating equations:
\begin{align}\label{eq:def-emp-AIPW-u}
\te \left [\frac{R\{1 - \hat{\pi}(\bm{X}) \}}{\hat{\pi}(\bm{X})} \left \{Y -\m \right \}  \bm{Z} +\left\{1 - \frac{R}{\hat{\pi}(\bm{X})}\right \}\left \{\hat{\phi}(\bm{X}) - \m \right\}  \bm{Z} \right ]=0.
\end{align}
For the PS and OR models, we adopt a similar construction as in Section \ref{sec:method}. Our AIPW estimator for $\beta^{0*}$, $\hat{\beta}^{0}$, is defined as the solution to the following estimating equations:
\begin{equation}
\tilde \E \{\tau^{0}(\bm{O}, \esa, \beta, \esg)\} = 0,
\end{equation}
where
\begin{align*}
 \tau^{0}(\bm{O}, \alpha, \beta, \gamma) = \left [\frac{R\{ 1-  \p \}}{\p} \{ Y - \m  \}  + \left\{1 - \frac{R}{\p} \right \}\left \{  \psin - \m \right \} \right ] \bm{Z}.
\end{align*}
The following result is an extension of Theorem \ref{thm:psc-h-p} to the estimation of $\beta^{0*}$.
\begin{thm} \label{thm:psc-h-u}
Under Assumptions \ref{ass:model}--\ref{ass:x}, if the PS model \eqref{eq:def-ps} is correctly specified with $\pi(\cdot;\bar\gamma) = \pi^*(\cdot)$, and $\ln\{(1 + p)/\epsilon\}/N < 1$,
then the following results hold.

(i) The estimator $\hat \beta^{0}$ is consistent and asymptotically normal, and
\begin{equation*}
    \sqrt{N}(\esb^{0} - \beta^{0*}) \xrightarrow{d} \N(0, \bm{\Sigma^{0}}),
\end{equation*}
where
$\bm{\Sigma^{0}} = \bm{\Gamma}^{0-1}\bm{\Lambda}^{0}\bm{\Gamma}^{0-1}$ with $\bm{\Gamma}^{0} = \E [\{1 - \pt\}  \psi_{1}(\beta^{0*\mathrm{\scriptscriptstyle T}}\bm{Z}) \bm{Z} \bm{Z}^{\T}]$ and
\[ \bm{\Lambda}^{0} =  \E\left \{ \tau^{0}(\bm{O}, \bar \alpha, \beta^*, \bar \gamma) \tau^{0}(\bm{O}, \bar \alpha, \beta^*, \bar \gamma)^\mathrm{\scriptscriptstyle T} \right \}. \]

(ii) A consistent estimator of $\bm{\Sigma^{0}}$ is $\hat{\bm{\Sigma}}^{0}= \hat{\bm{\bm{\Gamma}}}^{0-1}\hat{\bm{\Lambda}}^{0} \hat{\bm{\bm{\Gamma}}}^{0-1}$, where
$  \hat{\bm{\bm{\Gamma}}}^{0} = \tilde \E \{ \psi_{1}(\hat{\beta}^{0\mathrm{\scriptscriptstyle T}}\bm{Z})\bm{Z} \bm{Z}^{\T}\}$ and
 \begin{align*}
 \hat{\bm{\Lambda}}^{0} ={}& \tilde \E \left \{ \tau^{0}(\bm{O}, \hat \alpha, \hat{\beta}^{0\mathrm{\scriptscriptstyle T}}, \hat \gamma) \tau^{0}(\bm{O}, \hat \alpha, \hat{\beta}^{0\mathrm{\scriptscriptstyle T}}, \hat  \gamma)^\mathrm{\scriptscriptstyle T} \right \}.
 \end{align*}
Thus, for a constant vector $\bm{c}$ with the same dimension of $\beta$, an asymptotic $(1-\eta)$ confidence interval for $\bm{c}^{\T}\beta^{0*}$ is $\bm{c}^{\T}\hat{\beta}^{0\mathrm{\scriptscriptstyle T}}\pm z_{\eta/2}\sqrt{\bm{c}^\mathrm{\scriptscriptstyle T} \hat{\bm{\Sigma^{0}}} \bm{c}/N}$.
\end{thm}

Theorem \ref{thm:psc-h-u} shows that if the PS model is correct, regardless of the correctness of the OR working model, the proposed estimator $\hat{\beta}^{0}$ is consistent and asymptotically normal, and the proposed CIs based on $\hat{\bm{\Sigma}}^{0}$ are valid.
 Similarly to the estimation of $\beta^{*}$, the conclusions in Theorem \ref{thm:psc-h-u} also hold in low-dimensional settings with a reduced form of Assumptions \ref{ass:p}, \ref{ass:o} and \ref{ass:x}.

We point out that the method of \citet{liu2023} for CSTL can be viewed as an AIPW estimator of $\beta^{0*}$ under the stratified sampling setting, where the labeled and unlabeled datasets $\mathcal{L}$ and $\mathcal{U}$ are treated as two independent samples of fixed sizes $n$ and $N-n$.
They employed partial linear models for both PS and OR working models. By replacing their choices of semi-parametric nuisance models with our parametric models, the estimator of $\beta^{0*}$ in \citet{liu2023} can be reformulated as the solution to the following estimating equations:
\begin{equation} \label{eq:aipw-l}
\frac{1}{n} \sum_{i = 1}^{n} w(\bm{X}_{i}; \esg^{s}) \left [ \{ Y_{i} - \psi(\esa^{\T}\bm{G}_{i})  \} \bm{Z}_{i} \right ] + \frac{1}{N - n}\sum_{i=n+1}^{N} \left [\left \{\psi(\esa^{\T}\bm{G}_{i}) - \psi(\beta^{\T}\bm{Z}_{i})  \right \}\bm{Z}_{i} \right ] = 0.
\end{equation}
where $w( \bm{X}_i;\esg^{s}) = \exp(-
\esg^{s\T}\bm{F}_{i})$ and $\bm{F}_{i}$ is the abbreviation of $\bm{F(X_{i})}$; $\hat\gamma^s = (\hat\gamma^{s}_0, \hat\gamma^{s\T}_{1:p})^\T$ is an estimator of the parameter $\gamma^{s}$ in an exponential tilt model, defined as
\begin{equation}\label{eq:def-ept}
\dif G_{1} = \exp(\gamma^{s}_{0} + \gamma^{s\T}_{1:p}\bm{F}_{1:p}) \dif G_{0},
\end{equation}
where $G_0$ and $G_1$ are two probability distributions for the unlabeled and labeled data in $\bm{F}_{1:p}$ and $\gamma^{s}_{0} = - \log{\left \{\int\exp(\gamma^{s\T}_{1:p}\bm{F}_{1:p}) \dif G_{0}\right\}}$ to ensure that
$\int \,\dif G_1=1$.
The exponential tilt model \eqref{eq:def-ept} can be shown to be equivalent to the logistic PS model \eqref{eq:def-ps}, where the coefficients are related as follows \citep{prentice1979, qin1998,tian2023}:
\begin{align}\label{eq:exp-lr-r}
\gamma_{0} = \gamma^{s}_{0} + \ln\left (\frac{\rho_{m}}{1-\rho_{m}} \right ), \; \;\;\; \gamma_{1:p} =  \gamma^{s}_{1:p},
\end{align}
where $\rho_{m} = \P(R = 1)$, the true value of the proportion of missing data. When analyzing the asymptotic property in stratified sampling settings, we assume $n/N$ to be constant and, consequently, assume that $\rho_{m}= n/N$.
On the other hand, our estimating equations \eqref{eq:def-emp-AIPW-u} can be rewritten as
\begin{equation}\label{eq:aipw-c-t}
\frac{1}{N} \sum_{i = 1}^{n} w(\bm{X}_{i}; \esg) \left [ \{ Y_{i} - \psi(\esa^{\T}\bm{G}_{i})  \} \bm{Z}_{i} \right ] + \frac{1}{N}\sum_{i=n+1}^{N} [\left \{\psi(\esa^{\T}\bm{G}_{i}) - \psi(\beta^{\T}\bm{Z}_{i})  \right \}\bm{Z}_{i}] = 0.
\end{equation}
where $\hat\gamma = (\hat\gamma_0, \hat\gamma_{1:p})^\T$ is an estimator of the parameter $\gamma$ in logistic PS model \eqref{eq:def-ps}.
Suppose that the estimators $\hat\gamma^s$ and $\hat\gamma$ satisfy the same relationship as \eqref{eq:exp-lr-r}, i.e.,
$\hat{\gamma}_{0} = \hat{\gamma}^{s}_{0} + \ln(\frac{n}{N-n})$ and $\hat{\gamma}_{1:p} = \hat{\gamma}^{s}_{1:p}$.
Then, it is easily seen that $(N-n)/n\cdot w( \bm{X}_i;\esg^{s}) = \exp(-\hat{\gamma}^{\T}\bm{F}_{i}) =  w(\bm{X}_{i}; \esg)$,
and the two equations \eqref{eq:aipw-l} and \eqref{eq:aipw-c-t} match each other.
Therefore, the different forms of \eqref{eq:aipw-l} and \eqref{eq:aipw-c-t} can be explained by the relationship of the coefficient estimates
between the exponential tilt model \eqref{eq:def-ept} and the logistic regression model \eqref{eq:def-ps}.

\section{Summary}\label{sec:summary}

We present a new AIPW method for the inference of regression coefficients in (conditional) mean models in SSL and CSTL settings.
We demonstrate that various previous methods can be unified in our AIPW framework by suppressing detailed technical choices. By carefully exploiting the dependence of PS and OR models and designing estimating equations of nuisance parameters, our AIPW estimator achieves asymptotic normality, and valid CIs can be obtained, whether or not the OR working model is correctly specified, with high-dimensional data. Finite sample performances of the proposed method are confirmed by a simulation study and an application to a real-world dataset.

Currently, the proposed CIs can only achieve single robustness to the misspecification of the OR model.
Doubly robust CIs can be developed using the approach of \citet{Ghosh2022}, albeit at the cost of increasing technical and numerical complexities.
In addition, how to handle the case where $\lim_{n,N \rightarrow \infty} n/N \rightarrow 0$ under the random sampling process is also technically challenging, since the “positivity assumption” (Assumption \ref{assump2}) typical in missing data
theory is violated. New analysis needs to be developed to address the problem.

\bibliographystyle{apalike}
\bibliography{reference}

\clearpage

\setcounter{page}{1}

\setcounter{section}{0}
\setcounter{equation}{0}

\setcounter{figure}{0}
\setcounter{table}{0}

\renewcommand{\theequation}{S\arabic{equation}}
\renewcommand{\thesection}{\Roman{section}}
\renewcommand*{\theHsection}{\Roman{section}}

\renewcommand\thefigure{S\arabic{figure}}
\renewcommand\thetable{S\arabic{table}}

\setcounter{lem}{0}
\renewcommand{\thelem}{S\arabic{lem}}
\setcounter{ass}{0}
\renewcommand{\theass}{S\arabic{ass}}
\setcounter{cor}{0}
\renewcommand{\thecor}{S\arabic{cor}}
\setcounter{pro}{0}
\renewcommand{\thepro}{S\arabic{pro}}

\begin{center}
{\Large Supplementary Material for ``Semi-supervised Regression Analysis  \\
with Model Misspecification and High-dimensional Data"}

\vspace{.1in} {\large Ye Tian, Peng Wu and Zhiqiang Tan}
\end{center}

This  Supplementary Material consists of  Sections \ref{sec:sup-tec-tol}--\ref{sec:dp-cc}, where Section \ref{sec:sup-tec-tol} contains technical tools used in proofs of lemmas in Section \ref{sup:lem}. Section \ref{sup:lem} presents lemmas used in  proofs of Proposition \ref{prop:main-mix-p} and Theorem \ref{thm:psc-h-p}. Section \ref{sec:sup-pp-1} gives the technical proofs of Proposition \ref{prop:tan2017} and Proposition \ref{prop:main-mix-p}. Section \ref{sec:sup-pt} provides the proof of Theorem \ref{thm:psc-h-p}. Section \ref{sup:ess} includes contents of the extension to stratified sampling settings, including the proof of Proposition \ref{prop:ss} and the variance comparison. Section \ref{sec:dp-cc}  presents details of the application.

\section{Technical tools}\label{sec:sup-tec-tol}

We state the following concentration inequalities, to facilitate proofs of lemmas in Section \ref{sup:lem}, which can be obtained from \citet{buhlmann2011statistics} [Lemmas 14.11, 14.16 \& 14.9].

\begin{lem} \label{lem:sup-tec-tol-bound-b}
Let $(Y_{1}, \ldots, Y_{N})$ be independent variables such that $\E(Y_{i} )= 0$ for $i = 1, \ldots, n$ and $\max_{i = 1, \ldots, n} |Y_{i}| \leq c_{0}$ for some constant $c_{0}$. Then for any $t > 0$,
\begin{equation*}
 \P \left ( \left \vert \frac{1}{N} \sum_{i=1}^{N} Y_{i} \right \vert> t  \right ) \leq 2 \exp \left ( -\frac{nt^{2}}{2c^{2}_{0}} \right ).
\end{equation*}
\end{lem}

\begin{lem} \label{lem:sup-tec-tol-bound}
Let $(Y_{1}, \ldots, Y_{N})$ be independent variables such that $\E(Y_{i} )= 0$ for $i = 1, \ldots, n$ and $(Y_{1}, \ldots, Y_{N})$ are uniformly sub-gaussian: $\max_{i=1, \ldots, n} c^{2}_{1}\E \{ \exp(Y^{2}_{i}/c^{2}_{1}) -1 \} \leq c^{2}_{2}$. Then for any $t > 0$,
\begin{equation*}
    \P \left ( \left \vert \frac{1}{N} \sum_{i=1}^{N} Y_{i}\right \vert > t \right ) \leq 2 \exp \left\{ -\frac{nt^{2}}{8(c^{2}_{1} + c^{2}_{2})} \right \}.
\end{equation*}
\end{lem}

\begin{lem}\label{lem:sup-tec-cov-p}
Let $(Y_{1}, \ldots, Y_{N})$ be independent variables such that $\E(Y_{i} )= 0$ for $i = 1, \ldots, n$ and,
\begin{equation*}
\frac{1}{N} \sum_{i=1}^{N}\E(|Y_{i}|^{k}) \leq \frac{k!}{2}c^{k-2}_{3}c^{2}_{4}, \quad k = 2, 3, \ldots ,
\end{equation*}
for some constants $(c_{3}, c_{4})$. Then
for any $t > 0$,
\begin{equation*}
    \P \left ( \left \vert \frac{1}{N} \sum_{i=1}^{N} Y_{i}\right \vert > c_{3}t + c_{4}\sqrt{2t} \right ) \leq 2 \exp \left(- nt\right ).
\end{equation*}
\end{lem}

\begin{lem}\label{lem:sup-tec-cov-cond}
Suppose that $|X| \leq c_{5}$, $c_{5}$ is some constant, and $Y$ is sub-gaussian: $c^{2}_{1} \E\{ \exp{(X^{2}/c^{2}_{1})} $

\noindent $ -1\} \leq c^{2}_{2}$ for some constants $(c_{1}, c_{2})$. Then $Z=XY^{2}$ satisfies
\begin{equation*}
\E\{|Z - \E(Z)|^{k}\} \leq \frac{k!}{2}c^{k-2}_{6}c^{2}_{7}, \quad k = 2, 3, \ldots ,
\end{equation*}
for $c_{6} = 2c_{5}c^{2}_{1}$ and $c_{7} = 2c_{5}c_{1}c_{2}$.
\end{lem}

\begin{lem}\label{lem:sup-nom-ub}
Suppose that $Y$ is sub-gaussian: $c^{2}_{1} \E\{ \exp{(Y^{2}/c^{2}_{1})} -1\} \leq c^{2}_{2}$ for some constants $(c_{1}, c_{2})$. Then
\begin{equation*}
    \E(|Y|^{k}) \leq \Gamma(\frac{k}{2} + 1)(c^{2}_{1} + c^{2}_{2})c^{k-2}_{1}, \quad k = 2,3 \ldots.
\end{equation*}
\end{lem}

\section{Technical lemmas}\label{sup:lem}

\subsection{Lemmas for the parameter in the PS model}

The following Lemmas \ref{lem:ps}--\ref{eq:sup-o-cov-ub} will be used in proofs of Proposition \ref{prop:tan2017} and Theorem \ref{thm:psc-h-p}. Lemma \ref{lem:sup-p-2} would be used in proofs of lemmas  in Section \ref{sec:orl} and Theorem \ref{thm:psc-h-p}.

\begin{lem}\label{lem:ps} 
Under Assumptions \ref{ass:p}$(\mathrm{i})$ and \ref{ass:p}$(\mathrm{ii})$, the following statements hold:

$(\mathrm{i})$  Denoted by $\Omega_{00}$ the event that
\begin{equation*}
\sup_{j = 0, \ldots, p} \left \vert \te \left [ \left \{ - R \wt + (1 - R) \right \} f_{j}(\bm X) \right ] \right \vert \leq \lambda_{0}.
\end{equation*}
 If $\lambda_{0} \geq \sqrt{2}\{1 + \exp(-B_{0})\} C_{0} \sqrt{\ln\{(1 + p) / \epsilon\} / N}$, then $\P(\Omega_{00}) \geq 1 - 2\epsilon$.

$(\mathrm{ii})$ Denote by $\Omega_{01}$ the event that
\begin{equation}\label{eq:sup-ps-cov-ub}
\sup_{j,k = 0, \ldots, p} \left \vert (\Tilde{\bm{\Sigma}}_{\gamma})_{jk} - (\bm{\Sigma}_{\gamma})_{jk} \right \vert \leq \lambda_{0},
\end{equation}
where $\Tilde{\bm{\Sigma}}_{\gamma}$ is the empirical version of $\bm{\Sigma}_{\gamma}$.
 If $\lambda_{0} \geq 4 \exp(-B_{0}) C^{2}_{0} \sqrt{\ln\{(1 + p) / \epsilon\} / N}$, then $\P(\Omega_{01}) \geq 1 - 2\epsilon^{2}$.
\end{lem}

The result of Lemma \ref{lem:ps}$(\mathrm{ii})$ is taken from \citeauthor{tan2020b}~(\citeyear{tan2020b})[Lemma 1$(\mathrm{ii})$], the result of Lemma \ref{lem:ps}$(\mathrm{i})$ can be shown
similarly using Lemma \ref{lem:sup-tec-tol-bound-b} in Section \ref{sec:sup-tec-tol} and the union bound.

Let $\bm{\Sigma}_{m^{1}} = \E \{R \wt | Y - \psit | \bm{F}\bm{F}^{\T}\}$, and $\Tilde{\bm{\Sigma}}_{m^{1}}$ be the sample version of $\bm{\Sigma}_{m^{1}}$.

\begin{lem}\label{lem:sup-cov-m1-ub}
Let $\Omega_{02}$ denote the event that
\begin{align} \label{eq:sup-cov-m1-ub}
\sup_{j,k = 0, \ldots, p}|(\bm{\Sigma}_{m^{1}})_{jk} - (\Tilde{\bm{\Sigma}}_{m^{1}})_{jk}| \leq \sqrt{D^{2}_{0} + D^{2}_{1}}\lambda_{0}.
\end{align}
Under Assumptions \ref{ass:p}$(\mathrm{i})$, \ref{ass:p}$(\mathrm{ii})$ and \ref{ass:o}$(\mathrm{iv})$, if 
\begin{equation*}
\lambda_{0} \geq  4 \exp(-B_{0}) C^{2}_{0}\sqrt{\frac{\ln \left\{ \frac{(1 + p)}{\epsilon} \right \}}{N}},
\end{equation*}
then $\P(\Omega_{02}) \geq 1 - 2\epsilon^{2}$.
\end{lem}

\begin{prf}
Since $|R \wt f_{j}(\bm{X}) f_{k}(\bm{X})| \leq \exp(-B_{0})C^{2}_{0}$ for $j, k = 0, \ldots, p$ by Assumptions \ref{ass:p}$(\mathrm{i})$, and \ref{ass:p}$(\mathrm{ii})$, and $| Y - \psit |$ is uniformly sub-gaussian by Assumption \ref{ass:o}$(\mathrm{iv})$, $R \wt | Y - \psit |$

\noindent$  f_{j}(\bm{X})f_{k}(\bm{X})$ is uniformly sub-gaussian. By Lemma \ref{lem:sup-tec-tol-bound}, we have
\begin{equation*}
\P\left \{|(\bm{\Sigma}_{m^{1}})_{jk} - (\Tilde{\bm{\Sigma}}_{m^{1}})_{jk}| > t \right \} \leq \frac{2\epsilon^{2}}{(1 + p)^{^{2}}},
\end{equation*}
for $j,k = 0, \ldots, p$, where $t =  \exp(-B_{0})C^{2}_{0}\sqrt{8(D^{2}_{0} + D^{2}_{1})} \sqrt{\ln\{ (1 + p)^{2} /\epsilon^{2}\}/N}$. By union bounds, \eqref{eq:sup-cov-m1-ub} holds.
\end{prf}

Let $\bm{\Sigma}_{m^{2}} = \E [R \wt \{ Y - \psit \}^{2}\bm{F}\bm{F}^{\T}]$, the sample version of $\bm{\Sigma}_{m^{2}}$, $\Tilde{\bm{\Sigma}}_{m^{2}} = \te [R \wt $

\noindent$\{ Y - \psit \}^{2}\bm{F}\bm{F}^{\T}]$.

\begin{lem} \label{eq:sup-o-cov-ub}
Denote by $\Omega_{03}$ the event that
\begin{equation}
\sup_{j,k = 0, \ldots, p}|(\bm{\Sigma}_{m^{2}})_{jk} - (\Tilde{\bm{\Sigma}}_{m^{2}})_{jk}| \leq (D^{2}_{0} + D_{0} D_{1})\lambda_{0},
\end{equation}
 Under Assumptions \ref{ass:p}$(\mathrm{i})$,  \ref{ass:p}$(\mathrm{ii})$ and \ref{ass:o}$(\mathrm{iv})$, if
\begin{equation*}
(D^{2}_{0} + D_{0} D_{1})\lambda_{0} \geq 4 \exp(-B_{0}) C^{2}_{0}[D^{2}_{0}\ln\{(1 + p) / \epsilon\} / N + D_{0} D_{1} \sqrt{\ln\{(1 + p) / \epsilon\} / N}],
\end{equation*}
then, $\P(\Omega_{03}) \geq 1 - 2\epsilon^{2}$. Furthermore, if we assume that $\ln\{ (1 + p)/\epsilon \}/N < 1$, the above condition reduces to 
$
\lambda_{0} \geq 4 \exp(-B_{0}) C^{2}_{0} \sqrt{\ln\{(1 + p) / \epsilon\} / N}.
$
\end{lem}

\begin{prf}
For $j,k = 0, \ldots, p$, the variable $R \wt \{ Y - \psit \}^{2}f_{j}(\bm{X})f_{k}(\bm{X})$ is the product of $R \wt f_{j}(\bm{X})f_{k}(\bm{X})$ and $\{ Y - \psit \}^{2}$, where $|R \wt f_{j}(\bm{X}) f_{k}(\bm{X})|$

\noindent$\leq \exp(-B_{0})C^{2}_{0}$ by Assumptions \ref{ass:p}$(\mathrm{i})$ and  \ref{ass:p}$(\mathrm{ii})$; and $\{ Y - \psit \}$ is sub-gaussian by Assumption \ref{ass:o}$(\mathrm{iv})$. By Lemmas \ref{lem:sup-tec-cov-p} and \ref{lem:sup-tec-cov-cond} in Section \ref{sec:sup-tec-tol}, we have
\begin{align*}
&\P\left \{ |(\Tilde{\bm{\Sigma}}_{m^{2}})_{j,k} - \bm{\Sigma}_{m^{2}})_{j,k}| > 2e^{(-B_{0})} C^{2}_{0}D^{2}_{0}t + 2e^{(-B_{0})}  C^{2}_{0}D_{0}D_{1}\sqrt{2t} \right \} 
\leq 2 \frac{\epsilon^{2}}{(1 + p)^{2}},
\end{align*}
for $j, k = 0, \ldots, p$, where $t = \ln\{(1 + p)^{2} / \epsilon^{2}\} / N$. The result then follows from the union bound.
\end{prf}

\begin{lem}\label{lem:sup-p-2}
Under the conditions of Proposition \ref{prop:tan2017}, in the event $\Omega_{00} \cap \Omega_{01}$, we have
\begin{equation}\label{eq:sup-psh-rate}
    \te \left [R \wt \{(\esg - \tg)^{\T}\bm F \}^{2}\right] \leq \exp(\eta_{01})M_{0}|S_{\tg}|\lambda_{0}^{2},
\end{equation}
where $\eta_{01} = (A_{0}-1)^{-1}M_{0}C_{0}\zeta_{0}$.
\end{lem}
\begin{prf}
By Lemma \ref{lem:ps} and Proposition \ref{prop:tan2017}, in the event $\Omega_{00} \cap \Omega_{01}$, \eqref{eq:gamma-rate} holds, we obtain
\begin{equation} \label{eq:sup-ps-gm-rate}
    \| \esg - \tg \|_{1} \leq (A_{0} - 1)^{-1} M_{0} |S_{\tg}| \lambda_{0} \leq (A_{0} - 1)^{-1} M_{0} \zeta_{0},
\end{equation}
the second inequality holds due to Assumption \ref{ass:p}$(\mathrm{iv})$.
By the definition of $D^{\dagger}_{\mathrm{CAL}}(\cdot, \cdot)$, we obtain
\begin{align*}
D^{\dagger}_{\mathrm{CAL}}(\esg^{\T}\bm{F},\tg^{\T}\bm{F}) = & -\te [R \{\exp( - \esg^{\T}\bm{F}) - \exp( - \tg^{\T}\bm{F})\} \{ (\esg - \tg)^{\T}\bm{F} \}]\\
= & \te (R \wt \exp[-u\{ (\esg - \tg)^{\T}\bm{F} \}]\{ (\esg - \tg)^{\T}\bm{F} \}^{2})\\
\geq & \exp(-C_{0}  \| \esg - \tg \|_{1})\te [R \wt \{ (\esg - \tg)^{\T}\bm{F} \}^{2}]\\
\geq & \exp(-\eta_{01})\te [R \wt \{ (\esg - \tg)^{\T}\bm{F} \}^{2}].
\end{align*}
The second equality holds by the mean value theorem and $u$ is some scalar in $(0, 1)$. Combining the inequality with \eqref{eq:gamma-rate}, we obtain
\begin{equation*}
\te \left [ R \wt \{(\esg - \tg)^{\T}\bm{F}\}^{2}\right] \leq \exp(\eta_{01}) D^{\dagger}_{\mathrm{CAL}}(\esg^{\T}\bm{F}, \tg^{\T}\bm{F}) \leq \exp(\eta_{01}) M_{0}|S_{\tg}|\lambda_{0}^{2}.
\end{equation*}
\end{prf}

\subsection{Lemmas for the parameter in the OR model}\label{sec:orl}

The following Lemmas \ref{lem:sup-a}--\ref{lem:sup-comp} will be used in proofs of Proposition \ref{prop:main-mix-p} and Theorem \ref{thm:psc-h-p}.

\begin{lem}\label{lem:sup-a}
Let $\Omega_{10}$ denote the event that
\begin{equation}\label{eq:sup-lo-up}
    \sup_{j = 0, \ldots, q} \left \vert \te \left [  R \wt \{ Y - \psit \}  g_{j}(\bm{X})\right ]  \right \vert \leq \lambda_{1}.
\end{equation}
Under Assumptions \ref{ass:p},  \ref{ass:o}$(\mathrm{iii})$ and  \ref{ass:o}$(\mathrm{iv})$, if
$$\lambda_{1} \geq \exp(-B_{0}) C_{4} \sqrt{8(D_{0}^{2} + D^{2}_{1})} \sqrt{\ln\{(1 + q) / \epsilon\} / N},$$ 
then $\P(\Omega_{10}) \geq 1 - 2 \epsilon$.
\end{lem}
\begin{prf}
Let $S_{j} = R \wt \{ Y - \psit \}  g_{j}(\bm{X})$ for $j = 0, \ldots, q$. Then, $\E(S_{j}) = 0$ by the definition of $\ta$. Under Assumptions \ref{ass:p} and  \ref{ass:o}$(\mathrm{iii})$, $|S_{j}| \leq \exp(-B_{0}) C_{4}|R \{ Y - \psit \}|$. By Assumption \ref{ass:o}$(\mathrm{iv})$, the variables $(S_{0}, \ldots, S_{q})$ are uniformly sub-gaussian: $\max_{j = 0, \ldots, q} D^{2}_{2}\E\{ \exp(S^{2}_{j}/D^{2}_{2}) -1\}\leq D^{2}_{3}$, with $D_{2} = \exp(-B_{0}) C_{4}D_{0}$, and $D_{3} = e^{(-B_{0})}$

\noindent$C_{4}D_{1}$.
Therefore, by Lemma \ref{lem:sup-tec-tol-bound} in Section \ref{sec:sup-tec-tol} and the union bound, $\P(\Omega_{10}) \geq 1 - 2 \epsilon$, if $\lambda_{1} \geq \exp(-B_{0}) C_{4} \sqrt{8(D_{0}^{2} + D^{2}_{1})} \sqrt{\ln\{(1 + q) / \epsilon\} / N}$.
\end{prf}

\begin{lem}
Denote by $\Omega_{11}$ the event that
\begin{equation} \label{eq:sup-cov_a}
\sup_{j,k = 0, \ldots, q}|(\bm{\Sigma}_{\ta})_{jk} - (\Tilde{\bm{\Sigma}}_{\ta})_{jk}| \leq \lambda_{1},
\end{equation}
where $\Tilde{\bm{\Sigma}}_{\ta}$ is the empirical version of $\bm{\Sigma}_{\ta}$. Under Assumptions \ref{ass:p}$(\mathrm{ii})$, \ref{ass:o}$(\mathrm{i})$ and \ref{ass:o}$(\mathrm{iii})$, if $\lambda_{1} \geq 4 \exp (-B_{0})   C_{2} C^{2}_{4}\sqrt{\ln\{(1 + q) / \epsilon\} / N}$, then $\P(\Omega_{11}) \geq 1 - 2\epsilon^{2}$.
\end{lem}

\begin{prf}
Notice that $\left \vert R \wt \psi_{1}( \ta^{\T}\bm{G}) g_{j}(\bm{X}) g_{k}(\bm{X})\right \vert  \leq e^{(-B_{0})} C_{2}C^{2}_{4}$ for $j,k = 0, 1,\ldots, q$, by Assumptions \ref{ass:p}$(\mathrm{ii})$, \ref{ass:o}$(\mathrm{i})$, and \ref{ass:o}$(\mathrm{iii})$. Thus,
\begin{align*}
&\left \vert R \wt \psi_{1}( \ta^{\T}G) g_{j}(\bm{X}) g_{k}(\bm{X}) - \E \left \{ R \wt \psi_{1}( \ta^{\T}\bm{G}) g_{j}(\bm{X}) g_{k}(\bm{X}) \right \}\right \vert \\
\leq & 2\exp(-B_{0})C_{2}C^{2}_{4}.   
\end{align*}
By union bounds and applying Lemma \ref{lem:sup-tec-tol-bound-b} yields \eqref{eq:sup-cov_a}.
\end{prf}

\begin{lem}
For any $\alpha \in \mathbb{R}^{q + 1}$, we have
\begin{equation}\label{eq:sup-ineq-sbd}
\begin{aligned} 
& D^{\dagger}_{\mathrm{WL}} (\esa^{T} \bm{G}, \alpha^{\T}\bm{G}; \esg) + \lambda_{\alpha} \| \esa_{1:q} \|_{1} \\
\leq & (\esa - \alpha)^{\T} \te [R \we \{ Y - \psin \}\bm{G}] + \lambda_{\alpha} \| \alpha_{1:q} \|_{1}.
\end{aligned}    
\end{equation}
\end{lem}

\begin{prf}
For any $u \in (0, 1]$, by definition of $\esa$, we have
\begin{align*}
    \ell_{\mathrm{WL}}(\esa, \esg) + \lambda_{\alpha} \| \esa_{1:q} \|_{1} \leq \ell_{\mathrm{WL}}\{ (1-u) \esa + u \alpha; \esg\} + \lambda_{\alpha} \| (1-u) \esa_{1:q} + u \alpha_{1:q} \|_{1},
\end{align*}
which implies
\begin{equation*}
\ell_{\mathrm{WL}}(\esa, \esg) - \ell_{\mathrm{WL}}\{ (1-u) \esa + u \alpha; \esg\} + \lambda_{\alpha} u \| \esa_{1:q} \|_{1} \leq \lambda_{\alpha} u \| \alpha_{1:q} \|_{1},
\end{equation*}
by the convexity of $L_{1}$-norm. Dividing both sides of the preceding inequality by $u$ and letting $u \rightarrow 0_{+}$ leads to
\begin{align*}
    -\te \left [ R \we \left \{ Y - \psie \right \} \left \{(\esa - \alpha)^{\T}\bm{G} \right \}\right] + \lambda_{\alpha} \| \esa_{1:q} \|_{1} \leq \lambda_{\alpha} \| \alpha_{1:q} \|_{1},
\end{align*}
which yields \eqref{eq:sup-ineq-sbd} after rearranging using \eqref{eq:bd_or}.
\end{prf}

\begin{lem}\label{lem:hp-ub}
For any function $h(\bm{X})$, under the conditions of Proposition \ref{prop:tan2017}, in the event $\Omega_{00} \cap \Omega_{01}$,
\begin{equation}\label{eq:sup-o-bd-lb}
    D^{\dagger}_{\mathrm{WL}} \{\esa^{\T} \bm{G}, h(\bm{X}); \esg\} \geq \exp(-\eta_{01})D^{\dagger}_{\mathrm{WL}} \{\esa^{\T}\bm{G}, h(\bm{X}); \tg \}.
\end{equation}
\end{lem}

\begin{prf}
By the definition of $D^{\dagger}_{\mathrm{WL}}(\cdot, \cdot; \cdot)$,
\begin{align*}
& D^{\dagger}_{\mathrm{WL}} \{\esa^{\T} \bm{G}, h(\bm{X}); \esg\}  \\
= & \te \left (R \we [\psi(\esa^{\T}\bm{G}) - \psi\{h(\bm{X})\}] \{ \esa^{\T}\bm{G} - h(\bm{X}) \} \right )\\
= & \te \left ( R  \wt \exp\{-(\esg - \tg)^{\T}\bm{F}\}  [\psi(\esa^{T}\bm{G}) - \psi \{h(\bm{X})\}] \{ \esa^{\T}\bm{G} - h(\bm{X}) \} \right )\\
\geq & \te \left ( R \wt \exp(-\eta_{01})  [\psi(\esa^{\T}\bm{G}) - \psi \{h(\bm{X})\}] \{ \esa^{\T}\bm{G} - h(\bm{X}) \} \right )\\
= & \exp(-\eta_{01}) \te (R \wt [\psi(\esa^{\T}\bm{G}) - \psi\{h(\bm{X})\}] \{ \esa^{\T}\bm{G} - h(\bm{X}) \})\\
= & \exp(-\eta_{01}) D^{\dagger}_{\mathrm{WL}} \{\esa^{\T}\bm{G}, h(\bm{X}); \tg\}.
\end{align*}
The inequality holds since in the event $\Omega_{00} \cap \Omega_{01}$, \eqref{eq:gamma-rate} holds.
\end{prf}

For functions $h(\bm{X})$ and $h'(\bm{X})$, let $Q_{\mathrm{WL}}\{h(\bm{X}), h'(\bm{X}); \cdot\} = \te [R w(\bm{X}; \cdot)  \{ h(\bm{X}) - h'(\bm{X}) \}^{2}]$.

\begin{lem} \label{lem:sup-o-d-p-rate}
Suppose Assumption \ref{ass:o}$(\mathrm{iv})$ holds, in the event $\Omega_{00} \cap \Omega_{01} \cap \Omega_{03}$, \eqref{eq:sup-ineq-sbd} implies
\begin{equation*}
\begin{aligned}
 & \exp(-\eta_{01}) D^{\dagger}_{\mathrm{WL}}(\esa^{\T} \bm{G}, \ta^{\T}\bm{G}; \tg) + \lambda_{\alpha} \| \esa_{1:q} \|_{1}\\
 \leq &  (\esa - \ta)^{\T} \te \left [ R \wt \{ Y - \psit \}\bm{G} \right ] +  \lambda_{\alpha} \| \ta_{1:q}\|_{1}\\
& + \exp(\eta_{01})\sqrt{M_{1}|S_{\tg}|\lambda_{0}^{2}} \{ Q_{\mathrm{WL}}( \esa^{\T}\bm{G}, \ta^{\T}\bm{G};\tg  )\}^{1/2},
\end{aligned}   
\end{equation*}
where $M_{1} = \left [(2D_{0}^{2} + D^{2}_{1} + D_{0}D_{1})\left \{\frac{M^{2}_{0} \zeta_{0}}{(A_{0}-1)^{2}}\right \}  + (D^{2}_{0} + D^{2}_{1})\exp(\eta_{01})M_{0}\right ]$.
\end{lem}

\begin{prf}
Consider the following decomposition,
\begin{equation}\label{eq:sup-decom}
\begin{aligned} 
& (\esa - \ta)^{\T} \te \left [ R \we\{ Y - \psit \}\bm{G} \right ]\\
= & (\esa - \ta)^{\T} \te \left [ R \wt \{ Y - \psit \} \bm{G} \right ]\\
& + \te \left [ R \left  \{ \we - \wt \right \} \{ Y - \psit \} \{ (\esa - \ta)^{\T}\bm{G}\} \right ] ,
\end{aligned} 
\end{equation}
denoted as $\Delta^{0}_{0} + \Delta^{0}_{1}$. By the mean value theorem and Cauchy-Schwartz inequality,
\begin{equation}\label{eq:sup-q-ub}
\begin{aligned}
    \Delta^{0}_{1} \leq & \exp(C_{0}\| \esg - \tg \|_{1}) \te^{1/2} [R \wt  \{ (\esa - \ta)^{\T}\bm{G}\}^{2}]\\
    & \times \te^{1/2} [R \wt  \{ Y - \psit \}^{2} \{(\esg - \tg)^{\T} \bm{F} \}^{2}] \\
    \leq & \exp(\eta_{01})\{ Q_{\mathrm{WL}}( \esa^{\T}\bm{G}, \ta^{\T}\bm{G} ;\tg  )\}^{1/2} \\
    & \times \te^{1/2} [R \wt \{ Y - \psit \}^{2} \{(\esg - \tg)^{\T}\bm{F} \}^{2}].
\end{aligned}  
\end{equation}

We bound the third term in \eqref{eq:sup-q-ub}. By Assumption \ref{ass:o}$(\mathrm{iv})$ and Lemma \ref{lem:sup-nom-ub}, we have
\begin{equation*}
\E[\{Y - \psit\}^{2}|\bm{X}] \leq D^{2}_{0} + D^{2}_{1}.
\end{equation*}
Therefore,
\begin{equation*}
 \E [R \wt \{ Y - \psit \}^{2} \{ (\esg - \tg)^{\T} \bm{F} \}^{2}]   \leq (D_{0}^{2} + D_{1}^{2}) \E [R \wt \{ (\esg - \tg)^{\T} \bm{F} \}^{2}].
\end{equation*}
Let $(\te - \E)(\bm{U})$ denote $\te \{ \bm{U} - \E(\bm{U}) \}$ for $\bm{U}$, which is a function of $(\bm{X}, R, Y)$. Then in the event $\Omega_{01}$, by \eqref{eq:sup-ps-cov-ub}, we have
\begin{equation*}
    (\E - \te) [R \wt \{ (\esg - \tg)^{\T} \bm{F} \}^{2}] \leq \lambda_{0} \| \esg - \tg \|^{2}_{1}.
\end{equation*}
In the event $\Omega_{03}$, by \ref{eq:sup-cov_a}, we have
\begin{equation*}
 (\te - \E) [R \wt \{ Y - \psit \}^{2} \{ (\esg - \tg)^{\T} \bm{F} \}^{2}]   \leq (D_{0}^{2} + D_{0} D_{1}) \lambda_{0} \| \esg - \tg \|^{2}_{1}.
\end{equation*}
Combining preceding inequalities, we obtain in $\Omega_{00} \cap \Omega_{01} \cap \Omega_{03}$,
\begin{equation}\label{eq:sup-decom-d1-ub}
\begin{aligned} 
& \te [R \wt \{ Y - \psit \}^{2} \{ (\esg - \tg)^{\T}\bm{F} \}^{2}]\\
\leq & (D_{0}^{2} + D_{0} D_{1}) \lambda_{0} \| \esg - \tg \|^{2}_{1}\\
+ & (D_{0}^{2} + D^{2}_{1}) \left \{ \lambda_{0} \| \esg - \tg \|^{2}_{1} + \te [R \wt \{(\esg - \tg)^{\T}\bm{F}\}^{2}]   \right \} \\
\leq & (2D_{0}^{2} + D^{2}_{1} + D_{0}D_{1})\left \{\frac{M^{2}_{0} \zeta_{0}}{(A_{0}-1)^{2}}\right \} |S_{\tg}|\lambda^{2}_{0} + (D^{2}_{0} + D^{2}_{1})\exp(\eta_{01})M_{0} |S_{\tg}|\lambda^{2}_{0}\\
= & M_{1} |S_{\tg}|\lambda^{2}_{0},
\end{aligned}   
\end{equation}
where the last inequality holds due to \eqref{eq:gamma-rate} and \eqref{eq:sup-psh-rate}. Combining \eqref{eq:sup-decom} -- \eqref{eq:sup-decom-d1-ub}, we obtain
\begin{equation}\label{eq:sup-decom-ub}
\begin{aligned}
& (\esa - \ta)^{\T} \te \left [ R \we \{ Y - \psit \} \bm{G} \right ]\\
\leq & (\esa - \ta)^{\T} \te \left [ R \wt \{ Y - \psit \} \bm{G} \right ] \\
& + \exp(\eta_{01}) (M_{1} |S_{\tg}|\lambda^{2}_{0})^{1/2} \{ Q_{\mathrm{WL}}( \esa^{\T} \bm{G}, \ta^{\T} \bm{G} ;\tg  )\}^{1/2}.
\end{aligned}   
\end{equation}
The desired result follows by combining \eqref{eq:sup-ineq-sbd}, \eqref{eq:sup-o-bd-lb} and \eqref{eq:sup-decom-ub} in the event $\Omega_{00} \cap \Omega_{01} \cap \Omega_{03}$.
\end{prf}

\begin{lem} \label{lem:sup-ori-ineq}
Denote $b = \esa - \ta$. Suppose Assumption \ref{ass:o}$(\mathrm{iv})$ holds. In the event, $\Omega_{00} \cap \Omega_{01}\cap \Omega_{03} \cap \Omega_{10}$, we have
\begin{equation}\label{eq:sup-ori-ineq}
\begin{aligned}
 & \exp(-\eta_{01}) D^{\dagger}_{\mathrm{WL}}(\esa^{\T} \bm{G}, \ta^{\T} \bm{G}; \tg) + (A_{1}-1)\lambda_{1} \| b \|_{1}\\
 \leq &   2 A_{1}\lambda_{1} \sum_{j \in S_{\ta} }| b_{j}| + \exp(\eta_{01})\sqrt{M_{1}|S_{\tg}|\lambda_{0}^{2}} \{ Q_{\mathrm{WL}}( \esa^{\T} \bm{G}, \ta^{\T} \bm{G};\tg  )\}^{1/2}.
\end{aligned}  
\end{equation}
\end{lem}

\begin{prf}
 In the event $\Omega_{10}$ , we have
\begin{align*}
     b^{\T} \te [R \wt \{ Y - \psit \} \bm{G}] \leq \lambda_{1} \| b \|_{1},
\end{align*}
by which and Lemma \ref{lem:sup-o-d-p-rate}, we have in the event $\Omega_{00} \cap \Omega_{01} \cap \Omega_{03} \cap \Omega_{10}$,\
\begin{align*}
 & \exp(-\eta_{01}) D^{\dagger}_{\mathrm{WL}}(\esa^{\T} \bm{G}, \ta^{\T} \bm{G}; \tg) + A_{1}\lambda_{1} \| \esa_{1:q} \|_{1} \nonumber\\
 \leq & \lambda_{1} \| b \|_{1} +  A_{1}\lambda_{1} \| \ta_{1:q}\|_{1} + \exp(\eta_{01})\sqrt{M_{1}|S_{\tg}|\lambda_{0}^{2}} \{ Q_{\mathrm{WL}}(\esa^{\T} \bm{G}, \ta^{\T} \bm{G} ;\tg  )\}^{1/2}. \nonumber
\end{align*}
Applying to the preceding inequality the identity $\esa_{j} = |\esa_{j} - \ta_{j}|$ for $j \notin S_{\ta}$ and the triangle inequality
\begin{equation*}
    |\esa_{j}| \geq |\ta_{j}| - |\esa_{j} - \ta_{j}|, \quad j \in S_{\ta} \setminus \{0\},
\end{equation*}
and rearranging the result gives
\begin{align*}
 & \exp(-\eta_{01}) D^{\dagger}_{\mathrm{WL}}(\esa^{\T} \bm{G}, \ta^{\T} \bm{G}; \tg) + (A_{1}-1)\lambda_{1} \| b_{1:q} \|_{1}\\
 \leq & \lambda_{1} | b_{0} | +  2 A_{1}\lambda_{1} \sum_{j \in S_{\ta} \setminus \{ 0 \}}| b_{j}| + \exp(\eta_{01})\sqrt{M_{1}|S_{\tg}|\lambda_{0}^{2}} \{ Q_{\mathrm{WL}}( \esa^{\T} \bm{G}, \ta^{\T} \bm{G} ;\tg ) \}^{1/2}\\
 \leq & (A_{1} + 1)\lambda_{1} | b_{0} | +  2 A_{1}\lambda_{1} \sum_{j \in S_{\ta} \setminus \{ 0 \}}| b_{j}| + \exp(\eta_{01})\sqrt{M_{1}|S_{\tg}|\lambda_{0}^{2}} \{ Q_{\mathrm{WL}}( \esa^{\T} \bm{G}, \ta^{\T} \bm{G};\tg ) \}^{1/2}.
\end{align*}
By adding $(A_{1} - 1)\lambda_{1}|b_{0}|$ on both sides of the previous inequality, the conclusion follows.
\end{prf}

\begin{lem}\label{lem:sup-o-sd-lb}
Suppose Assumptions \ref{ass:o}$(\mathrm{ii})$ and \ref{ass:o}$(\mathrm{iii})$ hold. Then, for any $\alpha, \Tilde{\alpha} \in \mathbb{R}^{q+1}$,
 \begin{align*}
     D^{\dagger}_{\mathrm{WL}} ( \alpha^{\T} \bm{G}, \Tilde{\alpha}^{\T} \bm{G}; \tg) \geq \{b^{\T}\Tilde{\bm{\Sigma}}_{\alpha}(\Tilde{\alpha})b\}\frac{1 - \exp(-C_{40}\| b \|_{1})}{C_{40}\| b \|_{1}},
 \end{align*}
 where $b = \alpha - \Tilde{\alpha}$, $C_{40} = C_{3}C_{4}$ and $\Tilde{\bm{\Sigma}}_{\alpha}(\cdot) = \te \left [R \wt  \psi_{1}\{(\cdot)^{\T} \bm{G}\} \bm{G} \bm{G}^{\T} \right ]$. Throughout, set $\{1 - \exp(-c)\} / c = 1$, for $c = 0$.
\end{lem}

\begin{prf}
By the definition of $D^{\dagger}_{\mathrm{WL}} (\cdot,  \cdot; \cdot )$,
\begin{align*}
& D^{\dagger}_{\mathrm{WL}} (\alpha^{\T}\bm{G},  \Tilde{\alpha}^{\T}\bm{G}; \tg)\\
= & \te  \left [ R \wt \{\psi(\alpha^{\T}\bm{G}) - \psi(  \Tilde{\alpha}^{\T}\bm{G})\} ( \alpha^{\T}\bm{G} -  \Tilde{\alpha}^{\T}\bm{G}) \right ] \\
= & \te \left (R \wt \left [\int^{1}_{0} \psi_{1}\{ \Tilde{\alpha}^{\T}\bm{G} + u ( \alpha^{\T}\bm{G}- \Tilde{\alpha}^{\T}\bm{G} )\} \dif u \right ] \{ \alpha^{\T}\bm{G} -  \Tilde{\alpha}^{\T}\bm{G}\}^{2} \right ).
\end{align*}
By Assumptions \ref{ass:o}$(\mathrm{ii})$, \ref{ass:o}$(\mathrm{iii})$,  and the fact that $| \alpha^{\T}\bm{G} -  \Tilde{\alpha}^{\T}\bm{G}| \leq \{ \sup_{j = 0, \ldots, q} |g_{j}(\bm{X})|\} \| \alpha - \Tilde{\alpha} \|_{1} \leq C_{4} \| \alpha - \Tilde{\alpha} \|_{1}$, it follows that
\begin{align*}
 & D^{\dagger}_{\mathrm{WL}} (\alpha^{\T}\bm{G}, \Tilde{\alpha}^{\T}\bm{G}; \tg)\\
 \geq & \te \left [R \wt \left \{\int^{1}_{0} \psi_{1}(\Tilde{\alpha}^{\T}\bm{G}) \exp(-uC_{3} | \alpha^{\T}\bm{G} -  \Tilde{\alpha}^{\T}\bm{G}|)  \dif u \right \} (\alpha^{\T}\bm{G} -  \Tilde{\alpha}^{\T}\bm{G})^{2} \right ]\\
 \geq & \te  \{R \wt  \psi_{1}(\Tilde{\alpha}^{\T}\bm{G}) (\alpha^{\T}\bm{G} -  \Tilde{\alpha}^{\T}\bm{G})^{2} \}\left \{\int^{1}_{0} \exp(-uC_{40} \| \Tilde{\alpha} - \alpha \|_{1}) \dif u \right \},
\end{align*}
which gives the desired result since $\int^{1}_{0}\exp(-cu) \dif u = \{ 1 - \exp(-c) \}/c$ for $c \geq 0$.
\end{prf}

\begin{lem}\label{lem:sup-comp}
Suppose that Assumption \ref{ass:o}$(\mathrm{v})$ holds. In the event $\Omega_{11}$, Assumption \ref{ass:o}$(\mathrm{vi})$ implies a compatibility condition for $\Tilde{\bm{\Sigma}}_{\ta}$: for any vector $b = (b_{0}, \ldots, b_{q})^{\T}$ such that
 $\sum_{j \notin S_{\ta}} |b_{j}| \leq \xi_{1} \sum_{j \in S_{\ta}} |b_{j}|$, we have
 \begin{equation}\label{eq:sup-comp-g}
  (1 - \zeta_{1}) \nu^{2}_{1} \left(\sum_{j \in S_{\ta}} |b_{j}| \right)^{2} \leq |S_{\ta}| (b^{\T} \Tilde{\bm{\Sigma}}_{\ta} b).
 \end{equation}
\end{lem}

\begin{prf}
 In the event $\Omega_{11}$, we have $|b^{\T}(\tilde{\bm{\Sigma}}_{\ta} - \bm{\bm{\Sigma}}_{\ta})b| \leq \lambda_{1} \| b\|^{2}_{1}$. Then Assumption \ref{ass:o}$(\mathrm{v})$ implies that for any $b = (b_{0}, \ldots, b_{q})^{\T}$ satisfying
 $\sum_{j \notin S_{\ta}} |b_{j}| \leq \xi_{1} \sum_{j \in S_{\ta}} |b_{j}|$,
 \begin{align*}
 \nu_{1}^{2} \| b_{S_{\ta}} \|^{2}_{1} \leq &  |S_{\ta}| (b^{\T} \bm{\bm{\Sigma}}_{\ta} b) \leq |S_{\ta}| (b^{\T} \Tilde{\bm{\Sigma}}_{\ta} b + \lambda_{1}\|b  \|^{2}_{1}) \\
 \leq & |S_{\ta}| (b^{\T} \Tilde{\bm{\Sigma}}_{\ta} b)  +  |S_{\ta}| \lambda_{1} (1 + \xi_{1})^{2}  \| b_{S_{\ta}} \|^{2}_{1}\\
 \leq & |S_{\ta}| (b^{\T} \Tilde{\bm{\Sigma}}_{\ta} b)  +  \zeta_{1} \nu^{2}_{1} \| b_{S_{\ta}} \|^{2}_{1}
 \end{align*}
 where $ \| b_{S_{\ta}} \|_{1} = \sum_{j \in S_{\ta}} |b_{j}|$; and the last inequality holds due to  Assumption \ref{ass:o}$(\mathrm{vi})$, $(1 + \xi_{1})^{2} \nu^{-2}_{1}|S_{\ta}| \lambda_{1}$
 
\noindent$\leq \zeta_{1}$. Thus \eqref{eq:sup-comp-g} follows by rearrangement.
\end{prf}

\section{Proofs of Propositions \ref{prop:tan2017} and \ref{prop:main-mix-p}} \label{sec:sup-pp-1} 

\subsection{Proof of Proposition \ref{prop:tan2017}}
\begin{prf}\label{sup:pp1}
Let $c_{\gamma}$ in Assupmtion \ref{ass:o}$(\mathrm{iv}) = \max[\sqrt{2}\{1 + \exp(-B_{0})\} C_{0},4 \exp(-B_{0}) C^{2}_{0}]$.
This can be shown similarly to \citet{tan2020a} [Theorem 1] by Lemmas \ref{lem:ps}--\ref{eq:sup-o-cov-ub} and Lemmas similar to \ref{lem:sup-a}--\ref{lem:sup-comp}. The small difference in probability is due to extra constraints of $\lambda_{0}$ on $\Omega_{02}$ and $\Omega_{03}$ from the sequential estimate, which is also demonstrated in \citet{tan2020a} [Theorem 5]. 
\end{prf}

\subsection{Proof of Proposition \ref{prop:main-mix-p}}
\begin{prf}\label{sup:pp2}
To facilitate the proof, we first define some constants. 
Let $\nu_{2} = \nu_{1}(1 - \zeta_{1})^{1/2}$, $\xi_{2} = 1 - 2A_{1}/\{ (\xi_{1} + 1) (A_{1} - 1)\}$, $\xi_{3} = (\xi_{1} + 1) (A_{1} - 1)$, $c_{\alpha}$ in Assumption \ref{ass:o}$(\mathrm{vi})$ equals to $\max \{C_{4} \sqrt{8(D^{2}_{0}+D^{2}_{1})},$

\noindent$ 4  C_{2} C^{2}_{4}\}$ $\exp(-B_{0})$, $\Tilde{c}_{\alpha}$ and $\Tilde{c}_{\gamma}$ in Assumption \ref{ass:o}$(\mathrm{vii})$ equal to $C_{3}C_{4}\exp(\eta_{01}) (A_{1} - 1)^{-1} \xi^{2}_{3}\nu^{-2}_{2}$ and $C^{-1}_{1}C_{3}C_{4}(A_{1} - 1)^{-1} \xi^{-2}_{2}\exp(3 \eta_{01})M_{1}$, respectively.

Denote $b = \esa - \ta$, $D^{\dagger}_{\mathrm{WL}} = D^{\dagger}_{\mathrm{WL}}(\esa^{\T}\bm{G}, \ta^{\T}\bm{G}; \tg)$, $Q_{\mathrm{WL}} = Q_{\mathrm{WL}}( \esa^{\T}\bm{G}, \ta^{\T}\bm{G} ;\tg  )$ and $D^{\ddag}_{\mathrm{WL}} = \exp(-\eta_{01}) D^{\dagger}_{\mathrm{WL}} + (A_{1} - 1)\lambda_{1}\| b \|_{1}$.

In the event $\Omega_{00} \cap \Omega_{01} \cap \Omega_{03} \cap \Omega_{10} \cap \Omega_{11}$, \eqref{eq:sup-ori-ineq} in Lemma \ref{lem:sup-ori-ineq} leads to two possible cases: either
\begin{equation}\label{eq:sup-total-bound-c1}
    \xi_{2}D_{\mathrm{WL}}^{\ddag} \leq \exp(\eta_{01}) (M_{1}|S_{\tg}| \lambda^{2}_{0}Q_{\mathrm{WL}})^{1/2},
\end{equation}
or $(1 - \xi_{2})D^{\ddag}_{\mathrm{WL}} \leq 2A_{1}\lambda_{1}\sum_{j \in S_{\ta}}|b_{j}|$, i.e.,
\begin{equation}\label{eq:sup-total-bound-c2}
    D^{\ddag}_{\mathrm{WL}} \leq (\xi_{1} + 1)(A_{1} - 1) \lambda_{1} \sum_{j \in S_{\ta}}|b_{j}| = \xi_{3} \lambda_{1} \sum_{j \in S_{\ta}}|b_{j}|.
\end{equation}

By Lemma \ref{lem:sup-o-sd-lb}, we have
\begin{equation}\label{eq:sup-o-sd-lb-out}
D^{\dagger}_{\mathrm{WL}} \geq \{b^{\T}\Tilde{\bm{\Sigma}}_{\alpha}(\ta)b\}\frac{1 - \exp(-C_{40}\| b \|_{1})}{C_{40}\| b \|_{1}} =(b^{\T}\Tilde{\bm{\Sigma}}_{\ta}b)\frac{1 - \exp(-C_{40}\| b \|_{1})}{C_{40}\| b \|_{1}}.
\end{equation}

If \eqref{eq:sup-total-bound-c1} holds, notice that $D^{\dagger}_{\mathrm{WL}} \leq \exp(\eta_{01}) D_{\mathrm{WL}}^{\ddag}$ and by Assumption \ref{ass:o}$(\mathrm{i})$, $Q_{\mathrm{WL}} < C^{-1}_{1}(b^{\T}\Tilde{\bm{\Sigma}}_{\alpha}b)$, which together with \eqref{eq:sup-o-sd-lb-out} yields
\begin{equation}\label{eq:sup-o-sbd-ub-2}
D^{\ddag}_{\mathrm{WL}} \leq \exp(3\eta_{01}) \xi^{-2}_{2} C^{-1}_{1}(M_{1}|S_{\tg}|\lambda_{0}^{2}) \frac{C_{40}\| b \|_{1}}{1 - \exp(-C_{40}\| b \|_{1})}.
\end{equation}
Since $(A_{1} - 1)\lambda_{1} \| b \|_{1} \leq D^{\ddag}_{\mathrm{WL}}$ and Assumption \ref{ass:o}$(\mathrm{vii})$ holds, \eqref{eq:sup-o-sbd-ub-2} implies that
\begin{align*}
 1 - \exp(-C_{40}\| b \|_{1}) \leq (A_{1} - 1)^{-1} \exp(3\eta_{01}) \xi^{-2}_{2} C^{-1}_{1}(M_{1}|S_{\tg}|\lambda_{0}) C_{40} \leq \eta_{3} < 1.
\end{align*}
As a result, $C_{40} \| b \|_{1} \leq -\ln{(1 - \eta_{3})}$, which leads to
\begin{equation}\label{ineq:zeta2}
\frac{1 - \exp(-C_{40}\| b \|_{1})}{C_{40}\| b \|_{1}}  = \int^{1}_{0} \exp(-C_{40}\| b \|_{1} u) \dif u \geq \exp(-C_{40}\| b \|_{1}) \geq 1 - \eta_{3}.
\end{equation}
Combining the inequality with \eqref{eq:sup-o-sbd-ub-2}, we obtain $$D^{\ddag}_{\mathrm{WL}} \leq \exp(3\eta_{01}) \xi^{-2}_{2}\{ C_{1}(1 - \eta_{3})
 \}^{-1}(M_{1}|S_{\tg}|\lambda^{2}_{0})$$.

If \eqref{eq:sup-total-bound-c2} holds, then $\sum_{j \notin S_{\ta}}|b_{j}| \leq \xi_{1}\sum_{j \in S_{\ta}}|b_{j}|$, which, together with Assumptions \ref{ass:o}$\mathrm{(v)}$--\ref{ass:o}$\mathrm{(vi)}$, implies \eqref{eq:sup-comp-g} in Lemma \ref{lem:sup-comp}, that is,
\begin{equation}\label{eq:sup-comp-outcome}
\sum_{j \in S_{\ta}}|b_{j}| \leq (1 - \zeta_{1})^{-1/2}\nu^{-1}_{1} |S_{\ta}|^{1/2} \left(b^{\T} \Tilde{\bm{\Sigma}}_{\ta} b \right)^{1/2}.
\end{equation}

Since $D^{\dagger}_{\mathrm{WL}} \leq \exp(\eta_{01}) D^{\ddag}_{\mathrm{WL}}$, combining \eqref{eq:sup-total-bound-c2}, \eqref{eq:sup-o-sd-lb-out}  and \eqref{eq:sup-comp-outcome}
yields
\begin{equation}\label{eq:sup-o-sbd-ub-1}
D^{\ddag}_{\mathrm{WL}} \leq \exp(\eta_{01}) \xi^{2}_{3} (1 - \zeta_{1})^{-1} \nu_{1}^{-2} |S_{\ta}| \lambda^{2}_{1} \frac{C_{40}\| b \|_{1}}{1 - \exp(-C_{40}\| b \|_{1})}.
\end{equation}
Since $(A_{1} - 1)\lambda_{1} \| b \|_{1} \leq D^{\ddag}_{\mathrm{WL}}$ and Assumption \ref{ass:o}$(\mathrm{ii})$ holds, \eqref{eq:sup-o-sbd-ub-1} implies that
\begin{align*}
 1 - \exp(-C_{40}\| b \|_{1}) \leq (A_{1} - 1)^{-1} \exp(\eta_{01}) \xi^{2}_{3} (1 - \zeta_{1})^{-1} \nu_{1}^{-2} |S_{\ta}| \lambda_{1} C_{40} \leq \eta_{2} < 1.
\end{align*}
As a result, $C_{40} \| b \|_{1} \leq -\ln{(1 - \eta_{2})}$, which leads to
\begin{equation*}
\frac{1 - \exp(-C_{40}\| b \|_{1})}{C_{40}\| b \|_{1}}  = \int^{1}_{0} \exp(-C_{40}\| b \|_{1} u) \dif u \geq \exp(-C_{40}\| b \|_{1}) \geq 1 - \eta_{2}.
\end{equation*}
Combining the inequality with \eqref{eq:sup-o-sbd-ub-1}, we obtain $D^{\ddag}_{\mathrm{WL}} \leq \exp(\eta_{01}) \xi^{2}_{3} \nu_{2}^{-2}(1 - \eta_{2})^{-2}|S_{\ta}|\lambda_{1}^{2}$.

Therefore, we obtain
\begin{align*}
     & D^{\dagger}_{\mathrm{WL}}(\esa^{\T}\bm{G},  \ta^{\T}\bm{G}, \tg) + \exp(\eta_{01})(A_{1} - 1)\lambda_{1}\| \esa - \ta \|_{1} \\
    \leq{}& \exp(4\eta_{01}) \xi^{-2}_{2}\{ C_{1}(1 - \eta_{3})
 \}^{-1}(M_{1}|S_{\tg}|\lambda^{2}_{0}) + \exp(2\eta_{01}) \xi^{2}_{3}\nu_{2}^{-2}(1 - \eta_{2})^{-2}|S_{\ta}|\lambda_{1}^{2},
\end{align*}
Let $M_{11} =  \exp(4\eta_{01}) \xi^{-2}_{2}\{ C_{1}(1 - \eta_{3})
 \}^{-1}M_{1}$  and $M_{12} = \exp(2\eta_{01}) \xi^{2}_{3}\nu_{2}^{-2}(1 - \eta_{2})^{-2}$, then \eqref{eq:a-ub} holds. 

\end{prf}

\section{Proof of Theorem 1}\label{sec:sup-pt}

\subsection{Lemmas for the proposed estimator}

\begin{lem} \label{lem:sup-a-ub}
Suppose that Assumptions \ref{ass:p}$(\mathrm{i})$, \ref{ass:o}$(\mathrm{i})$, \ref{ass:o}$(\mathrm{ii})$, and \ref{ass:o}$(\mathrm{vi})$ hold, if $$\lambda_{0} \geq \sqrt{2}\{1 + \exp(-B_{0})\} C_{0} \sqrt{\frac{\ln(\frac{1 + P}{\epsilon})}{N}},$$ then, $\P(\Omega_{20}) \geq 1 - 2\epsilon$ for any $r \geq 0$, where $\Omega_{20}$ denotes the event
\begin{equation*}
 \sup_{\| \alpha - \ta \|_{1} \leq r; j = 0, \ldots, d}  \left \vert  (\te - \E) \left [ \left \{ \frac{R}{\pt} - 1 \right \} \left \{ \psie - \psit \right \} Z_{j}  \right ] \right \vert \leq B_{1} \lambda_{0}r,
\end{equation*}
where $B_{1}$ is a positive constant, depending on $(C_{0},C_{2}, C_{3}, C_{5})$.
\end{lem}
\begin{prf}
This can be shown similarly to Lemma 13 in the Supplement of \citet{tan2020a}.
\end{prf}

\begin{lem}\label{lem:op}
Suppose Assumptions \ref{ass:p}, \ref{ass:o} and \ref{ass:x} hold, if a function $h(\cdot)$ on a set of samples satisfying   $h[\{\bm{X}_{i} \}_{i=1}^{N}] \leq M\{|S_{\tg}|\lambda_{0}(\epsilon) + |S_{\ta}|\lambda_{1}(\epsilon)\}$ for some constant $M$ with probability $1 - c \epsilon$ for some constant $c > 0$ and any $\epsilon > 0$, then, $h[\{\bm{X}_{i} \}_{i=1}^{N}] = o_{p}(1)$.\footnote{We write $\lambda_{0}$, $\lambda_{1}$ as $\lambda_{0}(\epsilon)$, $\lambda_{1}(\epsilon)$, since we treat them as functions of $\epsilon$.} 
\end{lem}

\begin{prf}
For $\forall \epsilon > 0$, let $ \Omega_{\epsilon}$ be the event where $h[\{\bm{X}_{i} \}_{i=1}^{N}] \leq M\{|S_{\tg}|\lambda_{0}(\epsilon/c) + |S_{\ta}|\lambda_{1}(\epsilon/c)\}$ holds.
Suppose $|S_{\tg}|\lambda_{0}(\epsilon/c) < |S_{\ta}|\lambda_{1}(\epsilon/c)$, then, on $\Omega_{\epsilon}$,
\begin{equation}
\frac{h[\{\bm{X}_{i} \}_{i=1}^{N}]\sqrt{N}}{|S_{\ta}|\sqrt{\ln\{e(q + 1)\}}} \leq 2M \sqrt{\frac{\ln(\frac{q+1}{\epsilon/c})}{\ln\{e(q + 1)\}}} \leq 2M\sqrt{\frac{\ln(q+1)}{\ln(q+1)}  + \ln\left(\frac{c}{\epsilon}\right)} \leq 2M \sqrt{1 + \ln\left(\frac{c}{\epsilon}\right)},
\end{equation}
which implies that
\begin{equation*}
    \P \left [\frac{h[\{\bm{X}_{i} \}_{i=1}^{N}]\sqrt{N}}{|S_{\ta}|\sqrt{\ln\{e(q + 1)\}}} \leq 2M \sqrt{1 + \ln\left(\frac{c}{\epsilon}\right)}\right ] \geq \P(\Omega_{\epsilon}) = 1 - \epsilon.
\end{equation*}
Similarly, suppose $|S_{\ta}|\lambda_{1} <  |S_{\tg}|\lambda_{0}$, then, on $\Omega_{\epsilon}$, 
\begin{equation*}
    \P \left [\frac{h[\{\bm{X}_{i} \}_{i=1}^{N}]\sqrt{N}}{|S_{\tg}|\sqrt{\ln\{e(p + 1)\}}} \leq 2M \sqrt{1 + \ln\left(\frac{c}{\epsilon}\right)}\right ] \geq \P(\Omega_{\epsilon}) = 1 -\epsilon.
\end{equation*}
For $\forall \epsilon > 0$, we have
\begin{equation*}
    \P \left [\frac{h[\{\bm{X}_{i} \}_{i=1}^{N}]\sqrt{N}}{|S_{\ta}|\sqrt{\ln\{e(q + 1)\}} + |S_{\tg}|\sqrt{\ln\{e(p + 1)\}}} \leq 2M \sqrt{1 + \ln\left(\frac{c}{\epsilon}\right)} \right ] \geq 1 - \epsilon,
\end{equation*}
thus, $\frac{h[\{\bm{X}_{i} \}_{i=1}^{N}]\sqrt{N}}{|S_{\ta}|\sqrt{\ln\{e(q + 1)\}} + |S_{\tg}|\sqrt{\ln\{e(p + 1)\}}}  = O_{p}(1)$, and by Assumption \ref{ass:x}$(\mathrm{vi})$, it follows that
\begin{align*}
h[\{\bm{X}_{i} \}_{i=1}^{N}] = o_{p}\left[\frac{|S_{\ta}|\sqrt{\ln\{e(q + 1)\}} + |S_{\tg}|\sqrt{\ln\{e(p + 1)\}}}{\sqrt{N}} \right ] = o_{p}(1).
\end{align*}
\end{prf}

\begin{lem}\label{lem:op2}Suppose Assumptions \ref{ass:p}, \ref{ass:o} and \ref{ass:x} hold, if a function $h(\cdot)$ on a set of samples satisfying  $h[\{\bm{X}_{i} \}_{i=1}^{N}]\leq M(|S_{\tg}| + |S_{\ta}|)\lambda_{0}(\epsilon)\lambda_{1}(\epsilon)$ for some constant $M$ with probability $1 - c \epsilon$ for some constant $c$ and any $\epsilon > 0$, then, $h[\{\bm{X}_{i} \}_{i=1}^{N}] = o_{p}(1/\sqrt{N})$.
\end{lem}

\begin{prf}
By similar trick used in the proof of Lemma \ref{lem:op}, it can be easily shown that
 $h[\{\bm{X}_{i} \}_{i=1}^{N}] = o_{p}\left[\frac{(|S_{\tg}| + |S_{\ta}|)\sqrt{\ln\{e(q + 1)\}} \sqrt{\ln\{e(p + 1)\}}}{N} \right ] = o_{p}(1/\sqrt{N})$ by Assumption \ref{ass:x}$(\mathrm{vi})$.
\end{prf}

\begin{lem}\label{lem:awip}
Under Assumptions \ref{ass:p}, \ref{ass:o} and \ref{ass:x}, suppose either the PS model~\eqref{eq:def-ps} is correct or the OR model~\eqref{eq:def-or} is correct, the AIPW estimator $\esb \xrightarrow{\P} \tb$.
\end{lem}

\begin{prf}
First, we notice that when $\p$ is correct or $\phi(\bm{X};\alpha) = \psin$ is correct, $\tb$ is the unique solution to $\E \{\tau(\ta, \beta, \tg)\} = 0$. If $\p$ is correct, we obtain,
\begin{align*}
& \E \{\tau(\ta, \tb, \tg) \}\\
& = \E \left [\frac{R}{\pt} \{ Y - \mt  \} \bm{Z} +\left \{1 - \frac{R}{\pt}\right \} \left \{ \psit - \mt \right \} \bm{Z} \right ] \\
& =  \left. \E_{\bm{X}} \left ( \E \left [\frac{R \{ Y - \mt  \}}{\pt}  +\left \{1 - \frac{R}{\pt}\right \} \left \{ \psit - \mt \right \}  \right ] \bm{Z} \right \vert \bm{X} \right )\\
& = \E \left [\{ Y - \mt  \} \bm{Z} \right ]\\
& = 0.
\end{align*}
If $\psin$ is correct, we obtain,
\begin{align*}
& \E \{\tau(\ta, \tb, \tg) \}\\
& = \E \left [\frac{R}{\pt} \{ Y - \mt  \} \bm{Z} +\left \{1 - \frac{R}{\pt}\right \} \left \{ \psit - \mt \right \} \bm{Z} \right ] \\
& = \E \left [\{ \psit - \mt  \} \bm{Z} \right ]\\
& = \E \left [\{ Y - \mt  \} \bm{Z} \right ]\\
& = 0.
\end{align*}
The uniqueness is determined by the uniqueness of $\tb$.

Since Assumptions \ref{ass:x}$(\mathrm{ii})$ and \ref{ass:x}$(\mathrm{iii})$ hold, by standard argument of consistency (e.g. \cite{van2000asymptotic}), it suffices to show that
\begin{equation}\label{eq:consis-suf}
    \te \{ \tau (\ta, \esb, \tg) \} = o_{p}(1).
\end{equation}
We consider the following decomposition,
\begin{align*}
 & \te \{ \tau (\ta, \esb, \tg) \} =  \te \{ \tau (\ta, \esb, \tg) \} - \te \{\tau(\esa, \esb, \esg) \}\\
 & = \te \{ \tau(\ta, \esb, \esg) \} - \te \{ \tau(\esa, \esb, \esg) \}\\
 & + \te \{ \tau(\ta, \esb, \tg) \} - \te \{ \tau(\ta, \esb, \esg) \}\\
 & = \Delta^{1}_{0} + \Delta^{1}_{1},
\end{align*}
where
\begin{align*}
\Delta^{1}_{0} = &  \te \{ \tau(\ta, \esb, \esg) \} - \te \{ \tau(\esa, \esb, \esg) \}\\
& = \te \left[ \left \{ \frac{R}{\pe} - 1 \right \}\{\psie - \psit\} \bm{Z} \right ],\\
\Delta^{1}_{1} = & \te \{ \tau(\ta, \esb, \tg) \} - \te \{ \tau(\ta, \esb, \esg) \}\\
& = \te \left[ R \left \{ \frac{1}{\pt} - \frac{1}{\pe} \right \}\{Y - \psit\}\bm{Z} \right ].
\end{align*}

By Assumptions \ref{ass:p}$(\mathrm{iv})$ and \ref{ass:o}$(\mathrm{vi})$, we know that $\exists \; \zeta_{10} < 0$ a constant, such that $|S_{\tg}| \lambda_{0} + |S_{\ta}|\lambda_{1} \leq \zeta_{10}$. By \eqref{eq:a-ub} in Proposition \ref{prop:main-mix-p}, we know that  $\exists \; M^{0}_{0} > 0$ a constant, such that in the event $\Omega_{00} \cap \Omega_{01}\cap \Omega_{03}\cap \Omega_{10} \cap \Omega_{11}$, $\| \esa - \ta \|_{1} \leq M^{0}_{0} (|S_{\tg}| \lambda_{0} + |S_{\ta}|\lambda_{1})$.
Consider the $j$-th coordinate of $\Delta^{1}_{0}$,
\begin{align*}
|\Delta^{1}_{0,j}| = & \left \vert \te \left[ \left \{ \frac{R}{\pe} - 1 \right \}\{\psie - \psit\}Z_{j} \right ] \right \vert\\
\leq & \te^{1/2} \left[ \left \{ \frac{R}{\pe} - 1 \right \}^{2}Z_{j}^{2} \right ] \times  \te^{1/2} \{\psie - \psit\}^{2}\\
\leq & \te^{1/2} \left( \left [ R - 1 + R \wt \exp\{-(\esg - \tg)^{\T}\bm{F}\} \right ]^{2}Z_{j}^{2} \right )\\
& \times \sqrt{2}\te^{1/2} \left [ \{\psi(\Tilde{\alpha}^{\T}\bm{G}) - \psit\}\psi_{1}(\Tilde{\alpha}^{\T}\bm{G})(\esa - \ta)^{\T}\bm{G}\right ]\\
\leq &\sqrt{2} C_{5} \te^{1/2} \left [ (R - 1)^{2} + R w^{2}(\bm{X}; \tg) \exp\{-2(\esg - \tg)^{\T}\bm{F}\} \right ]\\
& \times \te^{1/2} \left [ \int^{1}_{0}\psi_{1}\{\ta^{\T}\bm{G} + u(\Tilde{\alpha} - \ta)^{\T}\bm{G}\} du \psi_{1}(\Tilde{\alpha}^{\T}\bm{G}) \{ (\esa - \ta)^{\T}\bm{G} \}^{2} \right ] \\
\leq & \sqrt{2} C_{5} \{1 + \exp(-2 B_{0} + 2 C_{0}\| \esg - \tg
\|_{1})\}^{\frac{1}{2}}\\
&\times \te^{1/2}[ \psi^{2}_{1}(\ta^{\T}\bm{G})\exp(2C_{3}|\Tilde{\alpha}^{\T}\bm{G} - \ta^{\T}\bm{G}|) \{ (\esa - \ta)^{\T}\bm{G} \}^{2}] \\
\leq &2 C_{5} \{1 + \exp(-2 B_{0} + 2 \eta_{01})\}^{\frac{1}{2}} \times  C_{2}C_{4}\exp(C_{3}C_{4}\|\esa - \ta\|_{1}) \|\esa - \ta\|_{1}\\
\leq &\sqrt{2} C_{2} C_{4} C_{5} \{1 + \exp(-2 B_{0} + 2 \eta_{01})\}^{\frac{1}{2}} \times  \exp(C_{3}C_{4}M^{0}_{0}\zeta_{10}) \|\esa - \ta\|_{1}\\
= & M^{1}_{0}(|S_{\tg}| \lambda_{0} + |S_{\ta}|\lambda_{1})\\
= & o_{p}(1)
\end{align*}
where $M^{1}_{0}$ is a constant, $\Tilde{\alpha} = u\esa + (1-u)\ta$ for some constant $u \in (0,1)$ and the last equality holds by Lemma \ref{lem:op}.

We consider the $j$-th coordinate of $\Delta^{1}_{1}$. By Cauchy-Schwarz inequality, Lemma \ref{eq:sup-o-cov-ub}, and Lemma \ref{lem:sup-nom-ub}, since Assumption \ref{ass:o}$(\mathrm{iv})$ holds, we obtain
\begin{align*}
|\Delta^{1}_{1,j}| = & \left \vert \te \left[ R \left \{\frac{1}{\pt} - \frac{1}{\pe} \right \}\{Y - \psit\}Z_{j} \right ]\right \vert \\
=  & \left \vert \te \left[ R \wt\exp\{-u(\esg-\tg)^{\T}\bm{F}\}(\tg - \esg)^{\T}\bm{F}\{Y - \psit\}Z_{j} \right ] \right \vert\\
\leq & \te^{1/2} \left[ R \wt \exp\{-2u(\esg-\tg)^{\T}\bm{F}\}\{(\tg - \esg)^{\T}\bm{F}\}^{2}Z_{j}^{2}\right ]\\
& \times \te^{1/2}[R \wt \{Y - \psit\}^{2}]  \\
\leq & \exp(-B_{0}/2 + C_{0} \| \esg - \tg \|_{1})C_{0}C_{5} \| \esg - \tg \|_{1}\\
& \times\sqrt{\E[R \wt \{Y - \psit\}^{2}] +  (D^{2}_{0} + D_{0}D_{1})\lambda_{0}}\\
\leq & \exp(-B_{0}/2 + \eta_{01})C_{0}C_{5} \sqrt{(D^{2}_{0} + D^{2}_{1})\{\lambda_{0} + \exp(-B_{0})\}} \| \esg - \tg \|_{1}\\
\leq & M^{0}_{1}(|S_{\tg}| \lambda_{0} + |S_{\ta}|\lambda_{1})\\
= & o_{p}(1),
\end{align*}
where $u \in (0,1)$ and $M^{0}_{1}$ are both constants.
Therefore, $\|\te \{ \tau (\ta, \esb, \tg) \}\|_{\infty} \leq \|\Delta^{1}_{0}\|_{\infty} $

\noindent $+ \|\Delta^{1}_{1}\|_{\infty} = o_{p}(1)$.  Hence, $\esb \xrightarrow{\P} \tb$.
\end{prf}

\subsection{Proof of Theorem \ref{thm:psc-h-p}$(\mathrm{i})$}
We show the asymptotic normality of $\sqrt{N}(\esb - \tb)$. First, we 
consider the following decomposition,
\begin{align*}
& \tau(\esa, \esb, \esg) -  \tau(\ta, \tb, \tg) \nonumber\\
= &  \left [ \frac{R}{\pe}  \{Y - \me\} - \left \{ \frac{R}{\pe} - 1 \right \} \{ \psie - \me \} \right ] \bm{Z} \\
& - \left [ \frac{R}{\pt}  \{Y - \mt\} - \left \{ \frac{R}{\pt} - 1 \right \} \{ \psit - \mt \} \right ] \bm{Z} \\
= & \{ \mt - \me \} \bm{Z} + R \left \{ \frac{1}{\pe} - \frac{1}{\pt} \right \} \{Y - \psit\} \bm{Z}\\
& + \left \{1 - \frac{R}{\pt} \right\}\left \{ \psie - \psit \right \} \bm{Z}\\
& +  \left \{\frac{R}{\pt} - \frac{R}{\pe} \right\}\left \{ \psie - \psit \right \} \bm{Z},
\end{align*}
denoted as $\delta^{0}_{0} + \delta^{0}_{1} + \delta^{0}_{2} + \delta^{0}_{3}$. Then, $ - \te \tau(\ta, \tb, \tg) = \Delta^{2}_{0} + \Delta^{2}_{1} + \Delta^{2}_{2} + \Delta^{2}_{3}$, with $\Delta^{2}_{i} = \te (\delta^{0}_{i})$, $i = 0, 1, 2, 3$.
First, we show that $\Delta^{2}_{1} + \Delta^{2}_{2} + \Delta^{2}_{3} = o_{p}(1/\sqrt{N})$.
To upper-bound $\Delta^{2}_{1}$, consider $\Delta^{2}_{1, j}$, the $j$-th coordinate of $\Delta^{2}_{1}$. By Taylor expansion in a neighborhood of $\tg$,
\begin{align*}
\Delta^{2}_{1,j} = &  \te \left[ \left \{  \frac{R}{\pe} - \frac{R}{\pt}   \right \} \{ Y - \psit \} Z_{j} \right] \\
= & - (\esg - \tg)^{\T}\te \left[ \bm{F} R \wt  \{ Y - \psit \} Z_{j} \right]  \\
& + \frac{1}{2}(\esg - \tg)^{\T} \te \left[\bm{F} R w(\bm{X}, \tilde{\gamma}_{j})  \{ Y - \psit \} Z_{j} \bm{F}^{\T}\right] (\esg - \tg),
\end{align*}
denoted as $\Delta^{2}_{10,j} + \Delta^{2}_{11,j}$ where $\tilde{\gamma}_{j} = u_{j}\esg + (1-u_{j})\tg$ for some $u_{j} \in (0, 1)$.

In the event $\Omega_{00} \cap \Omega_{01} \cap \Omega_{10}$, by \eqref{eq:gamma-rate} and Lemma \ref{lem:sup-a}, we obtain
\begin{align*}
|\Delta^{2}_{10,j}| \leq \| \esg - \tg \|_{1} \| \te \left[  \bm{F}R \wt  \{ Y - \psit \} Z_{j} \right]  \|_{\infty} \leq M^{1}_{10,j}|S_{\tg}| \lambda_{0} \lambda_{1},
\end{align*}
for some constant $M^{1}_{10,j} > 0$.
In the event $\Omega_{00} \cap \Omega_{01}$, by \eqref{eq:sup-psh-rate}, we obtain
\begin{align*}
|\Delta^{2}_{11,j}| = & \frac{1}{2}|(\esg - \tg)^{\T} \te \left[\bm{F} R w(\bm{X}; \tilde{\gamma}_{j})  \{ Y - \psit \} Z_{j} \bm{F}^{\T}\right] (\esg - \tg)|\\
\leq & \frac{1}{2}  C_{5} \exp(\|\esg - \tg\|_{1}C_{0})\te \left\{   R \wt  | Y - \psit |  |(\esg - \tg)^{\T}\bm{F}|^{2} \right\}.
\end{align*}
We bound $\te \left\{   R \wt  | Y - \psit |  |(\esg - \tg)^{\T}\bm{F}|^{2} \right\}$ by following steps. First,
by Lemma \ref{lem:sup-cov-m1-ub}, in the event $\Omega_{02}$,
\begin{align*}
(\te - \E)\left [ R \wt | Y - \psit | \{(\esg - \tg)^{\T}\bm{F}  \}^{2}  \right ] \leq \sqrt{D^{2}_{0} + D^{2}_{1}}\| \esg - \tg \|^{2}_{1}\lambda_{0}.
\end{align*}
Second, by Assumption \ref{ass:o}$(\mathrm{iv})$ and Lemma \ref{lem:sup-nom-ub}, we have
\begin{equation*}
\E[\{Y - \psit\}^{2}|\bm{X}] \leq D^{2}_{0} + D^{2}_{1}.
\end{equation*}
Therefore,
\begin{align*}
& \E \left [ R \wt | Y - \psit | \{(\esg - \tg)^{\T}\bm{F}  \}^{2}  \right ] \\
\leq & \E^{1/2} \left [ R \wt | Y - \psit |^{2} \{(\esg - \tg)^{\T}\bm{F}  \}^{2}  \right ] \times \E^{1/2} \left [ R \wt  \{(\esg - \tg)^{\T}\bm{F}  \}^{2}  \right ]\\
\leq & \sqrt{D^{2}_{0} + D^{2}_{1}} \E \left [ R \wt  \{(\esg - \tg)^{\T}\bm{F}  \}^{2}  \right ].
\end{align*}
Third, in the event $\Omega_{01}$, by \eqref{eq:sup-ps-cov-ub},
\begin{equation*}
(\E - \te)  \left [ R \wt  \{(\esg - \tg)^{\T}\bm{F}  \}^{2}  \right ] \leq \lambda_{0}\| \tg - \esg \|^{2}_{1}.
\end{equation*}
Combining preceding inequalities and \eqref{eq:sup-psh-rate} in Lemma\ref{lem:sup-p-2}, in the event $\Omega_{00} \cap \Omega_{01} \cap \Omega_{03} \cap \Omega_{02}$,
\begin{align*}
& \te \left\{   R \wt  | Y - \psit |  |(\esg - \tg)^{\T}\bm{F}|^{2} \right\}\\
\leq & \sqrt{D^{2}_{0} + D^{2}_{1}}\| \esg - \tg \|^{2}_{1}\lambda_{0} + \sqrt{D^{2}_{0} + D^{2}_{1}}\{\lambda_{0}\| \tg - \esg \|^{2}_{1} + \exp(\eta_{01}) M_{0}|S_{\tg}|\lambda_{0}^{2}\}\\
\leq &  2 \sqrt{D^{2}_{0} + D^{2}_{1}} \left (\frac{M_{0}}{A_{0} -1} \right)^{2}|S_{\tg}|^{2} \lambda_{0}^{3} + \sqrt{D^{2}_{0} + D^{2}_{1}}\exp(\eta_{01}) M_{0}|S_{\tg}|\lambda_{0}^{2}\\
\leq & 2 \zeta_{0}\sqrt{D^{2}_{0} + D^{2}_{1}} \left (\frac{M_{0}}{A_{0} -1} \right)^{2}|S_{\tg}| \lambda_{0}^{2} + \sqrt{D^{2}_{0} + D^{2}_{1}}\exp(\eta_{01}) M_{0}|S_{\tg}|\lambda_{0}^{2}.
\end{align*}
Therefore,
\begin{equation*}
\begin{aligned}
|\Delta^{2}_{11,j}|
\leq & \frac{1}{2}  C_{5} \exp(\|\esg - \tg\|_{1}C_{0})\te \left\{   R \wt  | Y - \psit |  |(\esg - \tg)^{\T}\bm{F}|^{2} \right\}\\
\leq & \frac{1}{2}  C_{5} \exp(\eta_{01})\\
& \times \left \{ 2 \zeta_{0}\sqrt{D^{2}_{0} + D^{2}_{1}} \left (\frac{M_{0}}{A_{0} -1} \right)^{2}|S_{\tg}| \lambda_{0}^{2} + \sqrt{D^{2}_{0} + D^{2}_{1}}\exp(\eta_{01}) M_{0}|S_{\tg}|\lambda_{0}^{2} \right \}\\
= &  M^{1}_{11} |S_{\tg}| \lambda_{0}^{2},
\end{aligned}
\end{equation*}
for some constant $M^{1}_{11} > 0$.
Hence, in the event $\Omega_{00} \cap \Omega_{01} \cap \Omega_{03} \cap \Omega_{02} \cap \Omega_{10}$,
\begin{equation}
\begin{aligned}
& \mathop{\sup}_{j = 0, \ldots, m-1} | \Delta^{2}_{1,j} | \\
\leq &\mathop{\sup}_{j = 0, \ldots,  m-1} (| \Delta^{2}_{10,j} | + | \Delta^{2}_{11,j} |)\\
\leq & \mathop{\sup}_{j = 0, \ldots,  m-1}M^{1}_{10} |S_{\tg}| \lambda_{0} \lambda_{1} + \mathop{\sup}_{j = 0, \ldots,  m-1} M^{1}_{11}|S_{\tg}|\lambda^{2}_{0} \\
= & M^{1}_{10} |S_{\tg}| \lambda_{0} \lambda_{1} + M^{1}_{11}|S_{\tg}|\lambda^{2}_{0} \\ \leq & M^{1}_{1} (|S_{\tg}| \lambda_{0} \lambda_{1} +|S_{\tg}|\lambda^{2}_{0}),
\end{aligned}    
\end{equation}
where $M^{1}_{1} = \max(M^{1}_{10}, M^{1}_{11})$.

To bound $\Delta^{2}_{2}$, consider $\Delta^{2}_{2,j}$, the $j$-th coordinate of $\Delta^{2}_{2}$, $\Delta^{2}_{2,j}$ can be decomposed as
\begin{align*}
\Delta^{2}_{2,j} = & (\te - \E) \left [ \left \{ 1 - \frac{R}{\pt} \right \} \left \{ \psie - \psit \right \} Z_{j}  \right ] \\
& + \E \left [ \left \{1 - \frac{R}{\pt}\right \} \left \{\psie - \psit \right \} Z_{j}  \right ],
\end{align*}
denoted as $\Delta^{2}_{20, j} + \Delta^{2}_{21, j}$. Since $\pt$ is correct, $\Delta^{2}_{21,j} = 0$.  By \eqref{eq:a-ub}, $\exists M_{\alpha} > 0$, such that $\|\esa - \ta\|_{1} \leq M_{\alpha}(|S_{\tg}|\lambda_{0} + |S_{\ta}|\lambda_{1})$\footnotemark\footnotetext{Note that $\lambda_{0} \leq \lambda_{1}$.}.  Take $r_{\alpha} =  M_{\alpha}(|S_{\tg}|\lambda_{0} + |S_{\ta}|\lambda_{1})$ in Lemma \ref{lem:sup-a-ub}, then in the event $\Omega_{00} \cap \Omega_{01} \cap \Omega_{03} \cap \Omega_{10} \cap \Omega_{11} \cap \Omega_{20}$, we have $\| \esa - \ta \|_{1} \leq r_{\alpha}$ and hence
\begin{equation}
    |\Delta^{2}_{20, j}| \leq B_{1} M_{\alpha}(|S_{\tg}|\lambda_{0} + |S_{\ta}|\lambda_{1})\lambda_{0}.
\end{equation}
Thus, in the event $\Omega_{00} \cap \Omega_{01} \cap \Omega_{03} \cap \Omega_{10} \cap \Omega_{11} \cap \Omega_{20}$,
\begin{equation}
\mathop{\sup}_{j = 0, \ldots, m-1} | \Delta^{2}_{2,j} | \leq  B_{1} M_{\alpha}(|S_{\tg}|\lambda_{0} + |S_{\ta}|\lambda_{1})\lambda_{0} = M^{1}_{2}(|S_{\tg}|\lambda_{0} + |S_{\ta}|\lambda_{1})\lambda_{0},
\end{equation}
for some positive constant $M^{1}_{2}$.

To deal with $\Delta^{2}_{3}$, first, by mean value theorem, we obtain for some $u \in (0,1)$,
\begin{equation} \label{eq:sup-pi-inv-ub}
\begin{aligned}
    \frac{1}{\pe} - \frac{1}{\pt} = & -\exp\{ -u \esg^{\T}\bm{F} - (1 - u) \tg^{\T}\bm{F}\} (\esg - \tg)^{\T} \bm{F} \\
    = & -\wt \exp\{ -u(\esg - \tg)^{\T}\bm{F}\}(\esg - \tg)^{\T}\bm{F}
\end{aligned}   
\end{equation}
and for some $\Tilde{\alpha}$ lies between $\ta$ and $\esa$,
\begin{align} \label{eq:sup-eta-dif-ub}
    \psit - \psie = -\psi_{1}(\Tilde{\alpha}^{\T}\bm{G}) (\esa - \ta)^{\T} \bm{G}.
\end{align}
Combining \eqref{eq:sup-pi-inv-ub} and \eqref{eq:sup-eta-dif-ub} and applying Cauchy-Schwartz inequality to $j$-th coordinate of $\Delta^{2}_{3}$ in the event $\Omega_{00} \cap \Omega_{01} \cap \Omega_{03} \cap \Omega_{10} \cap \Omega_{11} \cap \Omega_{20}$, we get
\begin{align}
 |\Delta^{2}_{3, j}| = & \left \vert\te \left [ \left \{ \frac{R}{\pe} - \frac{R}{\pt}  \right \} \left \{ \psit - \psie \right \} Z_{j} \right ] \right \vert \nonumber\\
 = & |\te ( [R \wt \exp\{ -u_{j}(\esg - \tg)^{\T}\bm{F}\}(\esg - \tg)^{\T} \bm{F}] \{\psi_{1}(\Tilde{\alpha}^{\T}_{j}\bm{G}) (\esa - \ta)^{\T} \bm{G}\})Z_{j}|\nonumber\\
 \leq & C_{5}\exp(\eta_{01}) \te^{1/2}|R\wt \{(\esg - \tg)^{\T}\bm{F}\}^{2} | \nonumber\\
 & \times \te^{1/2} |R \wt \psi^{2}_{1}(\Tilde{\alpha}^{\T}_{j}\bm{G}) \{(\esa - \ta)^{\T} \bm{G}\}^{2}| \\
 \leq & C_{5}\exp(\eta_{01}) \{\exp(\eta_{01})M_{0}|S_{\tg}|\lambda_{0}^{2}\}^{1/2}\nonumber\\
 & \times \te^{1/2}\left( R \wt [ \psi_{1} (\ta^{\T}_{j}\bm{G})\exp\{C_{3}|(\Tilde{\alpha} - \ta)^{\T}\bm{G}|\}]^{2} \{(\esa - \ta)^{\T}\bm{G}\}^{2}\right)\nonumber\\
 \leq & C_{5}\exp(\eta_{01}) \{\exp(\eta_{01})M_{0}|S_{\tg}|\lambda_{0}^{2}\}^{1/2} C^{\frac{1}{2}}_{2} \exp(C_{3}C_{4}r_{\alpha})\left \{\frac{M^{\dagger}(|S_{\tg}|\lambda_{0}^{2} +  |S_{\ta}|\lambda_{1}^{2})}{1 - \eta_{3}} \right \}^{\frac{1}{2}}\nonumber\\
 \leq & M^{1}_{3}\{|S_{\tg}|\lambda_{0}^{2} + (|S_{\tg}||S_{\ta}|)^{1/2}\lambda_{0}\lambda_{1}\}. \nonumber
\end{align}
The second inequality holds due to \eqref{eq:sup-psh-rate} in Lemma \ref{lem:sup-p-2}. The third inequality holds by Assumption \ref{ass:o}$(\mathrm{ii})$, \eqref{ineq:zeta2}, the facts that $ \te\left[ R \wt  \psi_{1} (\ta^{\T}_{j}\bm{G}) \{(\esa - \ta)^{\T}\bm{G}\}^{2}\right] = b^{\T}\Tilde{\bm{\Sigma}}_{\alpha}b$ in \eqref{eq:sup-o-sd-lb-out} and that by \eqref{eq:a-ub} in Proposition \ref{prop:main-mix-p}, $\exists M^{\dagger} > 0 $ a constant, such that $D^{\dagger}_{\mathrm{WL}} \leq M^{\dagger}(|S_{\tg}|\lambda_{0}^{2} +  |S_{\ta}|\lambda_{1}^{2})$.
Therefore,
\begin{equation}
\mathop{\sup}_{j = 0, \ldots, m-1} | \Delta^{2}_{3,j} | \leq M^{1}_{3}\{|S_{\tg}|\lambda_{0}^{2} + (|S_{\tg}||S_{\ta}|)^{1/2}\lambda_{0}\lambda_{1}\},
\end{equation}
for some constant $M^{1}_{3}$.
Thus, on the event $\Omega_{00} \cap \Omega_{01} \cap \Omega_{03} \cap \Omega_{02} \cap \Omega_{10} \cap \Omega_{11} \cap \Omega_{20}$,
\begin{align*}
\|\Delta^{2}_{1} + \Delta^{2}_{2} + \Delta^{2}_{3}\|_{\infty} \leq & \sup_{j = 0, \ldots, m-1} |\Delta^{2}_{1,j}| + \sup_{j = 0, \ldots, m-1} |\Delta^{2}_{2,j}| + \sup_{j = 0, \ldots, m-1} |\Delta^{2}_{3,j}|\\
\leq &M^{1}_{1} (|S_{\tg}| \lambda_{0} \lambda_{1} +|S_{\tg}|\lambda^{2}_{0}) + M^{1}_{2}(|S_{\tg}|\lambda_{0} + |S_{\ta}|\lambda_{1})\lambda_{0} \\
& + M^{1}_{3}\{|S_{\tg}|\lambda_{0}^{2} + (|S_{\tg}||S_{\ta}|)^{1/2}\lambda_{0}\lambda_{1}\}\\
\leq & M^{1} (|S_{\tg}|+ |S_{\ta}|)\lambda_{0}\lambda_{1}.
\end{align*}
By Lemma \ref{lem:op2}, $\|\Delta^{2}_{1} + \Delta^{2}_{2} + \Delta^{2}_{3}\|_{\infty} = o_{p}(1/\sqrt{N})$.

Then, we deal with $\Delta_{0}^{2}$. For the $j$-th coordinate of $\Delta_{0}^{2}$, $\Delta_{0,j}^{2}$,  by mean value theorem,
\begin{equation*}
\Delta^{2}_{0,j} =\te [\{ \mt - \me \} Z_{j}] = - \te \{ \psi_{1}(\Tilde{\beta}_{j}^{T}\bm{Z})Z_{j} \bm{Z}^{\T}(\esb - \tb)\},
\end{equation*}
where $\Tilde{\beta}_{j} = (1 - u_{j}) \tb + u_{j} \esb$ for some $u_{j} \in (0,1)$. We first show that $\te \{ \psi_{1}(\beta^{*\T}\bm{Z})Z_{j} \bm{Z}\}$

\noindent $ - \te \{ \psi_{1}(\Tilde{\beta}^{T}\bm{Z})Z_{j} \bm{Z}\} \xrightarrow{\P} 0$.
By Assumption \ref{ass:o}$(\mathrm{ii})$, we know that for $\forall u, u'$, if $\psi_{1}(u) > \psi_{1}(u')$, since $\psi_{1}(u') \geq \psi_{1}(u)\exp(-C_{3}|u - u'|)$, then $|\psi_{1}(u) - \psi_{1}(u')| \leq \psi_{1}(u) \{ 1 - \exp(-C_{3}|u - u'|)\}$; if $\psi_{1}(u) < \psi_{1}(u')$, $\psi_{1}(u') \leq \psi_{1}(u)\exp(C_{3}|u - u'|)$, $|\psi_{1}(u) - \psi_{1}(u')| \leq \psi_{1}(u) \{ \exp(C_{3}|u - u'|) - 1\}$; therefore, $|\psi_{1}(u) - \psi_{1}(u')| \leq \psi_{1}(u) \max\{1 - \exp(-C_{3}|u - u'|) , \exp(C_{3}|u - u'|) - 1\}$.
Consider the $i$-th element of the difference, if $C_{3} = 0$, $\psi_{1}$ is a constant, then 
\begin{align*}
|\te \{ \psi_{1}(\beta^{*\T}X)Z_{i} Z_{j}\} - \te \{ \psi_{1}(\Tilde{\beta}_{j}^{\T}\bm{Z})Z_{i} Z_{j}\}|
= 0.    
\end{align*}
Otherwise, 
\begin{align*}
&|\te \{ \psi_{1}(\beta^{*\T}X)Z_{i} Z_{j}\} - \te \{ \psi_{1}(\Tilde{\beta}_{j}^{\T}\bm{Z})Z_{i} Z_{j}\}|
\leq  \te \{|\psi_{1}(\beta^{*\T}\bm{Z}) - \psi_{1}(\Tilde{\beta}_{j}^{\T}\bm{Z})||Z_{i} Z_{j}|\} \\
\leq & C^{2}_{5}\te [\psi_{1}(\beta^{*\T}\bm{Z}) \max\{1 - \exp(-C_{3} |(\esb - \tb)^{\T}\bm{Z}|) , \exp(C_{3} |(\esb - \tb)^{\T}\bm{Z}|) - 1\} ]\\
= & C^{2}_{5} C_{6} \te \{C_{3} |(\esb - \tb)^{\T}\bm{Z}|+ o_{p}(|(\esb - \tb)^{\T}\bm{Z}|)\} \\
\leq & C_{3} C^{3}_{5} C_{6} \| \esb - \tb\|_{1} + o_{p}(\| \esb - \tb\|_{1})\\
= & o_{p}(1),
\end{align*}
which leads to $- \te \{ \psi_{1}(\Tilde{\beta}_{j}^{\T}\bm{Z})Z_{j} \bm{Z}^{\T}\} + \te \{ \psi_{1}(\beta^{*\T}\bm{Z})Z_{j} \bm{Z}^{\T}\} \xrightarrow{\P} 0$.
Therefore, we consider the following decomposition,
\begin{align*}
& - \te \{\psi_{1}(\Tilde{\beta}_{j}^{\T}\bm{Z})Z_{j} \bm{Z}\} + \E \{ \psi_{1}(\beta^{*\T}\bm{Z})Z_{j} \bm{Z}\}\\
= & - \te \{ \psi_{1}(\Tilde{\beta}_{j}^{T}\bm{Z})Z_{j} \bm{Z}\} + \te \{ \psi_{1}(\beta^{*\T}X)Z_{j} \bm{Z}\}\\
& - \te \{ \psi_{1}(\beta^{*\T}\bm{Z})Z_{j} \bm{Z}\} + E \{\psi_{1}(\beta^{*\T}\bm{Z})Z_{j} \bm{Z}\}\\
\xrightarrow{\P} & 0.
\end{align*}
Hence,
\begin{equation}\label{eq:sup-coefm-conv}
 \te \{\psi_{1}(\Tilde{\beta}_{j}^{\T}\bm{Z})Z_{j} \bm{Z}\} \xrightarrow{\P} \E \{\psi_{1}(\beta^{*\T}\bm{Z})Z_{j} \bm{Z}\} = \bm{\Gamma}_{j},
\end{equation}
where $ \bm{\Gamma}_{j}$ is the $j$-th row of $\bm{\Gamma}$, and \eqref{eq:sup-coefm-conv} holds for $j = 0, \ldots, m-1$.  
Hence, $\Delta^{2}_{0} = \te (\Tilde{\bm{Z}}\bm{Z}^{\T})(\esb - \tb)$, where $\Tilde{\bm{Z}}_{j} = -\psi_{1}(\Tilde{\beta}_{j}^{\T}\bm{Z})Z_{j}$, and $\te (\Tilde{\bm{Z}}\bm{Z}^{\T}) \xrightarrow{\P} -\bm{\Gamma}$. 
Suppose $\sqrt{N}(\esb - \tb) \xrightarrow{d} G_{2}$,
by continuous mapping theorem, 
\begin{align}
\sqrt{N}\Delta^{2}_{0}\xrightarrow{d} -\bm{\Gamma}G_{2}.
\end{align}
Besides, by central limit theorem, 
\begin{align}
\sqrt{N}\Delta^{2}_{0} = \sqrt{N} \te\left \{ \tau(\bm{O}, \bar \alpha, \beta^*, \bar \gamma)  \right \} + o_{p}(1) \xrightarrow{d} \N(0,\bm{\bm{\Lambda}}).
\end{align}
Therefore, 
\begin{align*}
\sqrt{N}(\esb - \tb) \xrightarrow{d} G_{2} \sim - \bm{\Gamma}^{-1}\N(0,\bm{\Lambda}) \sim \N(0,\bm{\Sigma}),
\end{align*}
where $\sim$ denotes "distributed as", i.e., for any two distributions $G_{0}$ and $G_{1}$, $G_{0} \sim G_{1}$ means the two distributions are the same.

\subsection{Proof of Theorem \ref{thm:psc-h-p}$(\mathrm{ii})$}
We show the consistency of $\hat{\bm{\bm{\Sigma}}}$.
First, if we let $\hat{ \bm{\Gamma}}_{j} = \te \{ \psi_{1}(\esb^{\T}\bm{Z})Z_{j} \bm{Z}\}$, then
\begin{align*}
\hat{ \bm{\Gamma}}_{j}  \xrightarrow{\P} & \E \{\psi_{1}(\beta^{*\T}\bm{Z})Z_{j} \bm{Z}\},
\end{align*}
i.e., $\hat{\bm{\Gamma}}_{j} \xrightarrow{\P} \bm{\Gamma}_{j}$, which can be shown in the way similar to the proof of \eqref{eq:sup-coefm-conv}. Then we get $\hat{\bm{\Gamma}} \xrightarrow{\P} \bm{\Gamma}$.

Next, we want to show that
$\hat{\bm{\Lambda}} \xrightarrow{\P} \bm{\Lambda}$. Since $\bm{\Lambda} =  \E\{\tau(\ta, \tb, \tg) \tau(\ta, \tb, \tg)^{\T}\}$ and $\hat{\bm{\Lambda}} =  \te \{\tau(\esa, \esb, \esg)$

\noindent $\tau(\esa, \esb, \esg)^{\T}\}$, it suffices to show that 
\begin{align*}
\te \{\tau(\esa, \esb, \esg)\tau(\esa, \esb, \esg)^{\T}\} - \te \{\tau(\ta, \tb, \tg) \tau(\ta, \tb, \tg)^{\T}\} = o_{p}(1).    
\end{align*}

We consider the $i,j$-th element of the difference above:
\begin{align*}
& |[\te \{\tau(\esa, \esb, \esg)\tau(\esa, \esb, \esg)^{\T}\} - \te \{\tau(\ta, \tb, \tg) \tau(\ta, \tb, \tg)^{\T}\}]_{ij}|\\
= & \biggl| \te \biggl ( \biggl [ \frac{R}{\pe}\{ Y - \psie \} + \{\psie - \me\} \biggr ]^{2} Z_{i}Z_{j}\\
& - \biggl [\frac{R}{\pt}\{ Y - \psit \} + \{\psit - \mt\}\biggr]^{2}Z_{i}Z_{j} \biggr)\biggr|\\
\leq & \te \biggl (  \biggl| \biggl [ \frac{R}{\pe}\{ Y - \psie \} + \{\psie - \me\} \biggr ]^{2}\\
& - \biggl [\frac{R}{\pt}\{ Y - \psit \} + \{\psit - \mt\}\biggr]^{2}\biggl| \biggr|Z_{i}Z_{j}\biggr| \biggr)\\
\leq & C^{2}_{5}\te \biggl (  \biggl| \biggl [ \frac{R}{\pe}\{ Y - \psie \} + \{\psie - \me\} \biggr ]^{2}\\
& - \biggl [\frac{R}{\pt}\{ Y - \psit \} + \{\psit - \mt\}\biggr]^{2} \biggr|\biggr)\\
\leq & C^{2}_{5}\te \biggl (   \biggl [ \frac{R}{\pe}\{ Y - \psie \} + \{\psie - \me\} \\
& - \frac{R}{\pt}\{ Y - \psit \} - \{\psit - \mt\}\biggr]^{2} \biggr)\\
& + 2 C^{2}_{5}\te \biggl (   \biggl [ \frac{R}{\pe}\{ Y - \psie \} + \{\psie - \me\} \\
& - \frac{R}{\pt}\{ Y - \psit \} - \{\psit - \mt\}\biggr]\\
& \times \biggl[ \frac{R}{\pt}\{ Y - \psit \} + \{\psit - \mt\}\biggr]
\biggr)\\
\leq & C^{2}_{5}\te \biggl (   \biggl [ \frac{R}{\pe}\{ Y - \psie \} + \{\psie - \me\} \\
& - \frac{R}{\pt}\{ Y - \psit \} - \{\psit - \mt\}\biggr]^{2} \biggr)\\
& + 2 C^{2}_{5}\te^{\frac{1}{2}} \biggl (\biggl [ \frac{R}{\pe}\{ Y - \psie \} + \{\psie - \me\} \\
& - \frac{R}{\pt}\{ Y - \psit \} - \{\psit - \mt\}\biggr]^{2}\biggr)\\
& \times\te^{\frac{1}{2}} \biggl ( \biggl[ \frac{R}{\pt}\{ Y - \psit \} + \{\psit - \mt\}\biggr]^{2}
\biggr),
\end{align*}
therefore, we only need to show that 
\begin{align*}
& \te \biggl (   \biggl [ \frac{R}{\pe}\{ Y - \psie \} + \{\psie - \me\} \\
& - \frac{R}{\pt}\{ Y - \psit \} - \{\psit - \mt\}\biggr]^{2} \biggr) = o_{p}(1).
\end{align*} 
Consider the following decomposition:
\begin{align*}
& \frac{R}{\pe}\{ Y - \psie \} + \{\psie - \me\}\\
& - \frac{R}{\pt}\{ Y - \psit \} - \{\psit - \mt\}\\
= & \{ \psie - \psit \} \left \{ 1 - \frac{R}{\pt} \right \}\\
& + R \left \{ Y - \psit \right \} \left \{ \frac{1}{\pe} - \frac{1}{\pt} \right \}\\
& + R\{ \psie - \psit \} \left\{\frac{1}{\pt} - \frac{1}{\pe}  \right \}\\
&+ \mt - \me,
\end{align*} 
denoted as $\delta^{1}_{0} + \delta^{1}_{1} + \delta^{1}_{2} + \delta^{1}_{3}$. Let $\Delta^{3}_{i} = \te \{(\delta^{1}_{i})^{2}\}$, $i = 0,\ldots, 3$, we only need to show that $\Delta^{3}_{i} = o_{p}(1)$, $i = 0,\ldots, 3$. 

By mean value theorem, 
\begin{align*}
\Delta^{3}_{0} = & \te \left [ \{ \psie - \psit \}^{2} \left \{ 1 - \frac{R}{\pt} \right \}^{2} \right ] \\
= & 2 \te  \left [ \psi_{1}(\Tilde{\alpha}^{\T}\bm{G}) \{ \psi(\Tilde{\alpha}^{\T}\bm{G}) - \psit \} (\esa - \ta)^{\T}\bm{G} \left \{ 1 - \frac{R}{\pt} \right \}^{2} \right ]\\
= & 2 \te  \left (\psi_{1}(\Tilde{\alpha}^{\T}\bm{G}) \left[ \int^{1}_{0}\psi_{1}\{\ta^{\T}\bm{G} + u(\Tilde{\alpha} - \ta)^{\T}\bm{G}\} \dif u \right ] \right.\\
& \left.\{(\esa - \ta)^{\T}\bm{G}\}^{2} \left \{ 1 - \frac{R}{\pt} \right \}^{2} \right )\\
 \leq &  2 \te  \left [\psi_{1}(\ta^{\T}\bm{G}) \exp(C_{3}C_{4}r_{\alpha}) \left\{ \int^{1}_{0}\psi_{1}(\ta^{\T}\bm{G}) \exp(C_{3}C_{4}r_{\alpha}) \dif u \right \} \right.\\
& \left.\{(\esa - \ta)^{\T}\bm{G}\}^{2} \left \{ 1 - \frac{R}{\pt} \right \}^{2} \right ]\\
 \leq &  2 C_{2}^{2}C_{4}^{2} \exp(2C_{3}C_{4}r_{\alpha}) \{ 1 + \exp(-B_{0}) \}^{2}\| \esa - \ta \|^{2}_{1} \\
 =& o_{p}(1). 
\end{align*}

\begin{align*}
\Delta^{3}_{1} = & \te \left [ R \left \{ Y - \psit \right \}^{2} \left \{ \frac{1}{\pe} - \frac{1}{\pt} \right \}^{2} \right ]\\
= & -2 \te \Bigl ( R \wt^{2}[ \exp\{-(\Tilde{\gamma} -\tg)^{\T}\bm{F}\} -1 ] \exp\{-(\Tilde{\gamma} -\tg)^{\T}\bm{F}\} \\
& \times  \left \{ Y - \psit \right \}^{2} (\esg - \tg)^{\T}\bm{G} \Bigr )\\
\leq &2 \{1 + \exp(\eta_{01})\}^{2}\exp(-B_{0} + \eta_{01}) C_{4} \| \esg - \tg \|_{1} \te \left [ R \wt\left \{ Y - \psit \right \}^{2} \right ]\\
\leq & 2\{1 + \exp(\eta_{01})\}^{2}\exp(-B_{0} + \eta_{01}) C_{4} \| \esg - \tg \|_{1}  \Bigl (\E \left [ R \wt\left \{ Y - \psit \right \}^{2} \right ] \\
& + ({D^{2}_{0}} + D_{0}D_{1})\lambda_{0}\Bigr)\\
= & o_{p}(1).
\end{align*}

By mean value theorem,
\begin{align*}
\Delta^{3}_{2} = & \te \left [ R\{ \psie - \psit \}^{2} \left\{\frac{1}{\pt} - \frac{1}{\pe}  \right \}^{2} \right]  \\
= & \te \left ( R\{ \psie - \psit \}^{2} \wt^{2}[ 1 - \exp\{ -(\esg - \tg)^{\T}\bm{F} \} ]^{2}\right)\\
\leq & \{ 1 + \exp(\eta_{01}) \}^{2}\exp(-2B_{0}) \te \left [ \{ \psie - \psit \}^{2} \right]\\
= & 2 \{ 1 + \exp(\eta_{01}) \}^{2}\exp(-2B_{0}) \te \left [ \{ \psi(\Tilde{\alpha}^{\T}\bm{G})- \psit \} \psi_{1}(\Tilde{\alpha}^{\T}\bm{G}) (\esa - \ta)^{\T} \bm{G} \right]\\
= & 2 \{ 1 + \exp(\eta_{01}) \}^{2}\exp(-2B_{0}) \\
& \times \te \biggl ( \psi_{1}(\Tilde{\alpha}^{\T}\bm{G}) \left[ \int^{1}_{0}\psi_{1}\{\ta^{\T}\bm{G} + u(\Tilde{\alpha} - \ta)^{\T}\bm{G}\} \dif u \right ]  \{(\esa - \ta)^{\T}\bm{G}\}^{2} \biggr) \\
\leq & 2C_{2}^{2}C_{4}^{2}  \{ 1 + \exp(\eta_{01}) \}^{2}\exp(2C_{3}C_{4}r_{\alpha}-2B_{0}) \| \esa - \ta \|^{2}_{1} \\
= & o_{p}(1).
\end{align*}

\begin{align*}
\Delta^{3}_{3} = &\te [\{ \mt - \me\}^{2} ]\\
= & 2  \te \left [ \{ \psi(\Tilde{\beta}^{\T}\bm{Z})- \mt \} \psi_{1}(\Tilde{\beta}^{\T}\bm{Z}) (\esb - \tb)^{\T} \bm{Z} \right]\\
= & 2 \te \biggl ( \psi_{1}(\Tilde{\beta}^{\T}\bm{Z}) \left[ \int^{1}_{0}\psi_{1}\{\beta^{*\T}\bm{Z} + u(\Tilde{\beta} - \tb)^{\T}\bm{Z}\} \dif u \right ] \{(\esb - \tb)^{\T}\bm{Z}\}^{2} \biggr) \\
\leq & 2C_{5}^{2}C_{6}^{2}  \{ 1 + \exp(\eta_{01}) \}^{2}\exp(2C_{3}C_{5}\| \esb - \tb \|_{1}) \| \esb - \tb\|^{2}_{1} \\
= & o_{p}(1).
\end{align*}
Therefore, $\hat{\bm{\Lambda}} \xrightarrow{\P} \bm{\Lambda}$. 
Then, by continuous mapping theorem, $\hat{\bm{\Gamma}}^{-1} \xrightarrow{\P} \bm{\Gamma}^{-1}$.
Thus, by continuous mapping theorem again, $\hat{\bm{\bm{\Sigma}}} \ \xrightarrow{\P} \bm{\bm{\Sigma}}$.

\section{Extension to the setting of stratified sampling} \label{sup:ess}

\subsection{Proof of Proposition \ref{prop:ss}}
Let
\begin{align}
\tau^{s}(\alpha, \beta, \cdot) = \frac{1}{N}\sum^{n}_{i=1} \frac{1}{ (\cdot) }\{ Y_{i} - \psi(\alpha^{\T}\bm{G}_{i})\}\bm{Z}_{i} + \frac{1}{N} \sum_{i=1}^{N} \{ \psi(\alpha^{\T}\bm{G}_{i}) - \psi(\beta^{\T}\bm{Z}_{i}) \}\bm{Z}_{i}.
\end{align}

We have 
\begin{align*}
& -\tau^{s}\{\ta, \tb, \pt\} = -\tau^{s}\{\ta, \tb, \pi^{*} (\bm{X})\} = \tau^{s}\{\esa, \hat{\beta}^{s} , \hat{\pi}(\bm{X})\} -\tau^{s}\{\ta, \tb, \pi^{*} (\bm{X})\}\\
= & \frac{1}{N}\sum_{i=1}^{N}[\{ \psi(\esa^{\T} \bm{G}_{i}) - \psi(\hat{\beta}^{s\T}\bm{Z}_{i}) \}\bm{Z}_{i}]  - \frac{1}{N}\sum_{i=1}^{N}[\{ \psi(\esa^{\T} \bm{G}_{i}) - \psi(\beta^{*\T}\bm{Z}_{i}) \}\bm{Z}_{i}]\\
& + \frac{1}{N}\sum_{i=1}^{N} [\{ \psi(\esa^{\T}\bm{G}_{i}) - \psi(\beta^{*\T}\bm{Z}_{i})\} \bm{Z}_{i}] - \frac{1}{N}\sum_{i=1}^{N} [\{ \psi(\ta^{\T}\bm{G}_{i}) - \psi(\beta^{*\T}\bm{Z}_{i})\} \bm{Z}_{i}]\\
& + \frac{1}{n} \sum_{i=1}^{n}  \{ Y_{i} - \psi(\esa^{\T}\bm{G}_{i}) \} \bm{Z}_{i} - \frac{1}{n} \sum_{i=1}^{n}  \{ Y_{i} - \psi(\ta^{\T}\bm{G}_{i}) \} \bm{Z}_{i},
\end{align*}
denoted as $\Delta^{0}_{5}$ + $\Delta^{1}_{5}$ + $\Delta^{2}_{5}$.     

We first deal with $\Delta^{0}_{5}$. Let $Z_{ij}$ denote the $j$-th co-ordinate of the $i$-th sample $\bm{Z}_{i}$. Then, we consider the $j$-th coordinate of $\Delta^{0}_{5}$, $\Delta^{0}_{5,j}$  
\begin{align*}
\Delta^{0}_{5,j} = &  \frac{1}{N}\sum_{i=1}^{N}[\{ \psi(\esa^{\T} \bm{G}_{i}) - \psi(\hat{\beta}^{s\T}\bm{Z}_{i}) \}Z_{ij}]  - \frac{1}{N}\sum_{i=1}^{N}[\{ \psi(\esa^{\T} \bm{G}_{i}) - \psi(\beta^{*\T}\bm{Z}_{i}) \}Z_{ij}]\\
= & \frac{1}{N}\sum_{i=1}^{N}[\{ \psi(\beta^{*\T}\bm{Z}_{i}) - \psi(\hat{\beta}^{s\T}\bm{Z}_{i})\}Z_{ij}]\\
= & \frac{1}{N}\sum_{i=1}^{N} \{ \psi_{1}(\Tilde{\beta}_{j}^{\T}\bm{Z}_{i}) Z_{ij} (\beta^{*} - \hat{\beta}^{s})^{\T}\bm{Z}_{i} \}\\
= & \frac{1}{N}\sum_{i=1}^{N} \{ \psi_{1}(\beta^{*\T}\bm{Z}_{i}) Z_{ij} (\beta^{*} - \hat{\beta}^{s})^{\T}\bm{Z}_{i} \} + O_{p}(\| \beta^{*} - \hat{\beta}^{s} \|^{2}_{1}), 
\end{align*}
where $\Tilde{\beta}_{j} = u_{j}\beta^{*} + (1 - u_{j})\hat{\beta}^{s}$ for some $u_{j} \in (0,1)$. The last equality holds since $\psi_{1}(\beta^{*\T}\bm{Z}_{i}) \exp$

\noindent $(-C_{3}C_{5}\| \beta^{*} - \hat{\beta}^{s} \|_{1}) \leq\psi_{1}(\Tilde{\beta}_{j}^{\T}\bm{Z}_{i})\leq \psi_{1}(\beta^{*\T}\bm{Z}_{i})\exp(C_{3}C_{5}\| \beta^{*} - \hat{\beta}^{s} \|_{1})$, which leads to $\psi_{1}(\Tilde{\beta}_{j}^{\T}\bm{Z}_{i}) $

\noindent $ = \psi_{1}(\beta^{*\T}\bm{Z}_{i}) + O_{p}(\| \beta^{*} - \hat{\beta}^{s} \|_{1})$.

Then we deal with $\Delta^{1}_{5}$ and $\Delta^{2}_{5}$ together.
Consider the $j$-th coordinate of $\Delta^{1}_{5} + \Delta^{2}_{5}$:
\begin{align*}
& |\Delta^{1}_{5,j} + \Delta^{2}_{5,j}|\\
= & \left | \frac{1}{n} \sum_{i=1}^{n} \left [  \left \{ \psi(\ta^{\T}\bm{G}_{i})  -  \psi(\esa^{\T}\bm{G}_{i}) \right \} Z_{ij}  \right ]\right.\\
& \left.+ \frac{1}{N}\sum_{i=1}^{N}[\{ \psi(\esa^{\T}\bm{G}_{i}) - \psi(\ta^{\T}\bm{G}_{i}) \}Z_{ij}] \right | \\
= & \left | \frac{1}{n} \sum_{i=1}^{n} \left \{  \psi_{1}(\Tilde{\alpha}_{j}^{\T}\bm{G}_{i})(\ta - \esa)^{\T}\bm{G}_{i}  Z_{ij}  \right \}\right.\\
& \left. + \frac{1}{N}\sum_{i=1}^{N}\{\psi_{1}(\Tilde{\alpha}_{j}^{\T}\bm{G}_{i})(\esa - \ta)^{\T}\bm{G}_{i}Z_{ij}\} \right | \\
\leq & \| \esa - \ta \|_{1}  \left \| \frac{1}{n} \sum_{i=1}^{n} \left \{ \psi_{1}(\Tilde{\alpha}_{j}^{\T}\bm{G}_{i})\bm{G}_{i}  Z_{ij}  \right \} - \frac{1}{N}\sum_{i=1}^{N}\{\psi_{1}(\Tilde{\alpha}_{j}^{\T}\bm{G}_{i})\bm{G}_{i}Z_{ij}\} \right \|_{\infty},
\end{align*}
where $\Tilde{\alpha}_{j} = u_{j}\esa + (1 - u_{j})\ta$, for some $0 < u_{j} < 1$, $j = 0, \ldots, m-1$.  Consider the $k$-th coordinate of $\bm{G}_{i}$. For technical convenience, we assume that $n$ is divisible by $N$. Let
\begin{align*}
V_{ijk}  \triangleq \left\{ \frac{n}{N} - 1  \right\} \psi_{1} (\Tilde{\alpha}_{j}^{\T}\bm{G}_{i})G_{ik}  Z_{ij} + \frac{n}{N}\sum^{n + i(N/n-1)}_{s = n + (N/n - 1)(i+1)+1}\psi_{1} (\Tilde{\alpha}_{j}^{\T}\bm{G}_{s})G_{sk}Z_{sj}, 
\end{align*}
then, $\exists$ constant $C_{7} > 0$, such that $|V_{ijk}| \leq C_{7}$. Moreover, we have $\E(V_{ijk}) = 0$; therefore, $\E(V^{2}_{ijk}) = \mathrm{Var}(V_{ijk}) =\{(n-N)^{2}/N^{2} + n^{2}/N^{2}(N/n - 1) \}\mathrm{Var}\{ \psi_{1} (\Tilde{\alpha}_{j}^{\T}\bm{G})G_{k}Z_{j} \}\leq C_{8}$ for some constant $C_{8} > 0$. Let $t = \sqrt{C_{8}\ln\{(q+1)/\epsilon\}/n}$, since $\ln (q + 1) = o(n)$, $\exists \; n$ large enough, such that $t^{2}/C_{8} \leq 3t/C_{7}$, then by Bernstein's inequality, we have 
\begin{align}
    \P\left (\left | \frac{1}{n} \sum_{i=1}^{n} V_{ijk} \right |\geq t \right ) = \P \left (\left | \sum_{i=1}^{n} V_{ijk}  \right |\geq nt  \right )  \leq \exp \left ( -\frac{nt^{2}}{C_{8}}\right ) = \frac{\epsilon}{q + 1}.
\end{align}
Let $V_{j}$ denote $\sup_{k = 0, \ldots, q} |1/n\sum_{i=1}^{n} V_{ijk}|$, and 
\begin{align*}
V_{j} < t \Rightarrow \frac{V_{j} \sqrt{n}}{\sqrt{C_{8}\ln\{e(q+1)\}}} \leq \sqrt{1 - \ln(\epsilon)}. 
\end{align*}
It follows that 
\begin{align*}
& \P\left [ \frac{V_{j} \sqrt{n}}{\sqrt{C_{8}\ln\{e(q+1)\}}} \leq \sqrt{1 - \ln(\epsilon)} \right] \geq \P(V_{j} < t ) =  \\
& 1 - \P(V_{j} \geq t) \geq 1 - 
\sum_{k = 0}^{q} \P\left (\left | \frac{1}{n} \sum_{i=1}^{n} V_{ijk} \right |\geq t \right ) = 1 - \epsilon.
\end{align*}
Therefore, $\frac{V_{j} \sqrt{n}}{\sqrt{C_{8}\ln\{e(q+1)\}}} = O_{p}(1)$, for $ j = 0, \ldots, m-1$, it follows that 
\begin{align*}
 & \left \| \frac{1}{n} \sum_{i=1}^{n} \left \{  \frac{1}{\pi^{s}(\tg)}\psi_{1}(\Tilde{\alpha}_{j}^{\T}\bm{G}_{i})\bm{G}_{i}  Z_{ij}  \right \} - \frac{1}{N}\sum_{i=1}^{N}\{\psi_{1}(\Tilde{\alpha}_{j}^{\T}\bm{G}_{i})\bm{G}_{i}Z_{ij}\} \right \|_{\infty}
 =  V_{j}  
 =  O_{p}(\sqrt{\ln(q+1)/n}). 
\end{align*}
Therefore, 
$$|\Delta^{1}_{5,j} + \Delta^{2}_{5,j}| \leq O_{p}(\sqrt{ln(q+1)/n})O_{p}(|S_{\ta}|\sqrt{ln(q+1)/n}) = o_{p}(1/\sqrt{n}),$$
it follows that $\Delta^{1}_{5} + \Delta^{2}_{5} = o_{p}(1/\sqrt{n})$.
Hence, by central limit theorem and continuous mapping theorem, we have 
\begin{align*}
    \sqrt{n}(\hat{\beta}^{s} - \beta^{*}) \xrightarrow{d}\sqrt{n} \Gamma^{-1} \tau^{s}\{\ta,\tb,\pi^{*}(\bm{X})\} \sim \N(0, \bm{\bm{\Sigma}}^{s}). 
\end{align*}
 
\subsection{Variance comparison}\label{sup:vc}
Because the following relationship holds,
\begin{align*}
\frac{1}{n}\bm{\Lambda}^{s} = & \frac{1}{n} \E \left [ \left \{Y - \frac{N-n}{N}\psi(\ta^{\T}\bm{G}) - \frac{n}{N} \psi(\beta^{*\T}\bm{Z})\right \}^{2} \bm{Z} \bm{Z}^{\T} \right ]\\
& + \frac{(N - n)}{N^{2}} \E \left [ \{ \psi(\ta^{\T}\bm{G}) - \psi(\beta^{*\T}\bm{Z}) \}^{2} \bm{Z} \bm{Z}^{\T} \right ]\\
= & \frac{1}{n}\E \left [ \{Y - \psi(\ta^{\T}\bm{G})\}^{2}\bm{Z} \bm{Z}^{\T} + \left( \frac{n}{N}\right )^{2}\{\psi(\ta^{\T}\bm{G}) - \psi(\beta^{*\T}\bm{Z})\}^{2}\bm{Z} \bm{Z}^{\T} \right]  \\
& +  \frac{2}{N} \E \left [\{Y - \psi(\ta^{\T}\bm{G})\} \{\psi(\ta^{\T}\bm{G}) - \psi(\beta^{*\T}\bm{Z})\} \bm{Z} \bm{Z}^{\T} \right ]\\
& + \frac{N - n}{N^{2}} \E \left [ \{ \psi(\ta^{\T}\bm{G}) - \psi(\beta^{*\T}\bm{Z}) \}^{2} \bm{Z} \bm{Z}^{\T} \right ]\\
= & \frac{1}{n}\E \left [ \{Y - \psi(\ta^{\T}\bm{G})\}^{2}\bm{Z} \bm{Z}^{\T} \right] +  \frac{2}{N} \E \left [\{Y - \psi(\ta^{\T}\bm{G})\} \{\psi(\ta^{\T}\bm{G}) - \psi(\beta^{*\T}\bm{Z})\} \bm{Z} \bm{Z}^{\T} \right ]\\
& + \frac{1}{N} \E \left [ \{ \psi(\ta^{\T}\bm{G}) - \psi(\beta^{*\T}\bm{Z}) \}^{2} \bm{Z} \bm{Z}^{\T} \right ]\\
= & \frac{1}{N} \bm{\Lambda},
\end{align*}
we obtain
\begin{align*}
\bm{\Sigma}^{s}/n = \bm{\Gamma}^{-1} \left ( \frac{1}{n}\bm{\Lambda}^{s} \right )\bm{\Gamma}^{-1} =  \bm{\Gamma}^{-1} \left( \frac{1}{N}\bm{\Lambda} \right)\bm{\Gamma}^{-1} =\bm{\Sigma}/N.     
\end{align*}

\section{Details of the application}\label{sec:dp-cc}
\subsection{Pre-processing details of the community and crime dataset}\label{sec:pre-p}
We pre-process the data in following steps:
\begin{itemize}
    \item [Step 1.] remove 22 covariates  missing 84\% of data and 2 variables missing roughly 59\% of data;

    \item [Step 2.] remove covariates with weak linear relationships to the response \texttt{ViolentCrimesPerPop} based on their correlation coefficients. 

    \item [Step 3.] remove covariates that exhibit multi-collinearity based on their values of variance inflation factors. 
\end{itemize}

After the process, we obtain 1993 observations of 26 covariates. 

\subsection{Test results of the covariate shift}\label{sec:test}
\subsubsection{Kernel two-sample test with maximum mean discrepancy}
\begin{itemize}
    \item Kernel: $\exp(-\| \cdot \|^{2}_{2})$
    \item MMD: $0.39227$
    \item P-value: $0.001$
\end{itemize}

\subsubsection{Bootstrap KS-tests for univariate covariates}
\begin{longtable}{| p{.245\textwidth} | p{.205\textwidth} | p{.15\textwidth} | p{.27\textwidth} |} 
\hline
\footnotesize Covariate & \footnotesize Bootstrap-KS P-value & \footnotesize KS-test Statistic& \footnotesize KS-test Approximate P-value   \\ \hline
    \footnotesize\texttt{racePctHisp} & 0.000 & 0.335 & 0.000   \\
    \footnotesize\texttt{pctWWage} & 0.000 & 0.230 & 0.000  \\
    \footnotesize\texttt{pctWInvInc} & 0.000 & 0.330 & 0.000  \\
    \footnotesize\texttt{blackPerCap} & 0.000 & 0.421 &  0.000 \\
    \footnotesize\texttt{PctLess9thGrade} & 0.010 & 0.119 & 0.010  \\
    \footnotesize\texttt{PctUnemployed} & 0.000 & 0.231 & 0.000  \\
    \footnotesize\texttt{PctOccupManu} & 0.000 & 0.205 & 0.000  \\
    \footnotesize\texttt{MalePctDivorce} & 0.000 & 0.373 & 0.000  \\
    \footnotesize\texttt{MalePctNevMarr} & 0.000 & 0.202 & 0.000  \\
    \footnotesize\texttt{PctTeen2Par} & 0.000 & 0.289 &  0.000 \\
    \footnotesize\texttt{PctIlleg} & 0.000 & 0.200 &  0.000 \\
    \footnotesize\texttt{NumImmig} & 0.000 & 0.267 & 0.000  \\
    \footnotesize\texttt{PctImmigRec10} & 0.001 & 0.141 & 0.001  \\
    \footnotesize\texttt{PctHousLess3BR} & 0.000 & 0.218 & 0.000  \\
    \footnotesize\texttt{MedNumBR} & 0.000 & 0.138 &  0.000 \\
    \footnotesize\texttt{HousVacant} & 0.000 & 0.184 &  0.000 \\
    \footnotesize\texttt{PctHousOccup} & 0.000 & 0.195 & 0.000  \\
    \footnotesize\texttt{PctHousOwnOcc} & 0.000 & 0.285 & 0.000  \\
    \footnotesize\texttt{PctVacantBoarded} & 0.797 & 0.040 & 0.923  \\
    \footnotesize\texttt{PctHousNoPhone} & 0.000 & 0.373 &  0.000 \\
    \footnotesize\texttt{PctWOFullPlumb} & 0.000 & 0.146 & 0.000  \\
    \footnotesize\texttt{RentLowQ} & 0.000 & 0.529 & 0.000  \\
    \footnotesize\texttt{MedRentPctHousInc} & 0.062 & 0.089 & 0.099  \\
    \footnotesize\texttt{NumInShelters} & 0.022 & 0.087 & 0.113  \\
    \footnotesize\texttt{NumStreet} & 0.001 & 0.101 & 0.043  \\
    \footnotesize\texttt{PopDens} & 0.000 & 0.266 & 0.000  \\    
\hline 
\caption{Bootstrap KS-tests for univariate covariates}
\end{longtable}
\subsection{Design of basis functions}\label{sec:dbf}
In this application, we design basis functions in the following way:
Given $\{ X_{ij}\}_{j=1}^{N}$, i.e., N samples of the $i$-th coordinate of $\bm{X}$, let $\{\xi_{ij}\}_{j=1}^{n_{k}}$ be the $n_{k}$ points equally spaced within the $[-a_i, b_i]$, where $a_i = \min_{j = 1, \ldots, N} X_{ij}$ and $b_i = \max_{j = 1, \ldots, N} X_{ij}$.
Let $f_{ij}(\bm{X})$ denote $(X_{i}-\xi_{ij})_{+}$, $i = 1, \ldots, d; j = 1, \ldots,n_{k}$.
Let $\bm{F} =  \{1, f_{11}(\bm{X}), \ldots, f_{1n_{k}}(\bm{X}), \ldots,  f_{d1}(\bm{X}), \ldots, f_{dn_{k}}(\bm{X})\}^{\mathrm{\scriptscriptstyle T}}$ be the basis functions of the PS model, and $\bm{G} = \{\bm{F}^{\mathrm{\scriptscriptstyle T}}, (\bm Z \bigotimes \bm{F})^{\mathrm{\scriptscriptstyle T}}\}^{\mathrm{\scriptscriptstyle T}}$. We choose $n_{k} = 4$ in this application.

\end{document}